\documentclass[10pt]{jfm}
\usepackage{graphicx}

\usepackage{multirow}

\newcommand{\enma}[1]   {\ensuremath{#1}}

\newcommand{\beq}{\begin{equation}}
\newcommand{\eeq}{\end{equation}}
\newcommand{\bseq}{\begin{subequations}}
\newcommand{\eseq}{\end{subequations}}
\newcommand{\beqn}{\begin{eqnarray}}
\newcommand{\eeqn}{\end{eqnarray}}
\newcommand{\ba}{\begin{array}}
\newcommand{\ea}{\end{array}}
\newcommand{\bct}{\begin{center}}
\newcommand{\ect}{\end{center}}
\newcommand{\btmz}{\begin{itemize}}
\newcommand{\etmz}{\end{itemize}}
\newcommand{\benum}{\begin{enumerate}}
\newcommand{\eenum}{\end{enumerate}}

\newcommand{\cF}{\enma{\mathcal F}}

\newcommand{\diag}      {\enma{\mathrm{diag}}}

\newcommand{\bv}{{\bf v}}

\newcommand{\matbegin}{
        \left[
}
\newcommand{\matend}{
        \right]
}

\newcommand{\thbo}[3]{
  \matbegin \begin{array}{c}
       #1 \\ #2 \\ #3
       \end{array} \matend }

\newcommand{\tbt}[4]{
  \matbegin \begin{array}{cc}
       #1 & #2 \\ #3 & #4
       \end{array} \matend }
\newcommand{\thbt}[6]{
  \matbegin \begin{array}{cc}
       #1 & #2 \\ #3 & #4 \\ #5 & #6
       \end{array} \matend }

\newcommand{\be}{\begin{equation}}
\newcommand{\ee}{\end{equation}}

\newcommand{\cplxs}{ C\kern -.35em \rule{0.03 em}{.7 ex}~   }

\def\complex{\hbox{C\kern -.45em \rule{0.03 em}{1.5 ex}}~}

\newcommand{\bi}{\begin{itemize}}
\newcommand{\ei}{\end{itemize}}

\usepackage{amsfonts,amssymb,latexsym,color,amsmath,pifont,epsfig,mathdots,mathtools,natbib}
\usepackage{setspace,subfigure}

\newcommand{\cA}{{\cal A}}

\newcommand{\cG}{{\cal G}}

\newcommand{\cX}{{\cal X}}

\newcommand{\cM}{{\cal M}}

\newcommand{\cO}{{\cal O}}

\newcommand{\bbZ}{\mathbb{Z}}

\newcommand{\non}{\nonumber}
\newcommand{\ds}{\displaystyle}
\newcommand{\const}{\mathrm{const.}}

\newcommand{\mrd}{\mathrm{d}}
\newcommand{\mre}{\mathrm{e}}
\newcommand{\mri}{\mathrm{i}}

\newcommand{\fvec}{{\bf f}}

\newcommand{\bu}{{\bf u}}
\newcommand{\bU}{{\bf U}}

\newcommand{\bphi}{\mbox{\boldmath$\phi$}}
\newcommand{\bpsi}{\mbox{\boldmath$\psi$}}

\newcommand{\bkappa}{\mbox{\boldmath$\kappa$}}

\newcommand{\p}{\partial}

\title{Model-based design of transverse wall oscillations for turbulent drag reduction}

\author{Rashad Moarref \and Mihailo R.\ Jovanovi\'c}

\affiliation{Department of Electrical and Computer Engineering, University of Minnesota, Minneapolis, MN 55455, USA}

\shortauthor{R.\ Moarref \& M.\ R.\ Jovanovi\'c}

\begin{document}

\maketitle

    \begin{abstract}
Over the last two decades, both experiments and simulations have demonstrated that transverse wall oscillations with properly selected amplitude and frequency can reduce turbulent drag by as much as $40\%$. In this paper, we develop a model-based approach for designing oscillations that suppress turbulence in a channel flow. We utilize {eddy-viscosity-enhanced} linearization of the {turbulent} flow with control in conjunction with turbulence modeling to determine skin-friction drag in a simulation-free manner. The Boussinesq eddy viscosity hypothesis is used to quantify the effect of fluctuations on the mean velocity in the flow subject to control. In contrast to the traditional approach that relies on numerical simulations, we determine the turbulent viscosity from the second order statistics of the linearized model driven by white-in-time stochastic forcing. The spatial power spectrum of the forcing is selected to ensure that the linearized model for the uncontrolled flow reproduces the turbulent energy spectrum. The resulting correction to the turbulent mean velocity induced by small amplitude wall movements is then used to identify the optimal frequency of drag reducing oscillations. In addition, the control net efficiency and the turbulent flow structures that we obtain agree well with the results of numerical simulations and experiments. This demonstrates the predictive power of our model-based approach to controlling turbulent flows and {is expected to} pave the way for successful flow control at higher Reynolds numbers than currently possible.
    \end{abstract}

\section{Introduction}
\label{sec.intro}

\subsection{{Background}}

Turbulent flows are ubiquitous in nature and engineering. Dissipation of kinetic energy by turbulent flow around airplanes, ships, and submarines increases resistance to their motion. This motivates design of  control strategies for enhancing performance of vehicles and other systems involving turbulent flows. Utility of different approaches for maintaining the laminar flow, reducing skin-friction drag, and preventing separation is surveyed in~\cite{jos98,gad00}. While traditional flow control techniques combine physical intuition with costly numerical simulations and experiments, model-based techniques utilize developments from control theory to improve flow manipulation. Recent research suggests that traditional strategies can be significantly enhanced by flow control design based on analytical models and optimization tools~\citep{kimbew07}.

The effectiveness of model-based feedback~\citep*{hogbewhen03} and sensor-less~\citep*{moajovJFM10,liemoajovJFM10} techniques for controlling the onset of turbulence at low Reynolds numbers stems from their ability to reduce receptivity and enhance robustness of the flow. Model-based approach to flow control design has been motivated by realization that a mechanism which initiates transition is governed by the degradation of robustness~\citep{tretrereddri93,sch07} and the consequential noise amplification~\citep{farioa93,bamdah01,mj-phd04,jovbamJFM05}. Consequently, the above mentioned techniques have utilized Navier-Stokes (NS) equations linearized around the laminar flow as a control-oriented model with the objective of reducing sensitivity to modeling imperfections. {In addition, a fully-developed turbulent channel flow at low Reynolds numbers was relaminarized via a gain-scheduled linear state-feedback controller~\citep*{hogbewhen03b}. Comparison between the impulse responses of the NS equations linearized around the laminar and turbulent mean velocities and the direct numerical simulations (DNS) of the turbulent flow subject to small amplitude impulsive perturbations has been provided by~\cite*{lucquazuc06}.}

{
The role of linear mechanisms in formation and maintenance of streamwise streaks in turbulent flows was examined by~\cite*{leekimmoi90}. It was shown that the streaks are formed by linear amplification of eddies that interact with the large mean shear. For homogenous flows subject to high shear rates,~\cite{leekimmoi90} demonstrated that the DNS-based instantaneous velocity is similar to the velocity that is predicted by the linearized equations in the rapid distortion limit~\citep{pop00}. Furthermore,~\cite{kimlim00} used DNS to demonstrate the importance of the linear vortex tilting mechanism in maintaining the streamwise vortices in a fully-developed turbulent channel flows.
}

{\cite{schhus02} conducted a thorough investigation of the role of near-wall streaks in generation of streamwise vortices. By examining the evolution of infinitesimal fluctuations around a streaky turbulent base flow, they determined the amplitude of streaks above which modal instability occurs. It was noted that only $20\%$ of the streaks in the fully developed turbulent buffer layer are strong enough to trigger modal instability. Consequently, a {\em secondary transient growth\/} mechanism for bypass transition in wall-bounded shear flows was proposed~\citep*{schhus02,hoebrahen05}; this mechanism appears to be capable of producing much larger transient growth rates than the secondary streak-instabilities. Furthermore,~\cite{chebai05} studied the origin of near-wall streaky patterns and demonstrated that the combination of linear effects including lift-up of the mean profile, tilting and stretching by the mean shear, and viscous dissipation induce formation of these patterns. Capability of the linearized NS equations to qualitatively predict both the streak spacing and its dependence on the wall-normal distance was also shown.
}

\subsection{{Previous studies on drag reduction by transverse wall oscillations}}

Several experimental and numerical studies have shown the effectiveness of sensor-less strategies for turbulence suppression in wall-bounded shear flows. The experiments of~\cite{brapon85} and DNS of~\cite{moishidriman90} showed that imposing a constant transverse strain on a turbulent boundary-layer can {\em transiently\/} reduce the turbulent kinetic energy and the Reynolds stresses. Motivated by this observation,~\citet*{junmanakh92} used DNS to establish a {\em sustained\/} turbulence suppression in a channel flow subject to transverse wall oscillations. For the flow with friction Reynolds number $R_\tau = 200$, skin-friction drag reduction of up to $40\%$ was reported with maximum drag reduction taking place for the period of oscillations $T^+ \approx 100$ (in viscous time units).
The numerical results of~\cite{junmanakh92} were experimentally verified by~\citet*{laaskamor94,chodebcla98,cho02,ric04}, where a drag reduction of up to $45\%$ was reported.~\cite{cho02} argued that wall oscillations induce negative spanwise vorticity, thereby suppressing turbulence by hampering the vortex stretching mechanism. In addition, the experiments of~\citet{ric04} showed that the near-wall flow is dragged laterally by wall oscillations which reduces the length of the streaks and increases the spacing between them.~\cite{toules12} also demonstrated that wall oscillations significantly distort the near-wall streaks and reduce the contribution of turbulence to the wall shear stress. Recent DNS study of~\cite{ricotthasqua12} further revealed that wall oscillations directly affect the turbulent dissipation. For $T^+ \lesssim 100$, it was shown that drag reduction scales linearly with the volume integral of an enstrophy production term caused by the spanwise shear layer.

Several alternative mechanisms for inducing transverse oscillations have also been investigated. For example,~\citet{berkimleelim00} used DNS of conductive fluids in a channel flow with $R_\tau = 100$
to show that time-periodic spanwise Lorentz force can reduce skin-friction drag up to $40\%$. The amount of drag reduction was found to decrease for larger $R_\tau$. In addition,~\citet*{dukar00,dusymkar02} studied the effect of Lorentz force in the form of spanwise traveling waves confined to the viscous sub-layer. For $R_\tau = 150$, their DNS showed a drag reduction of up to $30\%$. The drag-reducing mechanisms of transverse motions induced by spanwise oscillations, spanwise traveling waves, and riblets have been surveyed by~\cite{karcho03}. Recently, turbulent drag reduction by waves of spanwise velocity that travel in the streamwise direction has been examined using DNS~\citep*{quaricvio09}, experiments~\citep{autbarbelcamqua10}, and generalized Stokes layer theory~\citep{quaric11}. It was shown that upstream traveling waves reduce drag at any speed. On the other hand, downstream waves reduce drag only at speeds that are much larger or much smaller than the convecting speed of near-wall turbulent structures.

\cite{quasib00} used DNS to show that up to $40\%$ of drag reduction can be achieved by oscillating a cylindrical pipe along its longitudinal axis. In a series of papers,~\cite{barqua96,quaric03,quaric04,ricqua08} further studied drag reduction by transverse wall oscillations in a channel flow; also see recent review by~\cite{qua11}. In addition to quantifying the saved power associated with drag reduction, they accounted for the input power necessary for maintaining wall oscillations; for small oscillation amplitudes, it was established that a net power gain with drag reduction of up to $10\%$ can be achieved. {Furthermore, for the same values of oscillation amplitude and frequency, DNS of~\cite*{choxusun02} showed that the amount of drag reduction in a channel flow can drop by as much as $25\%$ as the friction Reynolds number $R_\tau$ increases from $100$ to $400$. More recent DNS studies of~\cite{ricqua08} and~\cite{toules12} confirmed deterioration in drag reduction by $15\%$ and $18\%$ with increase in $R_\tau$ from $200$ to $400$ and from $200$ to $500$, respectively. On the other hand, for boundary layers subject to plate oscillations with small periods ($T^+ \leq 83$), the experiments of~\cite{ricwu04} showed a weak dependence of drag reduction on the Reynolds number.~\cite{ricqua08} argued that this discrepancy may arise from high experimental uncertainty, effects of geometry, and smaller values of oscillation periods relative to DNS studies.

In contrast to the aforementioned experimental and numerical studies, there are relatively few theoretical developments regarding drag reduction by wall oscillations. We next briefly summarize notable exceptions~\citep{dhasi99,ban06,ricqua08}.~\cite{dhasi99} used exact solutions of the NS equations to study the interactions between the evolving streamwise vortices in a turbulent boundary layer and the Stokes layer induced by wall oscillations. Simulations of the resulting dynamical model showed that these interactions reduce the Reynolds stresses, the turbulence production, and the skin friction drag.~\cite{ban06} proposed a vorticity reorientation hypothesis and showed that the negative spanwise vorticity induced by the wall oscillations modifies the orientation of the total vorticity field in the near wall region and suppresses turbulence production. Furthermore, it was shown that the developed model yields drag reduction that agrees well with experimental data. Finally, in addition to conducting a thorough DNS study,~\cite{ricqua08} used the solution to the laminar Stokes problem to quantify the dependence of the drag and the input power necessary for maintaining wall oscillations on their period and amplitude.

\subsection{{Preview of key results}}

While most model-based efforts to date have considered the problem of maintaining the laminar flow {and relaminarization}, in this paper we show that turbulence modeling in conjunction with {eddy-viscosity-enhanced} linearization can extend utility of these methods to control of turbulent flows. Control-oriented turbulence modeling is challenging because of complex flow physics that arises from strong interactions between the turbulent fluctuations and the mean velocity. We build on recent research that demonstrates considerable predictive power of {nonmodal stability} analysis~{\citep{sch07}} even in turbulent flows~\citep{alajim06,cospujdep09,pajgarcosdep09}. These papers have shown that the equations linearized around turbulent mean velocity, with molecular viscosity augmented by turbulent viscosity, qualitatively capture features of turbulent flows with no control. For the flow with control, we examine the class of linearized models considered by~\citet{alajim06,cospujdep09,pajgarcosdep09} and use turbulent viscosity hypothesis to quantify the influence of turbulent fluctuations on the mean velocity. {We demonstrate the ability of this approach to quantitatively predict the effect of control on turbulent drag.}

The difficulty here arises from the fact that the turbulent viscosity of the flow with control has to be determined. Even though we use the {Boussinesq hypothesis} to capture the influence of control on turbulent viscosity, in contrast to current practice we do not rely on numerical simulations for finding
turbulent kinetic energy $k$ and its rate of dissipation $\epsilon$. Instead, we introduce a simulation-free method based on stochastically-forced linearized model of controlled flow to obtain $k$ and $\epsilon$ from the second-order statistics of velocity fluctuations. These statistics are used to determine the turbulent viscosity for the flow with control, and thereby to compute the effect of control on the turbulent mean velocity and on the skin-friction drag.

We utilize linearized equations subject to white-in-time stochastic forcing with appropriately selected second-order spatial statistics. Using analogy with homogenous isotropic turbulence~\citep{jovgeoAPS10}, we select these to be proportional to the two-dimensional energy spectrum of the uncontrolled flow. Note that while our approach takes advantage of the turbulent viscosity and the energy spectrum resulting from direct numerical simulations (DNS) of the uncontrolled flow~\citep*{kimmoimos87,moskimman99,deljim03,deljimzanmos04}, {\em we do not rely on numerical or experimental data for determining the effect of control on the turbulent flow.} To the best of our knowledge, the present work is the first to utilize publicly available DNS data of the uncontrolled flow to guide control-oriented modeling of turbulent flows. Even though the aforementioned databases provide coarse one-point correlations in the wall-normal direction, we demonstrate that they can be effectively employed for model-based flow control design.

In this paper, we use a model-based approach to examine the effect of transverse wall oscillations on the dynamics of a turbulent channel flow. We {start by showing} that the power necessary for maintaining wall oscillations increases quadratically with their amplitude, which is in agreement with DNS of~\citet{quaric04} {and theoretical study of~\cite{ricqua08}.} {Since} large control amplitudes yield poor net efficiency, we confine our study to small oscillation amplitudes and use perturbation analysis (in the amplitude of oscillations) to identify the period of oscillations that achieves largest drag reduction in a computationally efficient manner. In addition, we quantify the net efficiency, discuss the drag reduction mechanisms {and the effects of the Reynolds number}, and compare the dominant structures in flows with and without control. The close agreement between our results and the results obtained in experiments and DNS~\citep{junmanakh92,barqua96,cho02,quaric04,ricqua08} demonstrates the predictive power of our model-based approach to flow control design.

Our presentation is organized as follows: in \S~\ref{sec.problem}, we formulate the problem and provide a brief overview of the governing equations, turbulent mean velocity, skin-friction drag coefficient, net efficiency, and the {Boussinesq eddy viscosity hypothesis}. In~\S~\ref{sec.linearized-turbulent}, we use {eddy-viscosity-enhanced} stochastically forced linearized model to study the dynamics of infinitesimal fluctuations around the turbulent base flow. We also describe an efficient method for computing the second-order statistics of fluctuations. These statistics are used to determine the influence of control on turbulent viscosity. In~\S~\ref{sec.results}, we apply our theoretical developments to the problem of turbulent drag reduction with transverse wall oscillations, and provide a thorough analysis of the effect of control on skin-friction drag and net efficiency. In~\S~\ref{sec.control-structures}, we use characteristic eddy decomposition to visualize the effect of control on turbulent flow structures. We conclude the paper with a brief summary of our contributions and outlook for future research in~\S~\ref{sec.remarks}.

\section{Problem formulation}
\label{sec.problem}

      \begin{figure}
        \begin{center}
        \begin{tabular}{cc}
        \subfigure[]{\includegraphics[height=2.6cm]
                {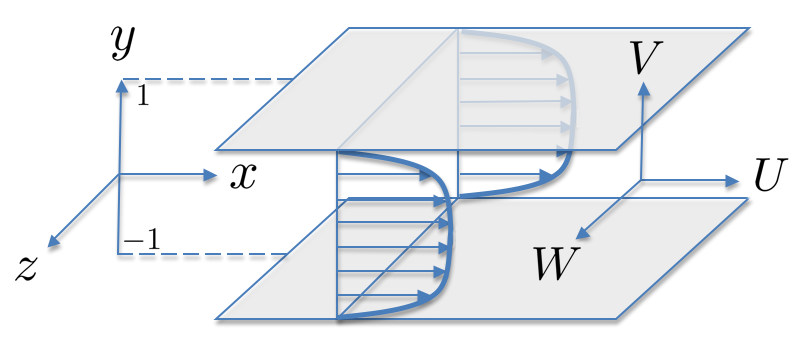}
        \label{fig.flow}}
        &
        \subfigure[]{\includegraphics[height=3cm]
                {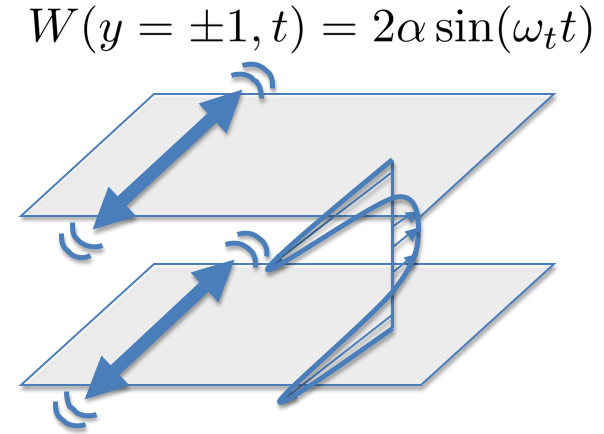}
        \label{fig.wall-osc}}
        \end{tabular}
        \end{center}
        \caption{(a) Pressure driven channel flow; and (b) Channel flow subject to transverse wall oscillations.}
      \end{figure}

The pressure-driven channel flow of incompressible Newtonian fluids, with geometry shown in figure~\ref{fig.flow}, is governed by the non-dimensional NS and continuity equations
    \be
    \ba{rcl}
    \bu_t
    &\!\! = \!\!&
    - (\bu \cdot \nabla) \bu
    \, - \,
    \nabla P
    \, + \, (1/R_\tau) \, \Delta \bu
    ,
    \\[0.1cm]
    0
    &\!\! = \!\!&
    \nabla \cdot \bu,
    \label{eq.NScts}
    \ea
    \ee
where $\bu$ is the velocity, $P$ is the pressure, $\nabla$ is the gradient, and $\Delta = \nabla \cdot \nabla$ is the Laplacian. The Reynolds number $R_\tau = u_\tau h / \nu$ is defined in terms of the channel's half-height $h$ and the friction velocity $u_\tau = \sqrt{\tau_w/\rho}$, $(x, y, z)$ are the streamwise, wall-normal, and spanwise coordinates, and $t$ is time. Here, $\tau_w$ is the wall-shear stress {(averaged over horizontal directions and time)}, $\rho$ is fluid density, and $\nu$ is kinematic viscosity. In~(\ref{eq.NScts}) {and throughout the paper}, spatial coordinates are non-dimensionalized by $h$, velocity by $u_\tau$, time by $h/u_\tau$, and pressure by $\rho u_\tau^2$. When normalized by $\nu/u_\tau$, the wall-normal coordinate is denoted $y^+ = R_\tau (1+y)$. The subscripts {$x$, $y$, $z$, and $t$} are used to denote the spatial and temporal derivatives, e.g., $\bu_t = \partial \bu/\partial t = \partial_t \bu$.

Throughout the paper we assume that the bulk flux, which is obtained by integrating the streamwise velocity over spatial coordinates {and time}, remains constant. This constraint is commonly imposed in experiments and DNS of turbulent flows and it can be enforced by adjusting the uniform streamwise pressure gradient $P_x$. In addition to the driving pressure gradient, which balances the wall-shear stress~\citep{mccomb91}, the flow is also subject to zero-mean transverse wall oscillations of amplitude $\alpha$ and frequency $\omega_t$; see figure~\ref{fig.wall-osc}. The period of oscillations normalized by $h/u_\tau$ (outer units) is given by $T = 2\pi/\omega_t$, which is equivalent to $T^+ = R_\tau \, T$ when normalized by $\nu/u_\tau^2$ (viscous units). The streamwise and wall-normal velocities satisfy no-slip and no-penetration boundary conditions at the walls.

{Reynolds decomposition separates} the velocity in a turbulent flow into the sum of the turbulent mean velocity, $\bU = [\,U~V~W\,]^T$, and the fluctuations around $\bU$, $\bv = [\,u~v~w\,]^T$,
	\[
	\bu
	\; = \;
	\bU
	\, + \,
	\bv,
	~~
	\bU
	\; = \;
	\left<\bu\right>,
	~~
	\left<\bv\right>
	\; = \;
	0.
	\]
{Here, $\left< \,\cdot\, \right>$ denotes the expectation operator,
	\be
	\langle \bu (x,y,z,t) \rangle
	\; = \;
	\underset{t_f \, \rightarrow \, \infty}{\lim} \,
	\dfrac{1}{t_f} \,
	\ds{\int_{0}^{t_f}} \bu (x,y,z,t + \tau) \, \mrd \tau.
	\non
	\ee}
This {decomposition} yields the Reynolds-averaged Navier-Stokes (RANS) equations for {the turbulent mean velocity}~\citep{mccomb91,durpet00,pop00},
	\be
       \ba{rcl}
	\bU_t
	& \!\! = \!\! &
	- \,
	\left(\bU \cdot \nabla \right) \bU
	\, - \,
	\nabla \left< P \right>
    \, + \,
    (1/R_\tau) \, \Delta {\bU}
    \, - \,
	\nabla \cdot \left< \bv \bv^T \right>,
       \\[0.15cm]
       0
    	& \!\! = \!\! &
    	\nabla \cdot \bU.
    	\ea
	\label{eq.turb-mean}
	\ee
{In a turbulent flow}, the second-order statistics of fluctuations $\langle \bv \bv^T \rangle$, i.e., the Reynolds stresses, introduce additional flux. The Reynolds stress tensor quantifies the transport of momentum arising from turbulent fluctuations and it has profound influence on the mean velocity, and thereby on skin-friction drag~\citep{mccomb91}. The difficulty in determining statistics of fluctuations arises from the nonlinearity in the NS equations which makes the $n$th velocity moment depend on the $(n+1)$th moment~\citep{mccomb91}.

\subsection{The turbulent mean velocity}
\label{sec.turbulent-viscosity}

The closure problem in~(\ref{eq.turb-mean}) can be overcome by expressing the higher order moments in terms of the lower-order moments. According to the Boussinesq eddy viscosity hypothesis~\citep{mccomb91,durpet00,pop00}, the turbulent momentum is transported in the direction of the mean rate of strain,
	\be
	\left< \overline{\bv \bv^T} \right>
	\, - \,
	\dfrac{1}{3} \, \mathrm{tr} \left( \left< \overline{\bv \bv^T} \right> \right) I
	\; = \;
	\, - \,
	\dfrac{\nu_T}{R_\tau} \,
	\left(
	\nabla \bU
	\, + \,
	(\nabla \bU)^T
	\right),
	\label{eq.turb-vis}
	\ee
where $\nu_T (y)$ is the turbulent viscosity normalized by $\nu$, overline denotes averaging over $x$ and $z$, $\mathrm{tr} \, (\cdot)$ is the trace of a given tensor, and $I$ is the identity tensor. In the flow subject to wall oscillations, $\nu_T$ is a function of both $y$ and $t$. However, in this paper, we consider the averaged effect of control (over one period of oscillations $T$) on $\nu_T$. This does not mean that the time-periodic flow is treated as time-independent. In fact, the approach outlined in~\S~\ref{sec.compute-correlations} and in Appendix~\ref{sec.variance} facilitates analysis of the time-periodic flow.

The steady-state solution (after the influence of initial conditions disappears) of the system~(\ref{eq.turb-mean})-(\ref{eq.turb-vis}) subject to a uniform pressure gradient, $P_x$, and the transverse wall oscillations,
    \[
    W (y \, = \, \pm 1, t)
	\; = \;
	2 \, \alpha \sin \left( \omega_t \, t \right),
    \]
is determined by $[\,U (y)~~0~~W (y,t)\,]^T$. It can be shown that the streamwise mean velocity averaged over one period of oscillations, $U (y)$, arises from the uniform pressure gradient, and that the spanwise mean velocity, $W (y,t)$, is induced by the wall oscillations,
	\begin{subequations}
   	\be
	\left\{
	\ba{l}
	0
	\; = \;
	\left(
	(1 + \nu_T (y)) \, U_y (y)
	\right)_y
	\, - \,
	R_\tau \, P_x,
	\\[0.15cm]
	U (y \, = \, \pm 1)
	\; = \;
	0,
	\ea
	\right.
	\label{eq.turb-mean-x}
	\ee
	\\[-0.75cm]
	\be
	\left\{
	\ba{l}
	R_\tau \, W_t (y,t)
	\; = \;
	\left(
	(1 + \nu_T (y)) \, W_y (y,t)
	\right)_y
	\\[0.15cm]
	W (y \, = \, \pm 1, t)
	\; = \;
	2 \, \alpha \sin \left( \omega_t \, t \right).
	\ea
	\right.
	\label{eq.turb-mean-z}
	\ee
	\label{eq.turb-mean-x-z}
	\end{subequations}
\hspace{-0.16cm}Here, $(1  \,+  \,\nu_T)$ represents an effective viscosity that accounts for both molecular and turbulent dissipation~\citep{pop00}.

For given turbulent viscosity $\nu_T$ and driving pressure gradient $P_x $,~(\ref{eq.turb-mean-x-z}) is an uncoupled system of equations for $U$ and $W$; $U$ can be obtained by solving~(\ref{eq.turb-mean-x}),
	\be
	U (y)
	\; = \;
	{R_\tau P_x}
	\, \ds{\int_{-1}^{y}} \,
	\dfrac{\xi}{1 + \nu_{T} (\xi)} \, \mrd \xi,
	\label{eq.U}
	\ee
and $W$ can be obtained by solving~(\ref{eq.turb-mean-z}). The imposed boundary conditions induce a $y$-dependent time-periodic spanwise velocity,
	\be
	W (y, t)
	\; = \;
	\alpha \, (
	{W}_{p} (y) \, \mre^{\mri \omega_t \, t}
	\, + \,
	{{W}_{p}^*} (y) \, \mre^{-\mri \omega_t \, t}
	),
	\label{eq.U_3}
	\ee
where $\mri = \sqrt{-1}$, {and $^*$ denotes the complex conjugate.} An equation for $W_p (y)$ is obtained by substituting~(\ref{eq.U_3}) in~(\ref{eq.turb-mean-z})
	\be
	\ba{rcl}
	\mri \, R_\tau \, \omega_t \, W_{p} (y)
	&\!\! = \!\! &
	\left( 1 + \nu_T (y) \right) W''_{p} (y)
	\, + \,
	\nu'_T (y) \, W'_{p} (y),
    	\\[0.15cm]
    	W_{p} (\pm1)
    	&\!\! = \!\! &
    	-\mri,
	\ea
	\label{eq.U3-bar}
	\ee
{where prime represents differentiation with respect to $y$.} The expression for $W_{p} (y)$ is readily obtained from a solution to the resulting two point boundary value problem~(\ref{eq.U3-bar}).

The difficulty in determining $U$ and $W$ from~(\ref{eq.turb-mean-x-z}) arises from the fact that $\nu_T$ depends on the fluctuations around the turbulent mean velocity, and thus it is not known {\em a priori\/}. A significant body of work has been devoted to finding an expression for $\nu_T$ that yields the turbulent mean velocity in the uncontrolled flow~\citep{mal56,ces58,reytie67}.~\cite{reytie67} extended the turbulent viscosity model, originally introduced by~\cite{ces58}, from the pipe flow to the channel flow,
	\be
	\nu_{T0} (y)
	\; = \;
	\dfrac{1}{2}
	\,
	\left( \left(1 \, + \, \left(
	\dfrac{c_2}{3} \, R_\tau \, (1 - y^2) \, (1 + 2 y^2) \,
	(1 - \mre^{-(1 - |y|) \, R_\tau / c_1}) \right)^2
	\right)^{1/2} - 1  \right).
	\label{eq.nuRT}
	\ee
This expression employs the law of the wall in conjunction with van Driest's damping function and Reichardt's middle law~\citep{reytie67}. The parameters $c_1$ and $c_2$ appear in the van Driest's wall law and in the von K{\'a}rm{\'a}n's log law~\citep{pop00}, respectively. These two parameters are selected to minimize least squares deviation between the mean streamwise velocity~(\ref{eq.U}) obtained with ${P_x = -1}$ and turbulent viscosity~(\ref{eq.nuRT}), and the mean streamwise velocity obtained in experiments or in simulations. Application of this procedure yields {$\{R_\tau = 186, c_1 = 46.2, c_2 = 0.61; R_\tau = 547, c_1 = 29.4, c_2 = 0.45; R_\tau = 934, c_1  = 27, c_2 = 0.43\}$} for the corresponding mean velocities in a turbulent channel flow resulting from DNS~\citep{deljim03,deljimzanmos04}.

Under the assumption that the turbulent viscosity~(\ref{eq.nuRT}) captures {the effect of background turbulence on the mean velocity}, the system of equations~(\ref{eq.turb-mean-x-z})-(\ref{eq.nuRT}) yields a solution $\bU_0 = [\,U_{0} (y)~0~W_{0} (y,t)\,]^T$. By construction, $U_0$ approximates the mean streamwise velocity in the uncontrolled turbulent flow, and $W_0$ is the spanwise velocity induced by the wall oscillations and obtained under the assumption that the turbulent viscosity is not modified by control. {As discussed below, this strong assumption is {\em only\/} used as a starting point for our analysis.} Figure~\ref{fig.Wp0-real-imag-Tplus-R186} shows that $W_{p,0} (y^+; T^+)$ in flow with $R_\tau = 186$ (solid curves) is localized in the viscous wall region, $y^+ < 50$. As expected from the analogy to the solution of the Stokes second problem (for example, see~\cite{pan96}), $W_{p,0}$ shifts away from the wall as $T^+$ increases. {In addition, for small values of $T^+$, we observe close correspondence between $W_{p,0}$ and the solution to the laminar Stokes problem (dashed curves) determined by~\citet{ricqua08}; both of these are in agreement with the DNS results of~\cite{quasib00,choxusun02,quaric03}. For larger values of $T^+$, the discrepancy between these solutions increases.~\citet{ricqua08} argued that in this case the rapidly-varying time evolution of the near-wall turbulent structures is not properly captured by the slowly-varying Stokes layer.
}

{Therefore,} if the turbulent viscosity of the uncontrolled flow $\nu_{T0}$ is used to model $\nu_T$, the oscillations induce $W_0$ but have no impact on $U_0$ (which, in this case, arises only from the uniform pressure gradient). The implications of this assumption for determining the skin-friction drag coefficient and the control net efficiency are discussed in~\S~\ref{sec.drag-power} {where we demonstrate the necessity of accounting for the effect of control on the turbulent viscosity.}

    \begin{figure}
    \begin{center}
    \begin{tabular}{cc}
    $\mbox{Re} (W_{p,0} (y^+; T^+))$
    &
    $\mbox{Im} (W_{p,0} (y^+; T^+))$
    \\[-0.2cm]
    \subfigure[]{\includegraphics[width=0.49\columnwidth]
    {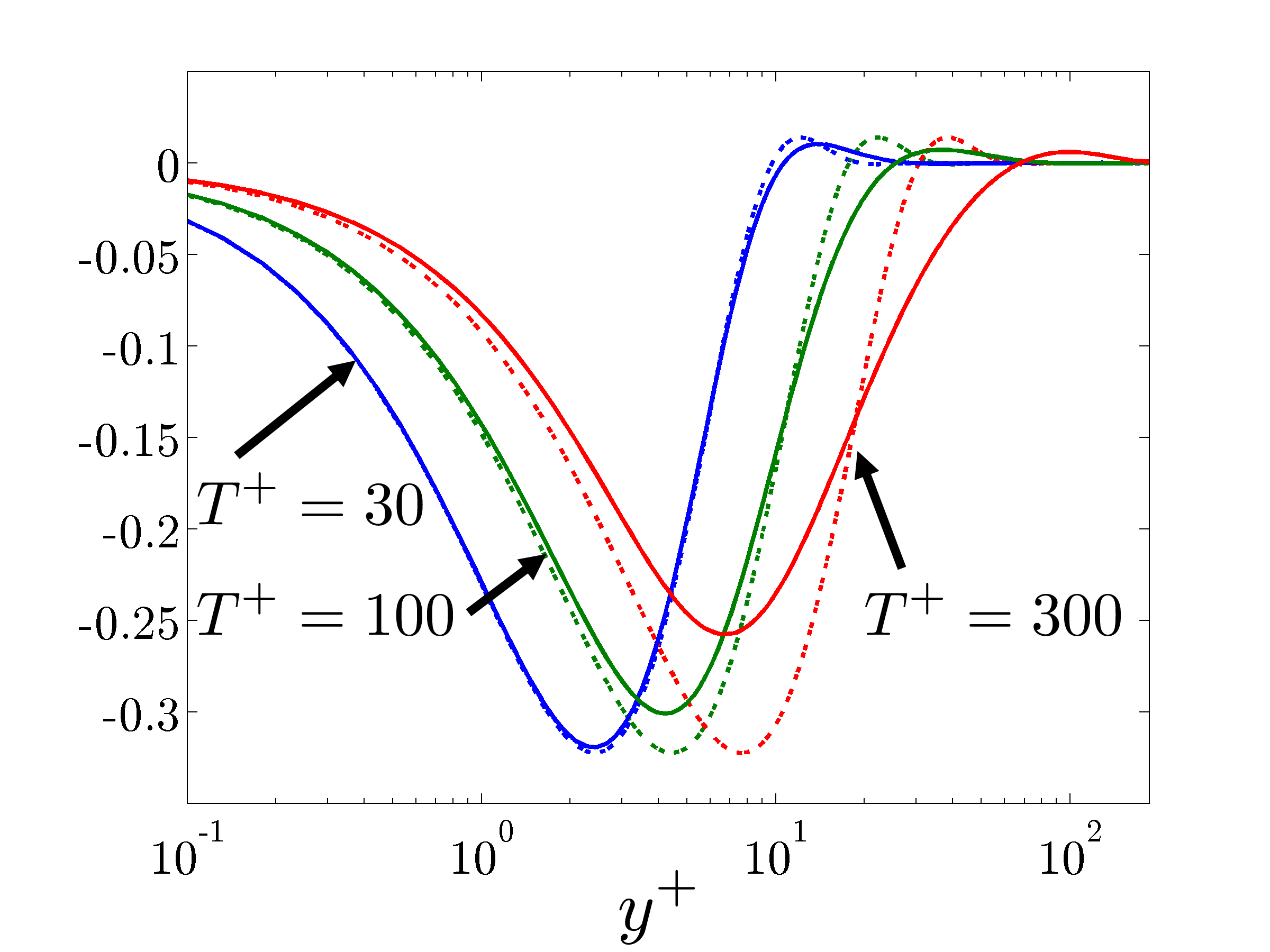}
    \label{fig.Wp0-real-Tplus-R186}}
    &
    \subfigure[]{\includegraphics[width=0.49\columnwidth]
    {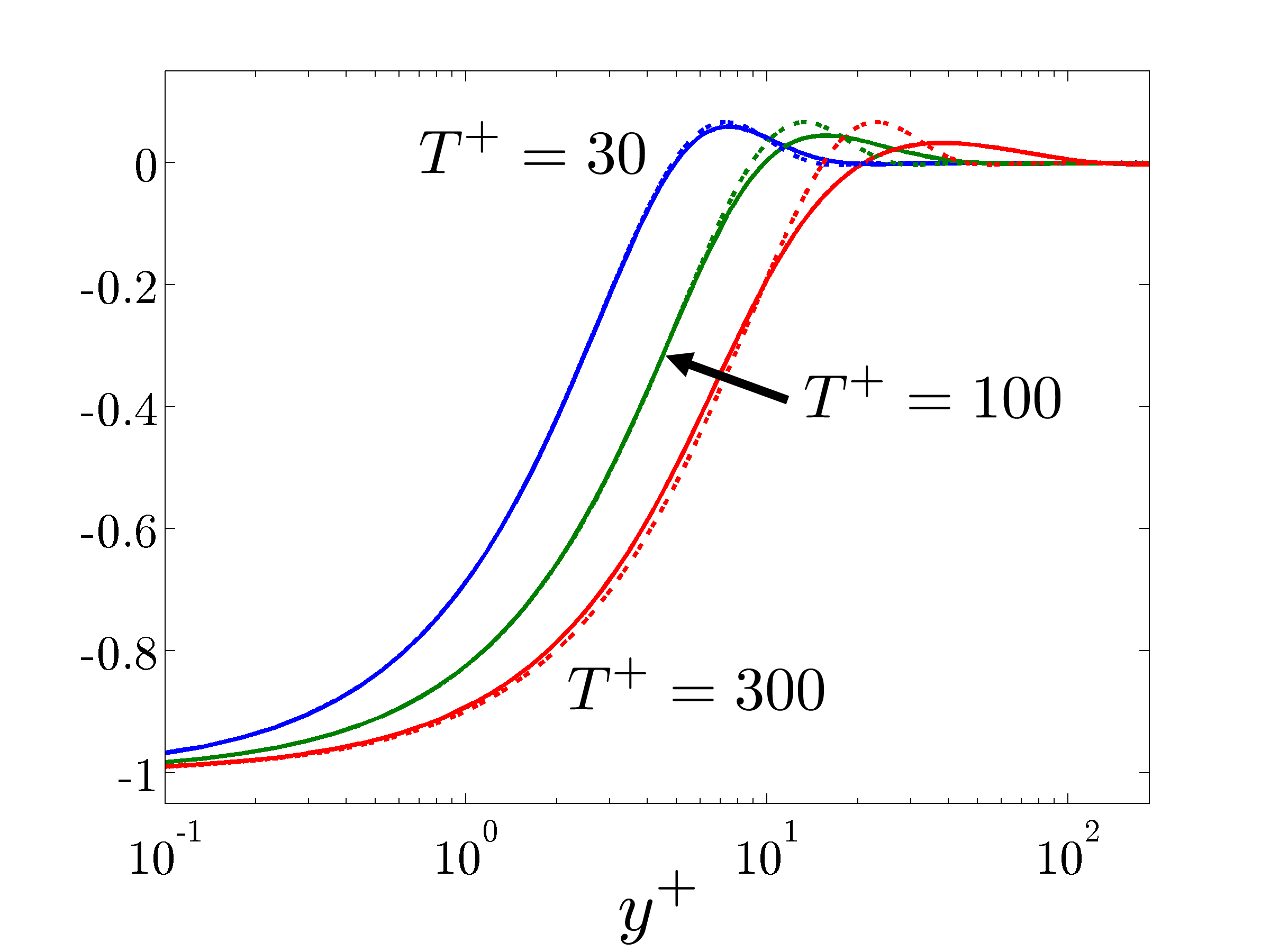}
    \label{fig.Wp0-imag-Tplus-R186}}
    \end{tabular}
    \end{center}
    \caption{
    (Color online available at {\sf journals.cambridge.org/flm}) (a) The real part; and (b) the imaginary part of the solution $W_{p,0} (y^+; T^+)$ to~(\ref{eq.U3-bar}) (solid curves) in the flows with $R_\tau = 186$ and $T^+ = 30$ (blue), $100$ (green), and $300$ (red). The solution $W_{p,0} (y^+; T^+)$ is obtained under the assumption that the turbulent viscosity of the uncontrolled flow captures the effect of fluctuations on the mean velocity. The solutions of the laminar Stokes problem are shown for comparison (dashed curves).
    }
    \label{fig.Wp0-real-imag-Tplus-R186}
    \end{figure}

\subsection{Skin-friction drag coefficient and net efficiency}
\label{sec.drag-power}
	
As mentioned in~\S~\ref{sec.problem}, the pressure gradient $P_x$ has to be adjusted in order to maintain the constant bulk flux,
    \[
	U_B
	~ = ~
	\dfrac{1}{2} \, \int_{-1}^{1} U (y) \, \mrd y
    ~ = ~
    \const
	\]
Since the skin-friction drag coefficient is proportional to $|P_x|$ and inversely proportional to $U_B^2$~\citep{mccomb91,pan96},
	\be
	C_f
	\; = \;
	2 \, {|P_x|}/U_B^2,
	\non
	\ee
for the flow with constant $U_B$, reduction (increase) in {$|P_x|$} induces drag reduction (increase).  The change in the skin-friction coefficient relative to the uncontrolled flow is given by
	\be
	\% C_f
	\; = \;
	100 \, \dfrac{C_{f,u} \, - \, C_{f,c}}{C_{f,u}}
	\; = \;
	100 \,
	{(1 \, + \, P_{x,c})},
	\label{eq.delta-Cf}
	\ee
where the subscripts $u$ and $c$ denote the quantities in the uncontrolled and controlled flows, respectively. Thus, the control leads to drag reduction when {$P_{x,c} > -1$}.

The drag reduction induces saving in power (per unit area of the channel surface and normalized by $\rho u_\tau^2$)~\citep{cur03}
	\be
	\Pi_{\mathrm{save}}
	\; = \;
	2 \, U_B \, {(1 \, + \, P_{x,c})}.
	\non
	\ee
Compared to the power required to drive the uncontrolled flow, $\Pi_u = 2 U_B$, the saved power is determined by the relative change in the skin-friction coefficient,
	\be
	\% \Pi_{\mathrm{save}}
	\; = \;
	100 \, \dfrac{\Pi_{\mathrm{save}}}{2 U_B}
	\; = \;
	100 \, {(1 \, + \, P_{x,c})}
    	\; = \;
    	\% C_f.
	\non
	\ee
On the other hand, an input power is required for balancing the spanwise shear stresses at the walls~\citep{cur03}. The required power exerted by wall oscillations expressed in fraction of the power necessary to drive the uncontrolled flow is given by (see Appendix~\ref{sec.Preq-A})
	\be
	\% \Pi_{\mathrm{req}}
	\; = \;
    	100 \,
	\dfrac{\alpha^2}{R_\tau U_B} \,
	\mbox{Im}
	\left(
	\left.
	W'_{p}
	\right|_{y \, = \, -1}
	\, - \,
	\left.
	W'_{p}
	\right|_{y \, = \, 1}
	\right),
	\label{eq.delta-Preq}
	\ee
where $\mbox{Im} \left( \cdot \right)$ denotes the imaginary part of a complex number. The net efficiency of control is quantified by the difference of the saved and required powers
	\be
	\% \Pi_{\mathrm{net}}
	\; = \;
	\% \Pi_{\mathrm{save}}
	\, - \,
	\% \Pi_{\mathrm{req}}.
	\non
	\ee
Since the net efficiency is obtained from $U$ and $W$, determining the turbulent mean velocities is essential for assessing the efficiency of wall oscillations.

    \begin{figure}
    \begin{center}
    \begin{tabular}{cc}
    {
    $\% \Pi_{\mathrm{req},0} (T^+)$; $\% \Pi_{\mathrm{req}} / \alpha^2$
    }
    &
    {
    $\% \Pi_{\mathrm{req},0} (T^+)$; $\% \Pi_{\mathrm{req}} / \alpha^2$
    }
    \\[-0.2cm]
    \subfigure[]{\includegraphics[width=0.49\columnwidth]
    {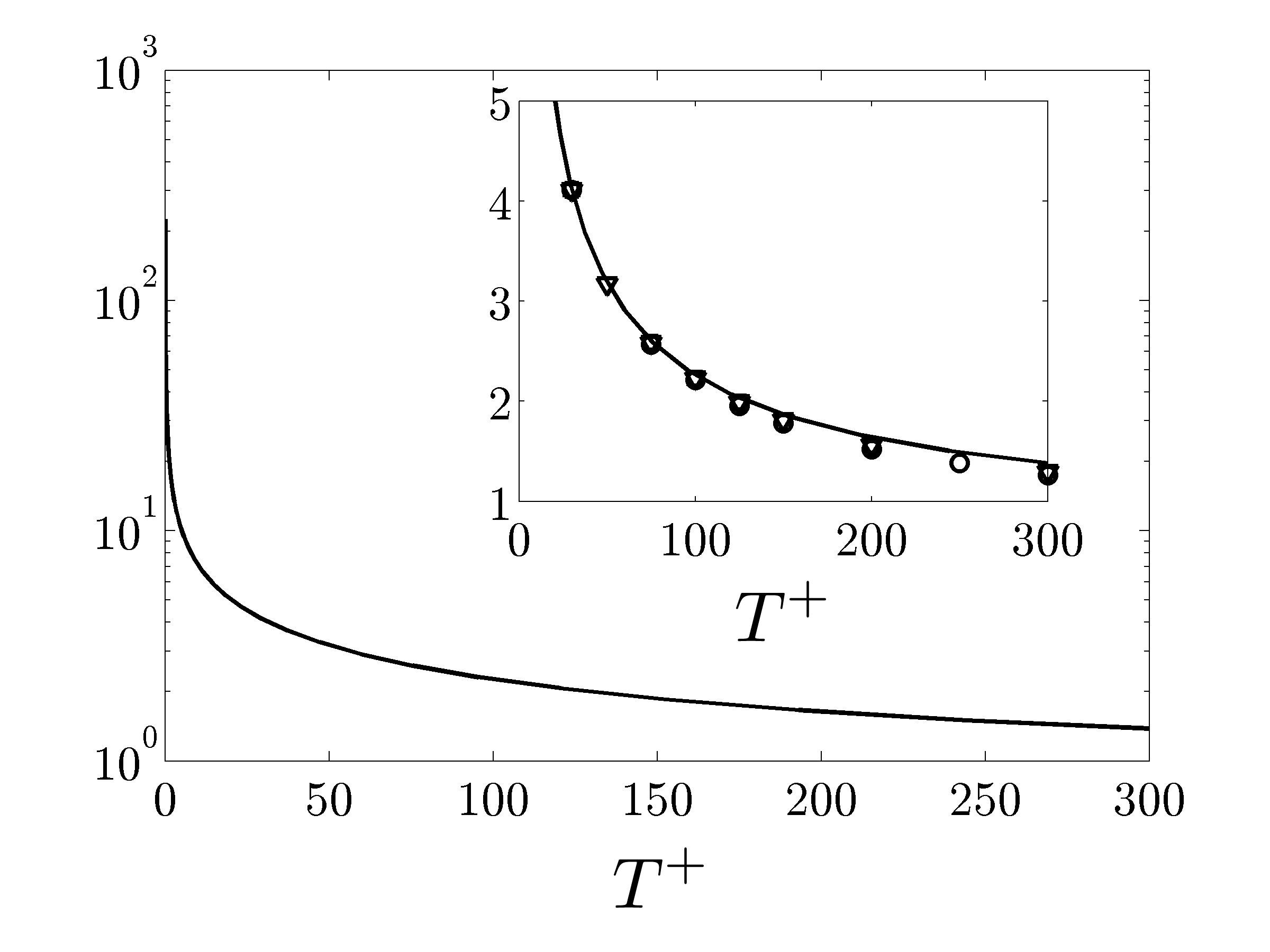}
    \label{fig.Preq-R186-quadrio-txt}}
    &
    \subfigure[]{\includegraphics[width=0.49\columnwidth]
    {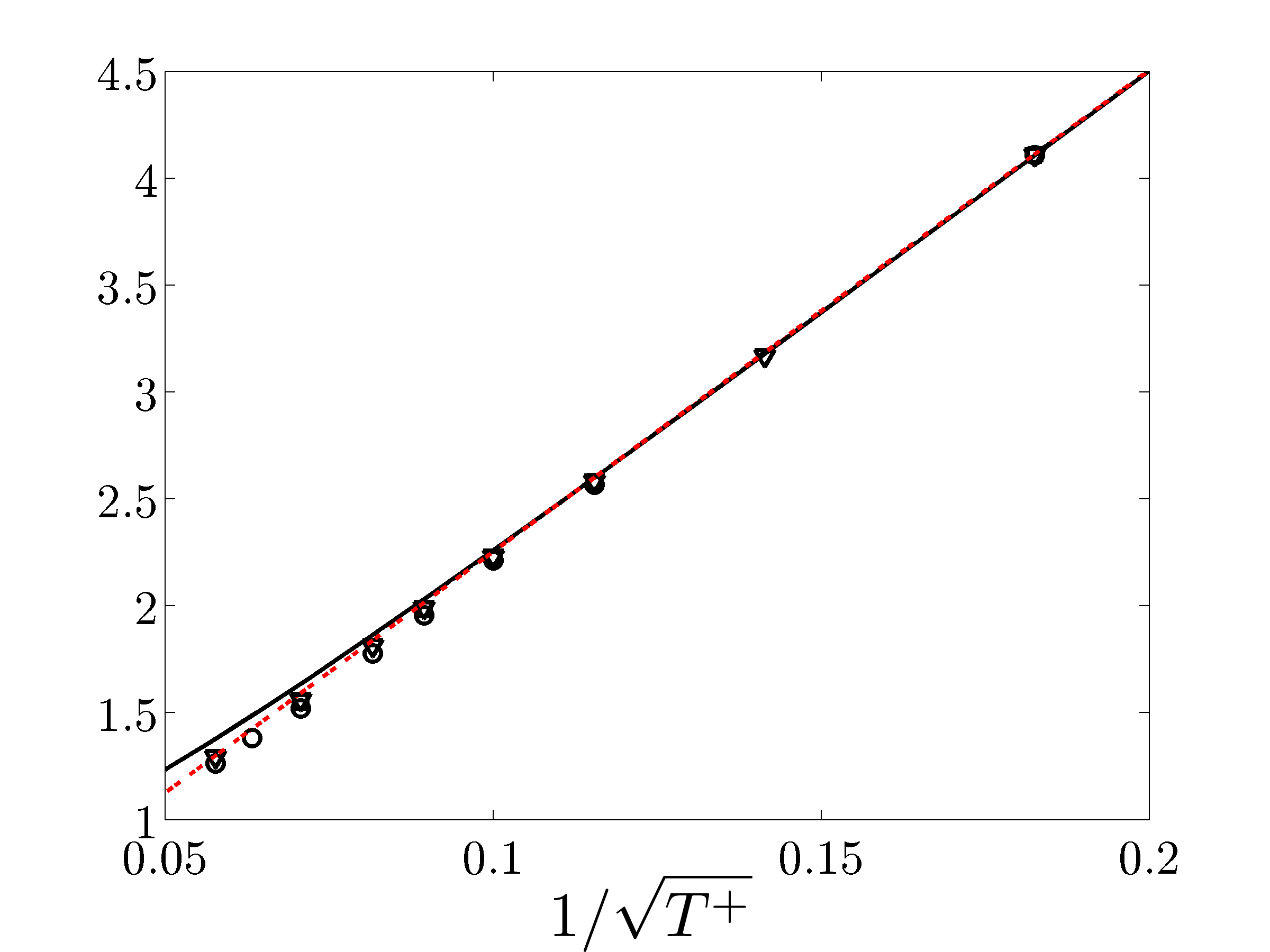}
    \label{fig.Preq-sqrt-R186-quadrio-txt}}
    \end{tabular}
    \end{center}
    \caption{
    (Color online) (a) (Solid curve) The required power, $\% \Pi_{\mathrm{req},0} (T^+)$, as a function of the period of oscillations $T^+$ for the flow with $R_\tau = 186$;
    (Symbols) DNS-based $\% \Pi_{\mathrm{req}} (T^+; \alpha)$ normalized by $\alpha^2$ at $R_\tau = 200$~\citep{quaric04} for oscillation amplitudes $\alpha = 2.25,~\circ$; $\alpha = 6,~\square$; and $\alpha = 9,~\triangledown$;
    (b) The solid curve and the symbols show the dependence on $1/\sqrt{T^+}$ for the same quantities as in figure~\ref{fig.Preq-R186-quadrio-txt}. The red dotted curve shows the required power obtained from the solution to the laminar Stokes problem~\citep{ricqua08}.
    }
    \label{fig.Preq0-R186-2}
    \end{figure}

For the spanwise mean velocity $W_0$ determined in \S~\ref{sec.turbulent-viscosity}, the required power grows quadratically with $\alpha$,
	\be
    \% \Pi_{\mathrm{req}}
	\; = \;
	\alpha^2 \, \% \Pi_{\mathrm{req},0},
    ~~
	\% \Pi_{\mathrm{req},0}
	\; = \;
    	100 \,
	\dfrac{1}{R_\tau U_B} \,
	\mbox{Im}
	\left(
	\left.
	W'_{p,0}
	\right|_{y \, = \, -1}
	\, - \,
	\left.
	W'_{p,0}
	\right|_{y \, = \, 1}
	\right).
	\label{eq.Preq0}
	\ee
Figure~\ref{fig.Preq-R186-quadrio-txt} shows that $\% \Pi_{\mathrm{req},0} (T^+)$ decreases monotonically with $T^+$ {and that small values of $T^+$ require prohibitively large input power. The inset demonstrates close correspondence between $\% \Pi_{\mathrm{req},0}$ (solid curve) and the DNS results of~\cite{quaric04} (symbols). Using the solution of the laminar Stokes problem,~\cite{ricqua08} showed that the required power scales as $\alpha^2/\sqrt{T^+}$. For $T^+ \lesssim 150$ (or, equivalently, for $1/\sqrt{T^+} \lesssim 0.08$), figure~\ref{fig.Preq-sqrt-R186-quadrio-txt} illustrates that our approach leads to similar scaling trends. For larger values of $T^+$, $\% \Pi_{\mathrm{req},0} (T^+)$ deviates from the value obtained using the laminar Stokes problem (red dotted curve). We note that for $T^+ \gtrsim 150$, the solution of~\cite{ricqua08} provides better agreement with the DNS results. This implies that the Reynolds stress that drives the evolution of $W$ in RANS is not as strongly correlated with $W_y$ as suggested by the turbulent viscosity of the uncontrolled flow. In~\S~\ref{sec.Preq} we show that accounting for the effect of control on the turbulent viscosity improves the prediction of the required power resulting from the use of $W_0$.
}

The apparent lack of influence of the wall movements on $U_0$, observed in \S~\ref{sec.turbulent-viscosity}, is at odds with experiments and simulations that have shown that properly designed oscillations can reduce drag by as much as $40\%$~\citep{junmanakh92,barqua96,laaskamor94,chodebcla98,cho02,quaric03,quaric04,ric04}. Thus, model-based control of turbulent flows requires thorough examination of the influence of control on $\nu_T$. For spanwise wall oscillations we address this problem in \S~\ref{sec.nuT-control}.

\subsection{{The model equation for $\nu_T$}}
\label{sec.k-eps-model}

Direct numerical simulations can be used to study the effect of control on turbulent flows. However, resolving all scales of motion at large Reynolds numbers may be prohibitively expensive, which motivates use of the Reynolds-averaged equations in conjunction with turbulence modeling. The challenge here is to establish a relation between $\nu_T$ and the second-order statistics of velocity fluctuations.
{
By choosing a velocity scale $k^{1/2}$ and a length scale $k^{3/2}/\epsilon$, turbulent viscosity can be expressed as~\citep{pop00}
	\be
	\nu_T (y)
	\; = \;
	c \, R_\tau^2 \, \dfrac{k^2 (y)}{\epsilon (y)},
	\label{eq.nuT}
	\ee	
where $c = 0.09$ is a multiplicative constant (for validity of this assumption in the near-wall region we refer the reader to~\cite{pop00}). Both the turbulent kinetic energy $k$ and its rate of dissipation $\epsilon$ are determined by averaging the second-order statistics of fluctuations over the horizontal directions and one period of oscillations
    	\be
	\ba{rcl}
	k (y)
	&\!\! = \!\!&
	\dfrac{1}{2T} \,
	\ds{\int_{0}^{T}}
	\left< \overline{u u} \, + \, \overline{v v} \, + \, \overline{w w} \right> (y,t) \,
	\mrd t,
	\\[0.35cm]
	\epsilon (y)
	&\!\! = \!\!&
	\dfrac{1}{T} \,
	\ds{\int_{0}^{T}}
	\, \left<
	2 \left(
	\overline{u_x u_x}
	\, + \,
	\overline{v_y v_y}
	\, + \,
	\overline{w_z w_z}
	\, + \,
	\overline{u_y v_x}
	\, + \,
	\overline{u_z w_x}
	\, + \,
	\overline{v_z w_y}
	\right)
	\, + \,
	\right.
	\\[0.2cm]
	&&
	~~~~~~~~~~~~
	\left.
	\overline{u_y u_y}
	\, + \,
	\overline{w_y w_y}
	\, + \,
	\overline{v_x v_x}
	\, + \,
	\overline{w_x w_x}
	\, + \,
	\overline{u_z u_z}
	\, + \,
	\overline{v_z v_z}
	\right>
	\, (y,t) \,
	\mrd t.
	\ea
	\label{eq.k-epsilon}
	\ee
The most widely used method for computing $k$ and $\epsilon$ in engineering flows is the $k$-$\epsilon$ model~\citep{jonlau72,lausha74} where $k$ and $\epsilon$ are determined by solving two transport equations~\citep{pop00}.}
Even though these are less complex than the NS equations, they are still computationally expensive, and {\em not convenient for control design and optimization\/}. In \S~\ref {sec.linearized-turbulent}, we instead develop a simulation-free method, which is computationally efficient and amenable to control design and optimization, for determining the effect of fluctuations on $\nu_T$ in the flow with control.
	
\section{Stochastically forced flow with control}
\label{sec.linearized-turbulent}

Since $\nu_T$ in~(\ref{eq.nuT}) is determined by the second-order statistics of velocity fluctuations, we use {\em the stochastically forced linearized NS equations\/} to compute $k$ and $\epsilon$ in the flow with control. Here, we utilize the fact that the second-order statistics of linear time-periodic systems can be obtained from the solution of the corresponding Lyapunov equation~\citep{jovfarAUT08}. It is well-known that the analysis of the steady-state variance of infinitesimal fluctuations around the laminar flow can be used to identify flow structures that initiate the onset of turbulence~\citep{farioa93,bamdah01,jovbamJFM05}. In this paper, we show that {eddy-viscosity-enhanced linearization of the turbulent flow with control in conjunction with turbulence modeling} can be used to {approximate the} turbulent viscosity in a computationally efficient way.

Next, we examine the effect of control on small-amplitude fluctuations around $\bU_0 = [\,U_{0} (y) ~0~W_{0} (y,t)\,]^T$. An equivalent expression for $\bU_0$ can be found from the steady-state solution of the modified NS equations subject to wall-oscillations,
	\be
       \ba{rcl}
	\bu_t
	& \!\! = \!\! &
	\, - \,
	\left(\bu \cdot \nabla \right) \bu
	\, - \,
	\nabla P
       \, + \,
       (1/R_\tau) \, \nabla \cdot \left( (1 + \nu_{T0}) \left(\nabla \bu \, + \, (\nabla \bu)^T \right) \right),
       \\[0.15cm]
       0
    	& \!\! = \!\! &
    	\nabla \cdot \bu.
    	\ea
	\label{eq.NS-mod-nuT}
	\ee
Model~(\ref{eq.NS-mod-nuT}) is obtained by augmenting the molecular viscosity in the NS equations~(\ref{eq.NScts}) with the turbulent viscosity $\nu_{T 0}$, and it facilitates analysis of the dynamics of turbulent flow fluctuations (up to first order),
	\be
       \ba{rcl}
	\bv_t
	& \!\! = \!\! &
	\, - \,
	\left(\bU_0 \cdot \nabla \right) \bv
	\, - \,
	\left(\bv \cdot \nabla \right) \bU_0
	\, - \,
	\nabla p
       \, + \,
       (1/R_\tau) \, \nabla \cdot \left( (1 + \nu_{T0}) \left(\nabla \bv \, + \, (\nabla \bv)^T \right) \right),
       \\[0.15cm]
       0
    	& \!\! = \!\! &
    	\nabla \cdot \bv.
    	\ea
	\label{eq.NS-lin}
	\ee
Recent research has demonstrated capability of~(\ref{eq.NS-lin}) to qualitatively predict {the spacing and length of near-wall turbulent structures observed in experiments and simulations~\citep{alajim06,cospujdep09,pajgarcosdep09}. This model was used by~\cite{reyhus72-3} as a method for obtaining closure in the equation for velocity fluctuations, and it can be traced back to~\citet{tow56}. These authors noted that the eddy viscosity represents the influence of fluctuations on the background turbulent stresses, and that this influence is best captured for the fluctuations that have larger wavelengths relative to the dominant turbulent scales. A different interpretation was provided by~\cite{alajim06} where the eddy viscosity was introduced to model the dissipative effects of small scale turbulent structures on the large scales.}

Our simulation-free design of drag-reducing transverse oscillations involves four steps:
    \bi
\item[(i)] {\em the turbulent mean velocity in the presence of control is obtained from the RANS equations~(\ref{eq.turb-mean-x-z}) where closure is achieved using the turbulent viscosity of the uncontrolled flow~(\ref{eq.nuRT})\/};

\item[(ii)] {\em $k$ and $\epsilon$ are determined from the second-order statistics of fluctuations that are obtained from the stochastically forced NS equations linearized around the turbulent mean velocity determined in\/} (i);

\item[(iii)] {\em for the flow with control, the modifications to $k$ and $\epsilon$ are used to determine the modification to the turbulent viscosity, $\nu_T$\/};

\item[(iv)] {\em the modified $\nu_T$ is used {in the RANS equations~(\ref{eq.turb-mean-x-z})} to determine the effect of fluctuations on the mean velocity, and thereby skin-friction drag and control net efficiency\/}.
    \ei

Figure~\ref{fig.diagram-nuT-update} represents these four steps using a block-diagram. We note that the slow time evolution of the mean flow (relative to the time evolution of fluctuations) is used to separate the update of the mean velocity (Steps~(i) and~(iv)) from the computation of the statistics~(Step (ii)) and the update of $\nu_T$~(Step~(iii)). Rather than updating the turbulent viscosity and the mean velocity in an iterative fashion, we update them only once; in~\S~\ref{sec.results}, we show that the resulting correction to the mean velocity reliably predicts the optimal period of drag-reducing oscillations. Also note that Step (i) amounts to finding the steady-state solution of system~(\ref{eq.NS-mod-nuT}) subject to wall oscillations, and that Step (ii) amounts to the analysis of the linearized model~(\ref{eq.NS-lin}) in the presence of stochastic forcing.

    \begin{figure}
    \centering
    {\includegraphics[width=0.75\columnwidth]
    {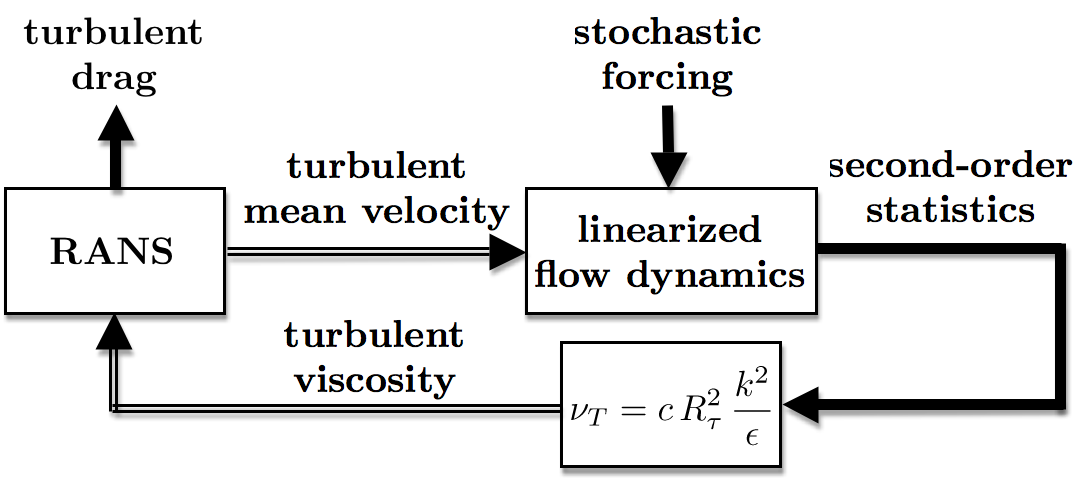}}
    \caption{
    Block diagram representing Steps (i)-(iv) of our simulation-free approach for determining the effect of control on skin-friction drag in turbulent flows. The hollow arrows indicate that some of the coefficients in the model of the output subsystems are determined by the outputs of the corresponding input subsystems. The turbulent mean velocity is updated only once in the present study.
    }
    \label{fig.diagram-nuT-update}
    \end{figure}

\subsection{Computation of the velocity correlations}
    \label{sec.compute-correlations}

The evolution form of the linearized model~(\ref{eq.NS-lin}) that governs the dynamics of fluctuations around $[\,U_{0} (y) ~0~W_{0} (y,t)\,]^T$ is given by
  \be
    \ba{rcl}
    \bpsi_t (y,\bkappa,t)
    & = &
    A (\bkappa,t) \, \bpsi(y,\bkappa,t)
    \; + \;
    \fvec (y,\bkappa,t),
    \\[0.1cm]
    \bv(y,\bkappa,t)
    &
    =
    &
    C (\bkappa) \, \bpsi(y,\bkappa,t),
    \ea
    \label{eq.lnse-turb}
    \ee
where $\bpsi = [\,v~\eta\,]^T$ is the state, $\eta = \mri \kappa_{z} u - \mri \kappa_{x} w$ is the wall-normal vorticity, {and $\fvec$ is the stochastic forcing with second-order statistics determined by~(\ref{eq.R})}. System~(\ref{eq.lnse-turb}) represents a $\bkappa$-parameterized family of PDEs in $y$ and $t$ with time-periodic coefficients. Here, $\bkappa$ denotes the streamwise and spanwise wavenumbers, $\bkappa = (\kappa_{x}, \, \kappa_{z})$, and the same symbol is used to denote the variables in physical and wavenumber domains {(when necessary, we will highlight the distinction by explicitly stating the dependence on $(x,z)$ and $(\kappa_x, \kappa_z)$, respectively)}. The operators $A$ and $C$ in~(\ref{eq.lnse-turb}) are given by
    \be
    \ba{rrl}
    A
    & \!\!=\!\! &
    \tbt{A_{11}}{0}{A_{21}}{A_{22}},
    ~~~
    C
    ~ = ~
    \thbo{C_u}{C_v}{C_w}
    \; = \;
    \dfrac{1}{\kappa^2}
    \thbt{\mri \kappa_{x} \p_y}{- \, \mri \kappa_{z}\\[-.3cm]}
    {\kappa^2}{0\\[-.3cm]}
    {\mri \kappa_{z} \p_y}{\mri \kappa_{x}},
    \\[0.75cm]
    A_{11}
    & \!\!=\!\! &
    {\Delta}^{-1}
    \left(
    (1/R_\tau)
    \left(
    (1+\nu_{T0})
    {\Delta}^2
    \,+\,
    2 \nu'_{T0} \Delta \p_y
    \,+\,
    \nu''_{T0} (\p_y^2+\kappa^2)
    \right)
    \,+\,
    \right.
    \\[0.3cm]
    &&
    \left.
    \mri \kappa_{x} \big( U''_{0} - U_{0} {\Delta} \big)
    \,+\,
    \mri \kappa_{z} \big( W''_{0} - W_{0} {\Delta} \big)
    \right),
    \\[0.2cm]
    A_{21}
    & \!\!=\!\! &
    - \, \mri \kappa_{z} U'_{0}
    \, + \,
    \mri \kappa_{x} W'_{0},
    \\[0.2cm]
    A_{22}
    & \!\!=\!\! &
    (1/R_\tau)
    \left(
    (1+\nu_{T0})  {\Delta}
    \, + \,
    \nu'_{T0} \p_y
    \right)
    \, - \,
    \mri \kappa_{x} U_{0}
    \, - \,
    \mri \kappa_{z} W_{0},
    \ea
    \label{eq.AC}
    \ee
where
    $
    \Delta
    \, = \,
    \p_y^2
    \, - \,
    \kappa^2
    $
is the Laplacian, $\Delta^2 \, = \, \p_y^4 \, - \, 2 \kappa^2 \, \p_y^2 \, + \, \kappa^4$,
    $
    \kappa^2
    \, = \,
    \kappa_{x}^2 \, + \, \kappa_{z}^2,
    $
and
    $
    v (\pm1,\bkappa,t)
    \, = \,
    v' (\pm1,\bkappa,t)
    \, = \,
    \eta (\pm1,\bkappa,t)
\, = \,
    0.
    $

We next briefly describe a method for determining the steady-state statistics of the linearized system~(\ref{eq.lnse-turb}) driven by a zero-mean temporally white stochastic forcing, with second-order statistics,
    \be
    \left<\fvec (\, \cdot \,,\bkappa,t_1) \otimes \fvec (\, \cdot \,,\bkappa,t_2)\right>
    \; = \;
    M (\bkappa) \, \delta (t_1 \, - \, t_2).
    \label{eq.R}
    \ee
Here, $\delta$ is the Dirac delta function, $\fvec \otimes \fvec$ is the tensor product of $\fvec$ with itself, and $M (\bkappa)$ is a spatial spectral-density of forcing. For homogeneous isotropic turbulence, the steady-state velocity correlation tensors can be reproduced by the linearized NS equations subject to white-in-time forcing with second-order statistics proportional to the turbulent energy spectrum~\citep{jovgeoAPS10}. Using this analogy, we select $M (\bkappa)$ to guarantee equivalence between the two-dimensional energy spectra of the uncontrolled turbulent flow and the flow governed by the stochastically forced NS equations linearized around $\bU_0 = [\,U_0 (y)~0~0\,]^T$. To this end, we use the DNS-based energy spectrum of the uncontrolled flow~\citep{deljim03,deljimzanmos04}, $E (y,\bkappa)$, to define
	\be
	\ba{c}
	M (\bkappa)
	\; = \;
    	\dfrac{\bar{E} (\bkappa)}{\bar{E}_0 (\bkappa)} \, M_0 (\bkappa),
	\\[0.35cm]
	{
	M_0 (\bkappa)
	\; = \;
	\tbt
	{\sqrt{E (y, \bkappa)} \, I}{0}
	{0}{\sqrt{E (y, \bkappa)} \, I}
	\tbt
	{\sqrt{E (y, \bkappa)} \, I}{0}
	{0}{\sqrt{E (y, \bkappa)} \, I}^+.
	}
	\ea
	\non
	\ee	
Here, $\bar{E} (\bkappa) = \int_{-1}^{1} E (y,\bkappa) \, \mrd y$ is the two-dimensional energy spectrum of the uncontrolled flow, $\bar{E}_0 (\bkappa)$ is the energy spectrum obtained from the linearized NS equations subject to a white-in-time forcing $\fvec$ with spatial spectrum $M_0 (\bkappa)$, {and $^+$ denotes the adjoint of an operator. Note that the adjoints of the operator appearing in the expression for $M_0 (\bkappa)$ and the operator $A$ should be determined with respect to the inner product that induces kinetic energy of flow fluctuations; for additional details, see~\cite{jovbamJFM05}.}

For the time-periodic system~(\ref{eq.lnse-turb}), the operator $A$ in~(\ref{eq.AC}) can be written as
	\be
	A (\bkappa,t)
	\; = \;
	A_0 (\bkappa)
	\, + \,
	\alpha
    \left(
    A_{-1} (\bkappa) \, \mre^{-\mri \, \omega_t \, t}
	\, + \,
	A_{1} (\bkappa) \, \mre^{\mri \, \omega_t \, t}
    \right),
	\label{eq.A-t}
	\ee
where the expressions for $A_0$, $A_{-1}$, and $A_1$ are given in Appendix~\ref{sec.A0-A1-Am1}. It is a standard fact that the response of the linear time-periodic system~(\ref{eq.lnse-turb}) subject to a stationary input is a cyclo-stationary process~\citep{gar90}, meaning that its statistical properties are periodic in time. For example, the auto-correlation operator of $\bpsi$ is given by
	\be
	\ba{l}
	X (\bkappa,t)
       	~ = ~
       	\left<
       	\bpsi(\, \cdot \,,\bkappa, t) \otimes \bpsi (\, \cdot \,,\bkappa, t)
       	\right>
       	~ = ~
        \\[0.2cm]
       	X_0 (\bkappa)
       	\, + \,
    	X_1 (\bkappa) \, \mre^{\mri \, \omega_t \, t}
       	\, + \,
       	X_1^+ (\bkappa) \, \mre^{-\mri \, \omega_t \, t}
       	\, + \,
       	X_2 (\bkappa) \, \mre^{\mri \, 2\, \omega_t \, t}
       	\, + \,
       	X_2^+ (\bkappa) \, \mre^{-\mri \, 2\, \omega_t \, t}
       	\, + \,
       	\ldots.
       	\ea
	\label{eq.X-t}
	\ee
The averaged effect of forcing (over one period $T$) is determined by the operator $X_0$
	\be
	\dfrac{1}{T} \,
	\ds{\int_{0}^{T}}
	X (\bkappa,t) \, \mrd t
	\; = \;
	X_0 (\bkappa).
	\label{eq.X-mean}
	\ee	

In the remainder of the paper, we consider small amplitude of wall oscillations $\alpha$. This choice is motivated by the observation that the power required to maintain the oscillations increases quadratically with $\alpha$ (cf.\ (\ref{eq.delta-Preq}) {and~\cite{ricqua08}}). Hence, large amplitudes may be prohibitively expensive from control expenditure point of view. Furthermore, for sufficiently small value of $\alpha$ the velocity correlations can be computed efficiently using perturbation analysis in $\alpha$~\citep{jovfarAUT08,jovPOF08}. We thus use perturbation analysis to identify wall oscillation periods that yield the largest drag reduction {and net efficiency; note that these do not necessarily coincide with each other}.

Up to second order in $\alpha$, the operator $X_0$ in~(\ref{eq.X-mean}) is given by
	\be
	X_0 (\bkappa)
	\; = \;
	X_{0,0} (\bkappa)
	\, + \,
	\alpha^2 \, X_{0,2} (\bkappa)
	\, + \,
	{\cal O}(\alpha^4),
	\label{eq.X0-pert}
	\ee
where $X_{0,0}$ and $X_{0,2}$ are obtained from the set of decoupled Lyapunov equations~\citep{jovfarAUT08,jovPOF08}; see Appendix~\ref{sec.variance} for details. The auto-correlation operator of the state $\bpsi$ of stochastically forced uncontrolled flow is determined by $X_{0,0}$. On the other hand, the operator $X_{0,2}$ represents the second-order correction to $X_{0,0}$ induced by wall oscillations. As shown in~\S~\ref{sec.nuT-control} and~\S~\ref{sec.drag-control}, $X_{0,2}$ determines the effect of fluctuations on $k$, $\epsilon$, $\nu_T$, and $C_f$ in the flow with control.
	
\subsection{Influence of fluctuations on turbulent viscosity}
\label{sec.nuT-control}

According to~(\ref{eq.nuT}), $\nu_T$ is determined by the second-order statistics of velocity fluctuations. By considering dynamics of infinitesimal fluctuations, these statistics can be obtained from the auto-correlation operator $X_0$. In the flow subject to small amplitude wall oscillations $X_0$ is given by~(\ref{eq.X0-pert}), implying that $k$ and $\epsilon$ can be expressed as
	\be
    \ba{rcl}
	k (y)
	&\!\!=\!\!&
	k_0 (y)
	\, + \,
	\alpha^2 \, k_2	(y)
	\, + \,
	\cO (\alpha^4),
	\\[0.2cm]
	\epsilon (y)
	&\!\!=\!\!&
	\epsilon_0 (y)
	\, + \,
	\alpha^2 \, \epsilon_2 (y)
	\, + \,
	\cO (\alpha^4).
	\ea
	\label{eq.turb-force}
	\ee
Here, the subscript $0$ denotes the corresponding quantities in the uncontrolled turbulent flow, and the subscript $2$ quantifies the influence of fluctuations in the controlled flow at the level of $\alpha^2$. A computationally efficient method for determining $k_2$ and $\epsilon_2$ from $X_{0,2}$ is provided in Appendix~\ref{sec.compute-k2-epsilon2}.

For small amplitude oscillations, substituting $k$ and $\epsilon$ from~(\ref{eq.turb-force}) into~(\ref{eq.nuT}) yields
	\be
	\nu_{T} (y)
	\; = \;
	c \,  R_\tau^2 \, \dfrac{k^2 (y)}{\epsilon (y)}
	\; = \;
	c \,  R_\tau^2 \,
	\dfrac{\left(k_0 (y) \, + \, \alpha^2 k_2 (y) \, + \, \cO (\alpha^4)\right)^2}
	{\epsilon_0 (y) \, + \, \alpha^2 \epsilon_2 (y) \, + \, \cO (\alpha^4)},
    \non
	\ee
which in conjunction with Neumann series expansion leads to
	\be
	\ba{rcl}
	\nu_T (y)
	&\!\! = \!\!&
	\nu_{T0} (y)
	\, + \,
	\alpha^2 \, \nu_{T2} (y)
	\, + \,
	\cO (\alpha^4),
    	\\[0.15cm]
	\nu_{T2} (y)
	&\!\! = \!\!&
	\nu_{T0} (y)
	\left(
	\dfrac{2 k_2 (y)}{k_0 (y)}
	\, - \,
	\dfrac{\epsilon_2 (y)}{\epsilon_0 (y)}
	\right).
	\ea
	\label{eq.nuT-f-pert}
	\ee
Therefore, up to second order in $\alpha$, the influence of fluctuations on turbulent viscosity in the flow with control is determined by second-order corrections to the kinetic energy $k_2$ and its rate of dissipation $\epsilon_2$.
	    	
\subsection{Skin-friction drag coefficient and net efficiency}
\label{sec.drag-control}

We next show how velocity fluctuations in the flow subject to small amplitude oscillations modify the skin-friction drag coefficient and the net efficiency. As discussed in~\S~\ref{sec.drag-power}, $C_f$ is determined by $U$ and $\% \Pi_{\mathrm{net}}$ is determined by both $U$ and $W$. The influence of fluctuations on $U$ and $W$ in the flow with control can be obtained by substituting $\nu_{T}$ from~(\ref{eq.nuT-f-pert}) into~(\ref{eq.turb-mean-x-z}), and thereby expressing $U$, $W$, and {$P_x$} as
	\be
	\ba{rcl}
	U (y)
	&\!\! = \!\!&
	U_{0} (y)
	\, + \,
	\alpha^2 \,
	U_{2} (y)
	\, + \,
	\cO (\alpha^4),
	\\[0.2cm]
	W_p (y)
	&\!\! = \!\!&
	W_{p,0} (y)
	\, + \,
	\alpha^2 \,
	W_{p,2} (y)
	\, + \,
	\cO (\alpha^4),
	\\[0.2cm]
	{P_x}
	&\!\! = \!\!&
	{-1
	\, + \,
	\alpha^2 \,
	P_{x,2}
	\, + \,
	\cO (\alpha^4)},
	\ea
	\label{eq.U1-P1-pert}
	\ee
where the expressions for $U_2$, $W_{p,2}$, and {$P_{x,2}$} are provided in Appendix~\ref{sec.compute-U2-tau2-W2-Psave2-Preq2}.

An expression for the saved power is obtained by substituting {$P_x$} from~(\ref{eq.U1-P1-pert}) into~(\ref{eq.delta-Cf})
	\be
	\% C_{f}
	\; = \;
	\% \Pi_{\mathrm{save}}
	\; = \;
	\alpha^2 \, \% \Pi_{\mathrm{save},2}
	\, + \,
	\cO (\alpha^4),
	~~~
	\% \Pi_{\mathrm{save},2}
	\; = \;
	{100\,  P_{x,2}}.
	\label{eq.delta-Cf-pert}
	\non
	\ee
In flows subject to small amplitude oscillations, the above equation shows that a positive (negative) value of $\% \Pi_{\mathrm{save},2}$ signifies drag reduction (increase). On the other hand, the required power can be obtained by substituting $W_p$ from~(\ref{eq.U1-P1-pert}) into~(\ref{eq.delta-Preq})
	\be
	\% \Pi_{\mathrm{req}}
	\, = \,
	\alpha^2 \,
	\left(
	\% \Pi_{\mathrm{req},0}
	\, + \,
	\alpha^2 \, \% \Pi_{\mathrm{req},2}
	\right)
	\, + \,
	\cO (\alpha^6),
	\label{eq.delta-Preq-pert}
	\ee
where $\% \Pi_{\mathrm{req},0}$ is given in~(\ref{eq.Preq0}), and $\% \Pi_{\mathrm{req},2}$ is provided in Appendix~\ref{sec.compute-U2-tau2-W2-Psave2-Preq2}. We note the contrast in the way fluctuations influence saved and required powers; while fluctuations make $\cO (\alpha^2)$ contribution to $\% \Pi_{\mathrm{save}}$, they affect $\% \Pi_{\mathrm{req}}$ only at the level of $\alpha^4$. This explains the close agreement, observed in~\S~\ref{sec.drag-power}, between $\% \Pi_{\mathrm{req},0}$ and the DNS results of~\citet{quaric04}. Finally, the net efficiency is given by
	\be
	\% \Pi_{\mathrm{net}}
	\, = \,
	\alpha^2 \,
	\% \Pi_{\mathrm{net},2}
	\, + \,
	\cO (\alpha^4),
	~~~
	\% \Pi_{\mathrm{net},2}
	\; = \;
	\% \Pi_{\mathrm{save},2}
	\, - \,
	\% \Pi_{\mathrm{req},0}.
	\non
	\ee

The developments of this section are used in~\S~\ref{sec.results} to determine the skin-friction drag coefficient and the net efficiency in the flow subject to wall oscillations.	
	
\section{Turbulent drag reduction}
\label{sec.results}

In this section, we examine the effect of transverse wall oscillations on skin-friction drag and net efficiency in flows with $R_\tau = 186$, $547$, and $934$. For these Reynolds numbers, the second-order statistics of the uncontrolled turbulent flow were obtained using DNS~\citep{deljim03,deljimzanmos04}. As explained in \S~\ref{sec.compute-correlations}, we use available DNS data to determine the spatial spectrum of the stochastic forcing~(\ref{eq.R}) in the evolution model~(\ref{eq.lnse-turb}). The results of~\citet{deljim03,deljimzanmos04} also provide the turbulent kinetic energy in the uncontrolled flow, $k_0$, and thereby its rate of dissipation,
	\be
	\epsilon_0 (y)
	\; = \;
	c R_\tau^2 \, \dfrac{k_0^2 (y)}{\nu_{T0} (y)},
	\non
	\ee
where $\nu_{T0}$ is defined by~(\ref{eq.nuRT}).

The differential operators in the wall-normal direction are discretized using $N_y$ collocation points~\citep{weired00}. In horizontal directions, we use $N_x \times N_z$ wavenumbers with $0 < \kappa_x < \kappa_{x,\mathrm{max}}$ and $0 < \kappa_z < \kappa_{z,\mathrm{max}}$, where $\kappa_{x,\mathrm{max}}$ and $\kappa_{z,\mathrm{max}}$ are the largest wavenumbers used in the DNS of~\cite{deljim03,deljimzanmos04}; table~\ref{table.dom} provides summary of parameters used in our study. The DNS-based energy spectrum taken from
    {\sf http://torroja.dmt.upm.es/ftp/channels/data/}
is interpolated on the $N_y$ collocation points in $y$.

	\begin{table*}
         \centering
         \begin{tabular}{lrrrrrrrr}
         \hline
         \\[-0.075cm]
            $R_\tau$ \hspace{0.5cm}  & \hspace{0.5cm}$N_x$  & \hspace{0.5cm}$N_y$ & \hspace{0.5cm}$N_z$ & \hspace{0.5cm}$\kappa_{x,\mathrm{max}}$ & \hspace{0.5cm}$\kappa_{z,\mathrm{max}}$ & \hspace{0.9cm}$c_1$ & \hspace{0.9cm}$c_2$  & \hspace{0.9cm}$U_B$ \\[.1cm]
         \hline
         \\[-0.075cm]
            $186$ & $50$  & $101$   & $51$ &  $42.5$ & $84.5$   & $46.2$ & $0.61$ & $15.73$\\[.3mm]
            $547$ & $50$  & $151$   & $51$ &  $128$  & $255.5$ & $29.4$ & $0.45$ & $18.38$\\[.3mm]
            $934$ & $50$  & $201$   & $51$ &  $255$  & $511.3$ & $27.0$ & $0.43$ & $19.86$\\[.3mm]
         \end{tabular}
         \caption{The parameters used in our study. At each $R_\tau$, $c_1$ and $c_2$ are selected to minimize the least squares deviation between the mean streamwise velocity obtained from~(\ref{eq.U})-(\ref{eq.nuRT}) and the mean velocity obtained in DNS~\citep{deljim03,deljimzanmos04}. The bulk flux $U_B$ is kept constant by adjusting the pressure gradient.
         }
         \label{table.dom}
         \hrulefill
      \end{table*}

    \begin{figure}
    \begin{center}
    \includegraphics[width=0.49\columnwidth]
    {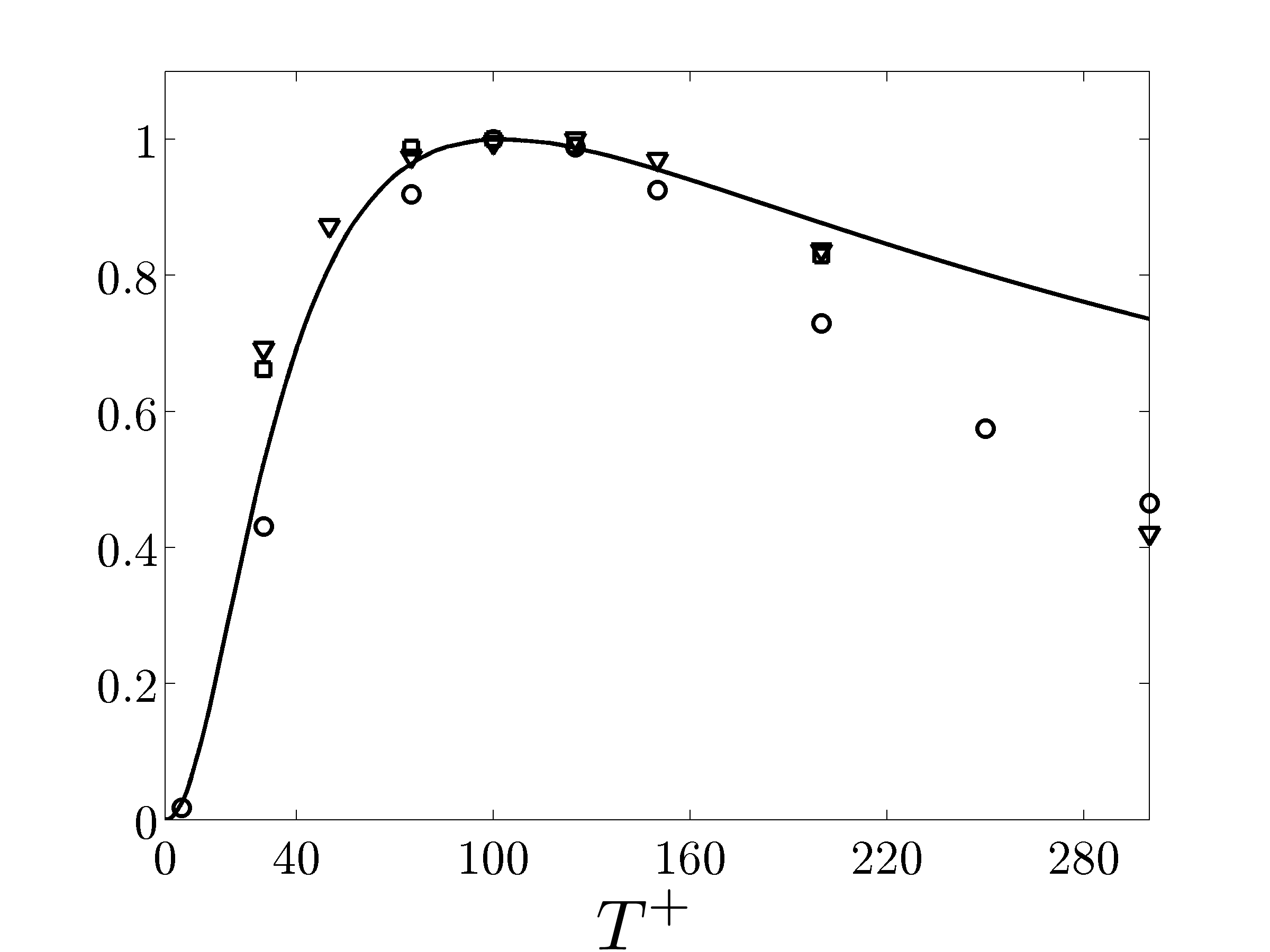}
    \caption{
    {
    (Solid curve) Second-order correction to the saved power, $\% \Pi_{\mathrm{save},2} (T^+)$ normalized by $\max \, (\% \Pi_{\mathrm{save},2} (T^+)) = 2.8$, as a function of the period of oscillations $T^+$ for the flow with $R_\tau = 186$;
    (Symbols) DNS-based $\% \Pi_{\mathrm{save}} (T^+)$ normalized by the corresponding largest values at $R_\tau = 200$~\citep{quaric04} for control amplitudes $\alpha = 2.25,~\circ$; $\alpha = 6,~\square$; and $\alpha = 9,~\triangledown$.
    }
    }
    \label{fig.Psave2-R186}
    \end{center}
    \end{figure}

\subsection{Saved power}
    \label{sec.Psave}

We first examine the effect of period of oscillations $T^+$ on the turbulent drag reduction and the saved power. {Solid curve in} figure~\ref{fig.Psave2-R186} shows the second-order correction to the saved power $\% \Pi_{\mathrm{save},2} (T^+)$, {normalized by its maximum value $\max \, (\% \Pi_{\mathrm{save},2} (T^+)) = 2.8$,} for the controlled flow with $R_\tau = 186$. The positive value of $\% \Pi_{\mathrm{save},2}$ indicates that drag is reduced for all values of $T^+$ that we considered, with the largest drag reduction taking place at $T^+ = 102.5$. Our theoretical predictions are in close agreement with DNS at $R_\tau = 200$~\citep{quaric04}, where it was shown that $T^+ \approx 100$ yields the largest drag reduction for control amplitudes $\alpha = 3.1$ and $6.2$.

Figure~\ref{fig.Psave2-R186} also compares $\% \Pi_{\mathrm{save},2}$ (solid curve) obtained using our analysis with $\% \Pi_{\mathrm{save}}$ (symbols) obtained using DNS at $R_\tau = 200$~\citep{quaric04}; {both sets of results are normalized by their corresponding maximal values.} In DNS, the largest drag reduction takes place at $T^+ \approx 100$ for $\alpha = 2.25$ and $\alpha = 6$, and at $T^+ \approx 125$ for $\alpha = 9$. Thus, for small control amplitudes, perturbation analysis up to second order in $\alpha$ reliably predicts optimal period of drag reducing oscillations. For the optimal period of oscillations and $\alpha = 2.25$, our perturbation analysis predicts $13.6\%$ drag reduction, whereas $17.4\%$ drag reduction is obtained in DNS~\citep{quaric04}. The quantitative difference between the DNS results and the results of perturbation analysis may be attributed to the effects of higher order corrections. Another factor that warrants further scrutiny is modeling of the spatial spectrum of stochastic forcing. Analysis of these effects is beyond the scope of the current study.

For small oscillation periods, $T^+ \lesssim 0.5$, the second-order correction to the saved power is negligible. Recent research has determined necessary conditions on the amplitude and period of oscillations for drag reduction. For $T^+ > 30$,~\cite{ricqua08} used the solution to the Stokes second problem to show that the smallest oscillation amplitude that yields drag reduction is approximately equal to $1.8$ (in viscous units). It was also shown that the amplitudes of oscillations that achieve drag reduction become prohibitively large as the Stokes layer thickness $\delta_S \sim \sqrt{T^+}$ approaches zero.
In addition,~\cite{quaric11} showed that the smallest value of the Stokes layer thickness that results in drag reduction by traveling waves of spanwise wall motion is approximately equal to $1$ in viscous units.
       	
\subsection{Required control power}
    \label{sec.Preq}

We next study the power required to maintain wall oscillations. From~(\ref{eq.delta-Preq-pert}) it follows that, up to second order in $\alpha$, $\% \Pi_{\mathrm{req},0}$ defined by~(\ref{eq.Preq0}) determines the required power. From \S~\ref{sec.drag-power} we also recall that {the discrepancy between $\% \Pi_{\mathrm{req},0}$ and the DNS-based required power gets larger for $T^+  \gtrsim 150$; cf.\ figure~\ref{fig.Preq-R186-quadrio-txt}. Here, we show that accounting for the effect of fluctuations in the flow with control reduces this discrepancy. As shown in figure~\ref{fig.Preq2-R186-quadrio}, the fourth-order correction to the required power, $\% \Pi_{\mathrm{req},2} (T^+)$, is negative for $T^+ \gtrsim 45$ and it decreases with $T^+$. This is in agreement with our earlier observation that $\% \Pi_{\mathrm{req},0}$ overestimates the required power obtained in DNS. The filled symbols in figure~\ref{fig.Preq2-R186-quadrio} represent the difference between the required power obtained by~\cite{quaric04} using DNS (shown by open symbols in figure~\ref{fig.Preq-R186-quadrio-txt}) and the required power that we obtain using $\% \Pi_{\mathrm{req},0}$ (shown by solid curve in figure~\ref{fig.Preq-R186-quadrio-txt}). For $\alpha = 2.25$, $6$, and $9$, $\% \Pi_{\mathrm{req}} - \alpha^2 \% \Pi_{\mathrm{req},0}$ is in close agreement with the fourth-order correction to the required power, $\% \Pi_{\mathrm{req},2}$ (solid curve in figure~\ref{fig.Preq2-R186-quadrio}).}

\begin{figure}
    \begin{center}
    \begin{tabular}{c}
    \includegraphics[width=0.49\columnwidth]
    {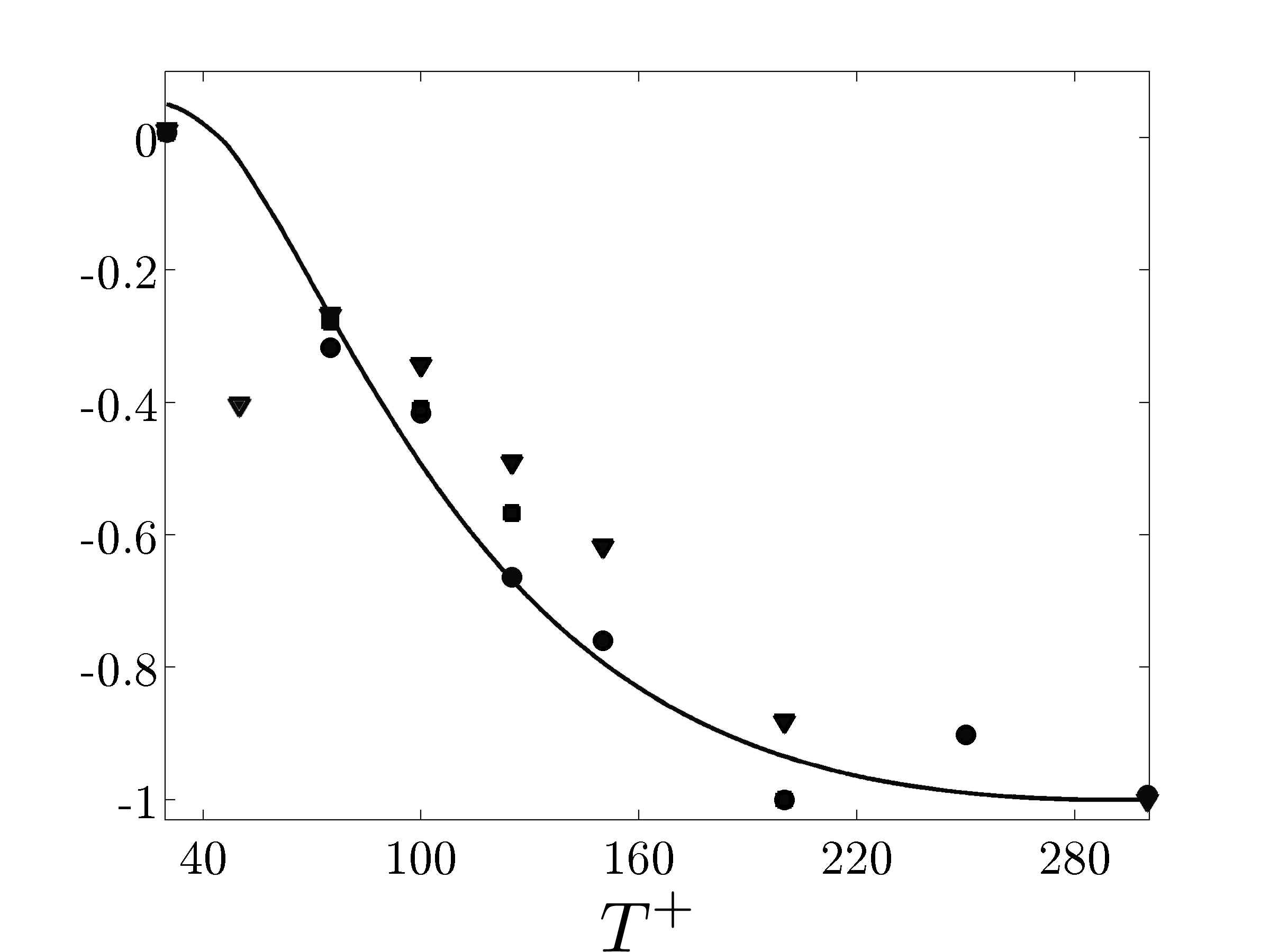}
    \end{tabular}
    \end{center}
    \caption{
    {
    (Solid curve) Fourth-order correction to the required power, $\% \Pi_{\mathrm{req},2} (T^+)$, as a function of the period of oscillations $T^+$ for the flow with $R_\tau = 186$;
    (Filled symbols) the difference between the DNS-based required power $\% \Pi_{\mathrm{req}}$ (shown by open symbols in figure~\ref{fig.Preq-R186-quadrio-txt}) and the required power $\% \Pi_{\mathrm{req},0}$ (shown by solid curve in figure~\ref{fig.Preq-R186-quadrio-txt}) that is obtained from the solution $W_0 (y,t)$, $\% \Pi_{\mathrm{req}} - \alpha^2  \% \Pi_{\mathrm{req},0} (T^+)$. $\% \Pi_{\mathrm{req}}$ is obtained using DNS at $R_\tau = 200$~\citep{quaric04} for $\alpha = 2.25,~\bullet$; $\alpha = 6,~\blacksquare$; and $\alpha = 9,~\blacktriangledown$. The results are normalized by $\max \, (|\% \Pi_{\mathrm{req},2} (T^+)|) = 0.0126$ and the largest values of $| \% \Pi_{\mathrm{req}} - \alpha^2  \% \Pi_{\mathrm{req},0} (T^+) |$.
    }
    }
    \label{fig.Preq2-R186-quadrio}
    \end{figure}
            	
    \begin{figure}
    \begin{center}
    $\% \Pi_{\mathrm{net},2} (T^+); \% \Pi_{\mathrm{net}}/\alpha^2$
    \\[0.1cm]
    \includegraphics[width=0.49\columnwidth]
    {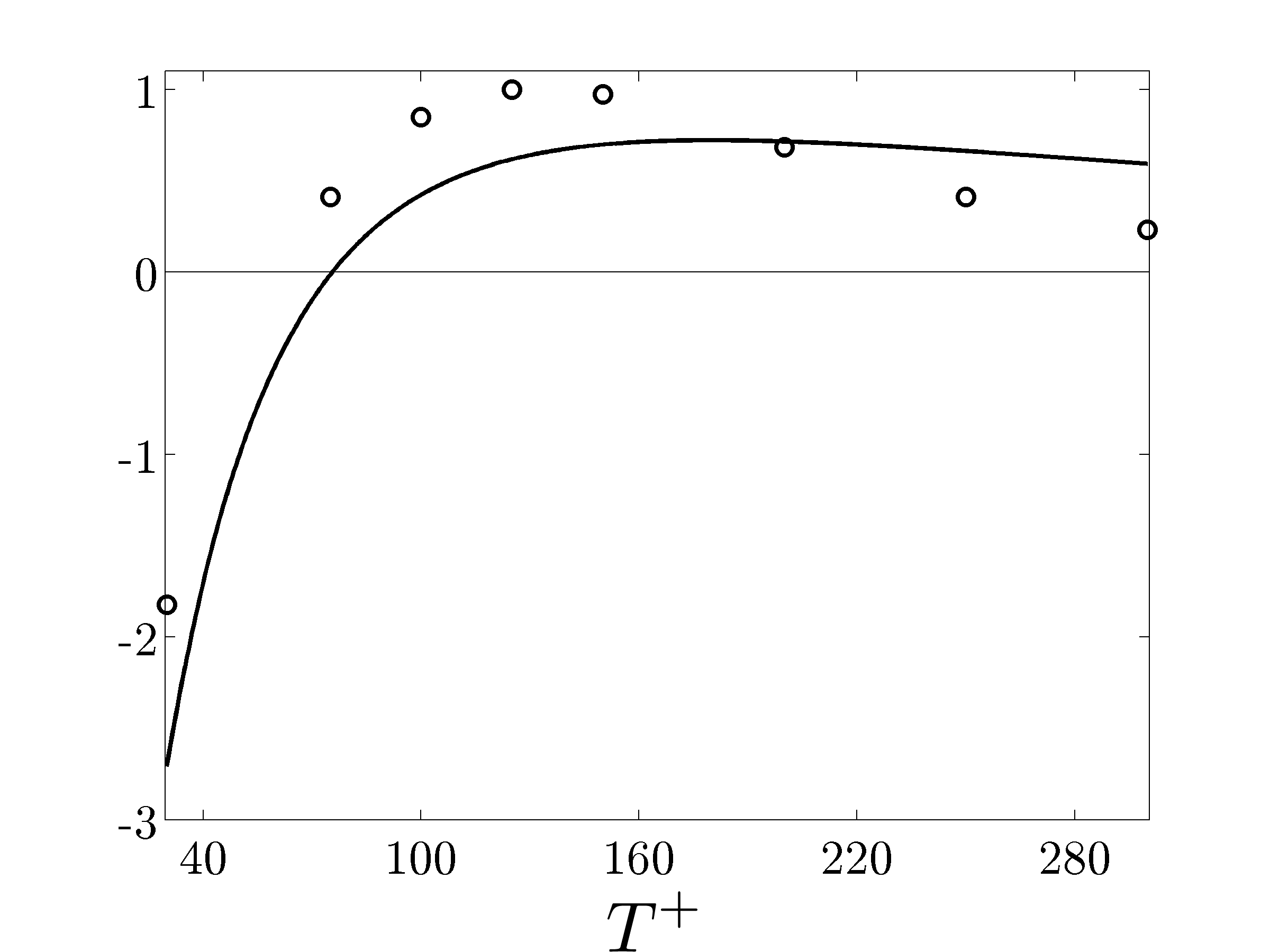}
    \end{center}
    \caption{
    Comparison between second-order correction to the net efficiency $\% \Pi_{\mathrm{net},2} (T^+)$ (solid curve) for the flow with $R_\tau = 186$ (solid curve), and $\% \Pi_{\mathrm{net}}/\alpha^2$ (symbols). Symbols show DNS data at $R_\tau = 200$~\citep{quaric04} for $\alpha = 2.25,~\circ$.
    }
    \label{fig.Pnet2-R186-quadrio}
    \end{figure}

\subsection{Net efficiency}
    \label{sec.Pnet}

The net efficiency in the flow subject to wall oscillations is determined by the difference between the saved and required powers. The solid curve in figure~\ref{fig.Pnet2-R186-quadrio} shows the second-order correction to the net efficiency, $\% \Pi_{\mathrm{net},2} (T^+)$, in the flow with $R_\tau = 186$. We see that $\% \Pi_{\mathrm{net},2} > 0$ for $T^+ > 75$, indicating that, for small control amplitudes, a positive net efficiency can be achieved if the period of oscillations is large enough. Our prediction is in close agreement with DNS at $R_\tau = 200$~\citep{quaric04} where positive net efficiency of oscillations with $\alpha = 2.25$ is obtained for $T^+ > 70$. Furthermore, up to second order in $\alpha$, the largest net efficiency takes place at $T^+ = 180$. This value differs from the value of $T^+$ that yields the largest saved power, $T^+ = 102.5$ (cf.\ \S~\ref{sec.Psave}). This difference can be explained by the fact that the peak of $\% \Pi_{\mathrm{net},2} = \% \Pi_{\mathrm{save},2} - \% \Pi_{\mathrm{req},0}$, takes place at a value of $T^+$ where the derivatives (with respect to $T^+$) of $\% \Pi_{\mathrm{save},2} $  and $\% \Pi_{\mathrm{req},0} (T^+)$ are equal to each other. Since $\% \Pi_{\mathrm{req},0} (T^+)$ is a monotonically decreasing function, its derivative is negative for all $T^+$. Therefore, the largest net efficiency takes place at some $T^+ > 102.5$ where the slope of the curve $\% \Pi_{\mathrm{save},2} (T^+)$ is also negative. (Our analysis shows that this happens at $T^+ = 180$.) The symbols in figure~\ref{fig.Pnet2-R186-quadrio} show $\% \Pi_{\mathrm{net}}/\alpha^2$ obtained from DNS at $R_\tau = 200$~\citep{quaric04} for $\alpha = 2.25$. Even though the essential trends are captured by $\% \Pi_{\mathrm{net},2}$ (solid curve), the DNS net efficiency peaks at $T^+ = 125$, which is approximately $30 \%$ smaller than the value of $T^+$ predicted by our perturbation analysis. This discrepancy may be attributed to a slower rate of decay of $\% \Pi_{\mathrm{save},2}$ relative to $\% \Pi_{\mathrm{save}}$ obtained in DNS for $T^+ > 100$; cf.\ figure~\ref{fig.Psave2-R186}.

We note that in DNS the net efficiency becomes negative for large control amplitudes. At $R_\tau = 200$,~\citet{quaric04} showed that $\% \Pi_{\mathrm{net}}$ becomes negative for all $T^+$ if $\alpha \gtrsim 3.5$. Therefore, the positive net efficiency predicted by perturbation analysis is only valid for small control amplitudes. This can be explained by noting that perturbation analysis predicts quadratic increase of both saved and required powers with $\alpha$. On the other hand, \citet{quaric04} showed that saved power exhibits slower than linear growth with $\alpha$ for large control amplitudes.

    \begin{figure}
    \begin{center}
    \begin{tabular}{cc}
    $\% \Pi_{\mathrm{save},2} (T^+)$
    &
    $\% \Pi_{\mathrm{req},0} (T^+)$
    \\[-0.2cm]
    \subfigure[]{\includegraphics[width=0.49\columnwidth]
    {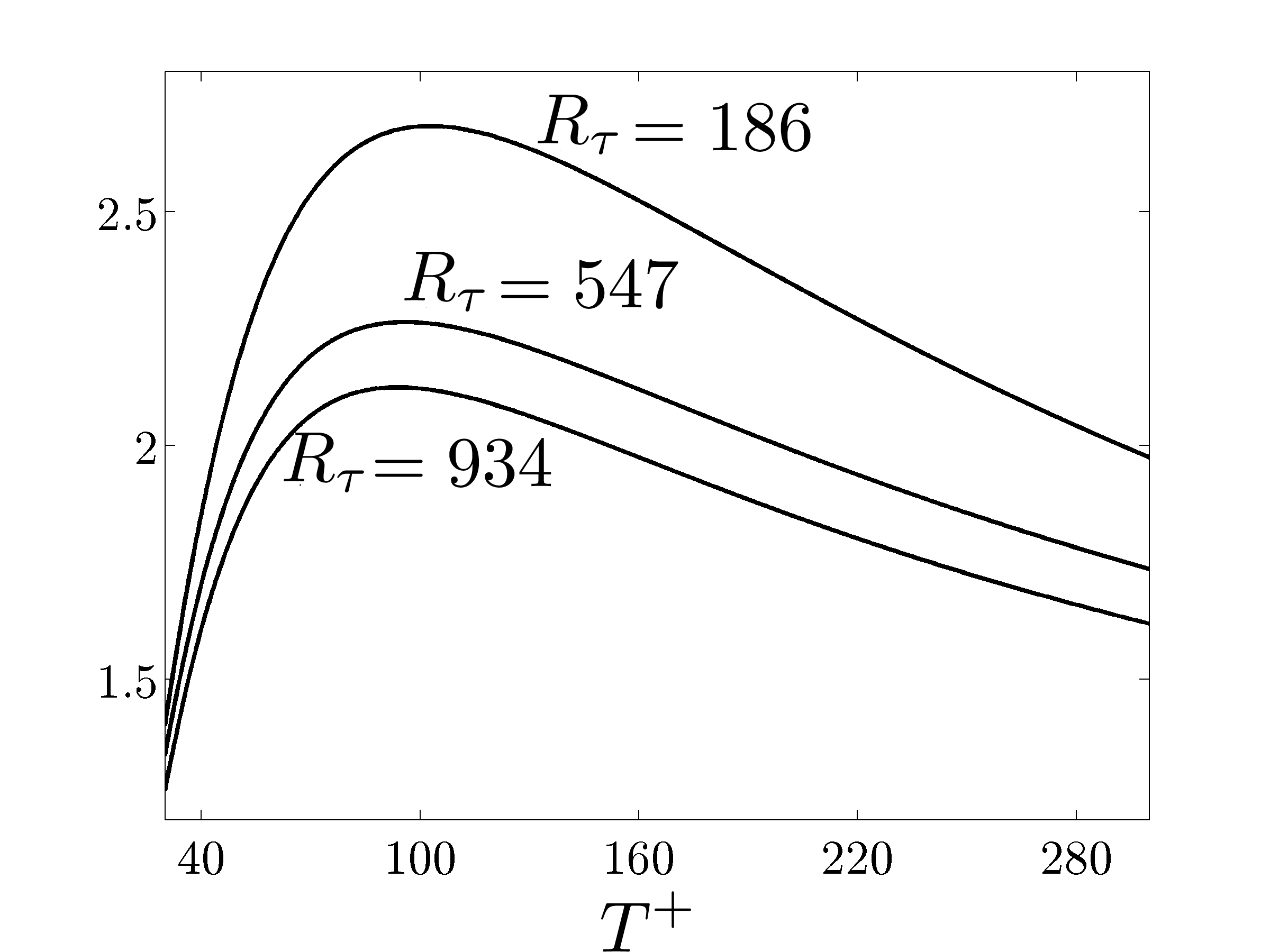}
    \label{fig.Psave2-R186-547-934}}
    &
    \subfigure[]{\includegraphics[width=0.49\columnwidth]
    {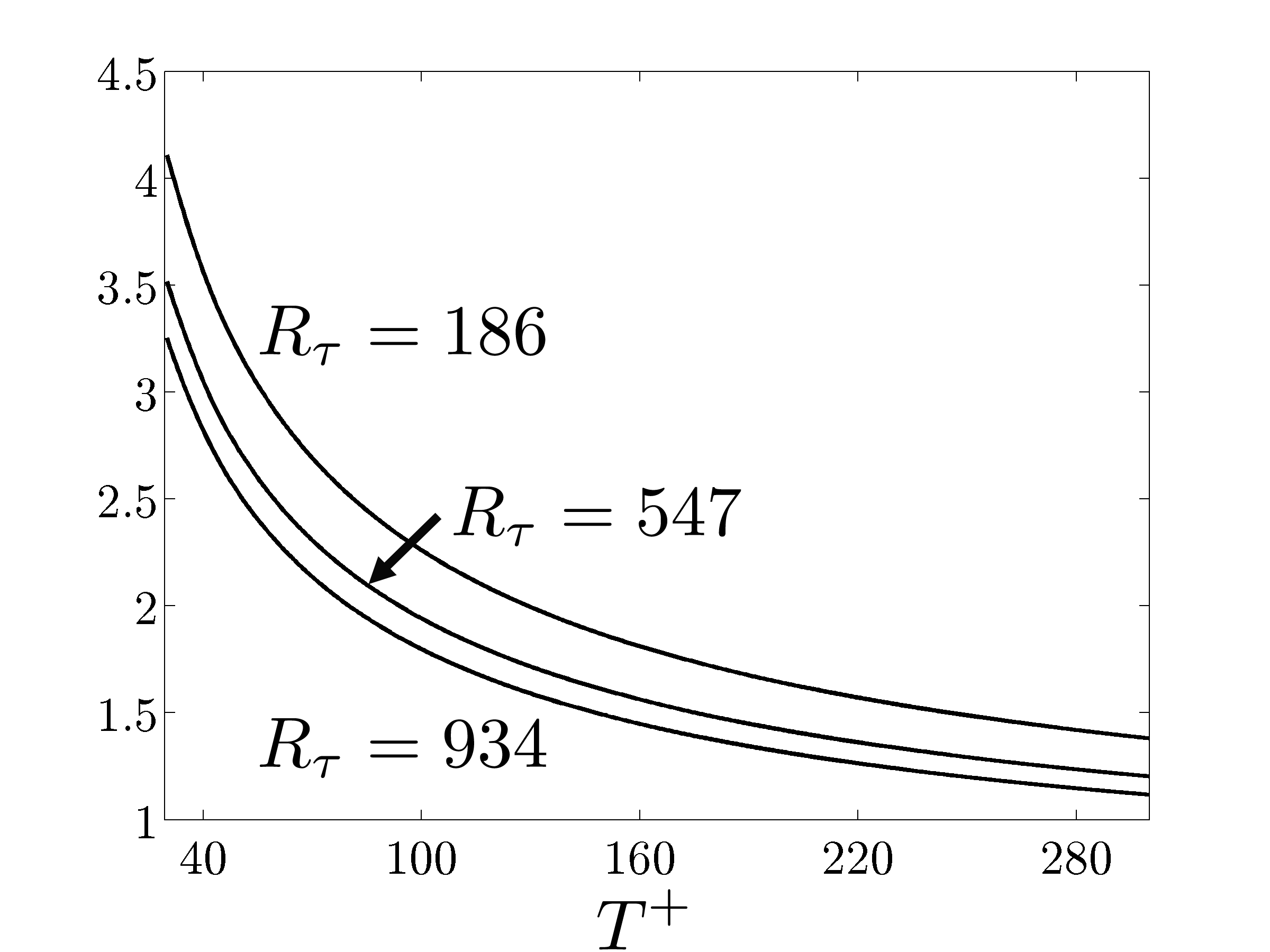}
    \label{fig.Preq0-R186-547-934}}
    \\
    $\% \Pi_{\mathrm{req},2} (T^+)$
    &
    $\% \Pi_{\mathrm{net},2} (T^+)$
    \\[-0.2cm]
    \subfigure[]{\includegraphics[width=0.49\columnwidth]
    {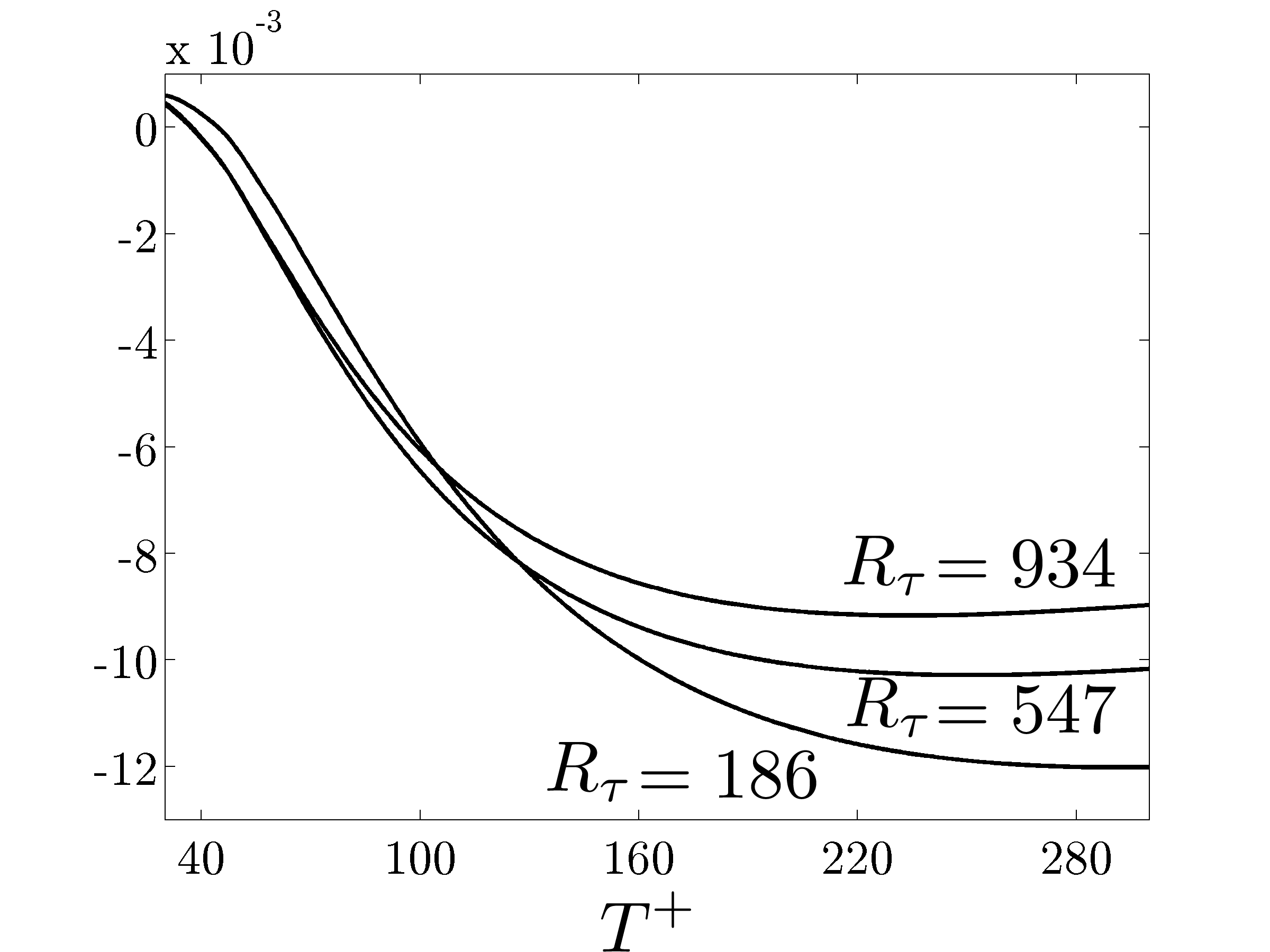}
    \label{fig.Preq2-R186-547-934}}
    &
    \subfigure[]{\includegraphics[width=0.49\columnwidth]
    {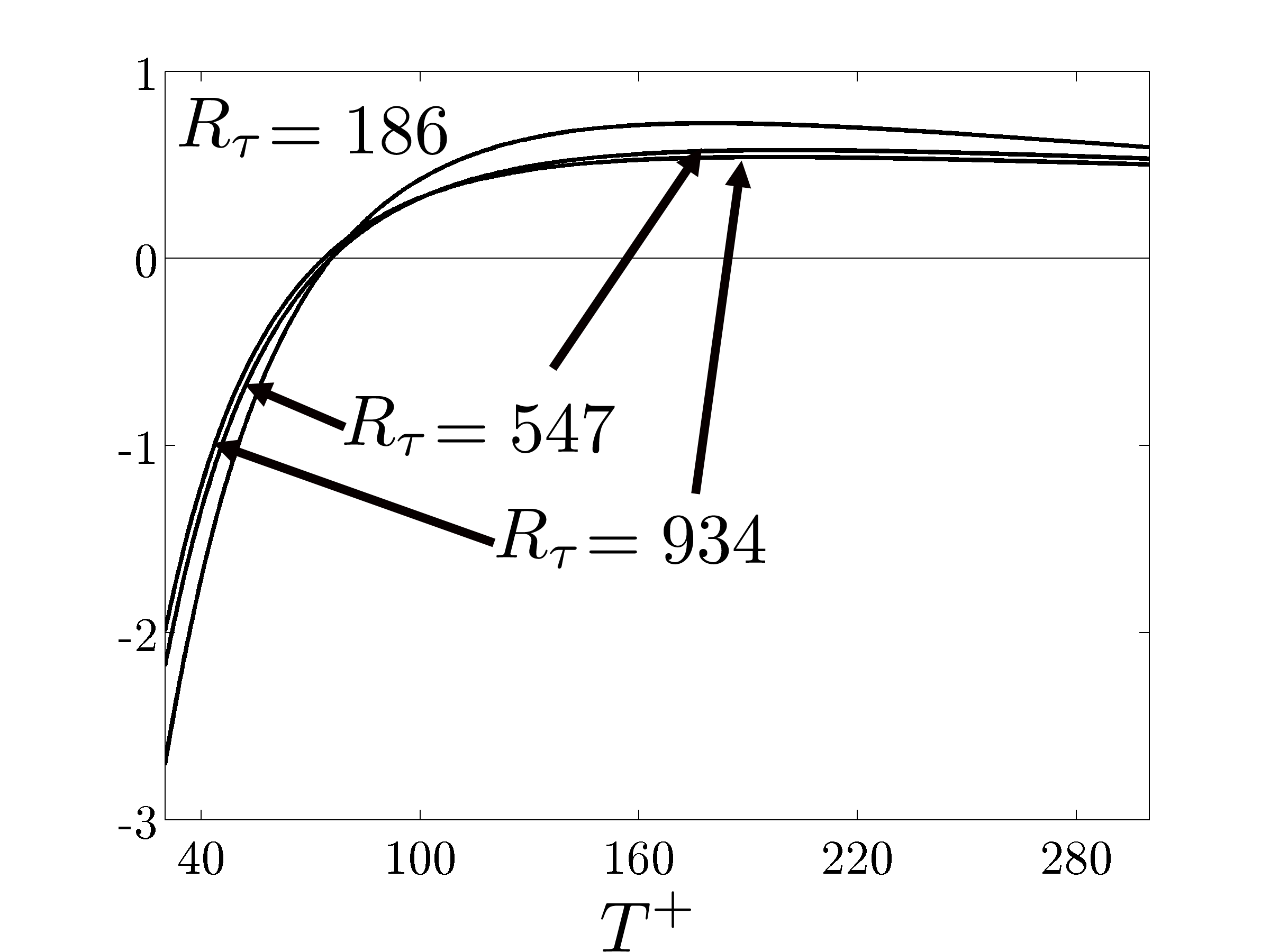}
    \label{fig.Pnet2-R186-547-934}}
    \end{tabular}
    \end{center}
    \caption{
    The second-order correction to
    (a) the saved power, $\% \Pi_{\mathrm{save},2} (T^+)$;
    (b) the required power, $\% \Pi_{\mathrm{req},0} (T^+)$; and
    (d) the net efficiency, $\% \Pi_{\mathrm{net},2} (T^+)$; and
    (c) the fourth-order correction to the required power, $\% \Pi_{\mathrm{req},2} (T^+)$,
    as a function of the period of oscillations $T^+$ for the flows with $R_\tau = 186$, $547$, and $934$.
    }
    \label{fig.Psave2-req0-req2-net2-R186-547-934}
    \end{figure}

\subsection{Drag reduction in flows with larger Reynolds numbers}
    \label{sec.larger-Rtau}

After assessing utility of perturbation analysis for flows with $R_\tau = 186$, we turn our attention to the effect of control at $R_\tau = 547$ and $934$. Figure~\ref{fig.Psave2-R186-547-934} shows that the second-order correction to the saved power, $\% \Pi_{\mathrm{save},2}$, is positive for all $R_\tau$, and that the optimal $T^+$ slightly decreases with $R_\tau$ ($T^+ = 102.5$, $96$, and $94$ for $R_\tau = 186$, $547$, and $934$, respectively). The corresponding periods of oscillations in outer units, $T = T^+/R_\tau$, are $T = 0.55$, $0.18$, and $0.10$. Therefore, as $R_\tau$ increases, larger frequency of oscillations is required for the optimal drag reduction. In addition, since $\% \Pi_{\mathrm{save},2}$ decreases with $R_\tau$, the drag-reducing ability of wall oscillations deteriorates at higher Reynolds numbers, which is in agreement with the observations of~\cite{choxusun02,ricwu04,ricqua08,toules12}. Our results suggest that the amount of drag reduction scales as $R_\tau^{-0.15}$, which agrees fairly well with the decline in drag reduction $R_\tau^{-0.2}$ found by~\cite{choxusun02,toules12}. Our analysis also demonstrates that for $T^+ = \{ 40$, $100$, $160$, $220 \}$ the second order correction to the saved power reduces by $\{ 10.4\%$, $20.2\%$, $20.8\%$, $18.8\% \}$ when $R_\tau$ increases from $186$ to $547$, and by $\{ 16.6\%$, $27.2\%$, $28.1\%$, and $26.1\% \}$ when $R_\tau$ increases from $186$ to $934$, respectively. This is in line with deterioration of $18\%$ observed in the DNS of~\cite{toules12} when $R_\tau$ increases from $200$ to $500$. In addition, our predictions qualitatively agree with the results of~\cite{ricqua08} where it was shown that deteriorating effect of $R_\tau$ on drag reduction becomes stronger with increase in $T^+$.

Figure~\ref{fig.Preq0-R186-547-934} shows monotonic decrease of $\% \Pi_{\mathrm{req},0} (T^+)$ with both $T^+$ and $R_\tau$. We also note that a product between the bulk flux $U_B$ and $\% \Pi_{\mathrm{req},0}$ does not change with $R_\tau$. This demonstrates invariance under change in $R_\tau$ of the second-order correction to the required power (before normalization is done).
{~\cite{ricqua08} noted that since $U_B$ approximately scales as $R_\tau^{0.136}$, the required power should approximately scale as $R_\tau^{-0.136}$. For $T^+ = \{ 40$, $100$, $160$, $220 \}$ and $r = \{ 0.145$, $0.141$, $0.138$, $0.135 \}$, respectively, we indeed obtain very weak variations of $\% \Pi_{\mathrm{req},0}/R_\tau^{-r}$ with $R_\tau$.}
The fourth-order correction to the required power, $\% \Pi_{\mathrm{req},2}$, is also reduced as $R_\tau$ increases; see figure~\ref{fig.Preq2-R186-547-934}. The difference between the quantities shown in figures~\ref{fig.Psave2-R186-547-934} and~\ref{fig.Preq0-R186-547-934} determines the second-order correction to the net efficiency, $\% \Pi_{\mathrm{net},2}$. From figure~\ref{fig.Pnet2-R186-547-934} we see that the largest net efficiency reduces with $R_\tau$, and that $\% \Pi_{\mathrm{net},2}$ saturates for large Reynolds numbers.

    \begin{figure}
    \begin{center}
    \begin{tabular}{cc}
    $\nu_{T0} (y^+)$
    &
    $U_{0} (y^+)$
    \\[-0.2cm]
    \subfigure[]{\includegraphics[width=0.49\columnwidth]
    {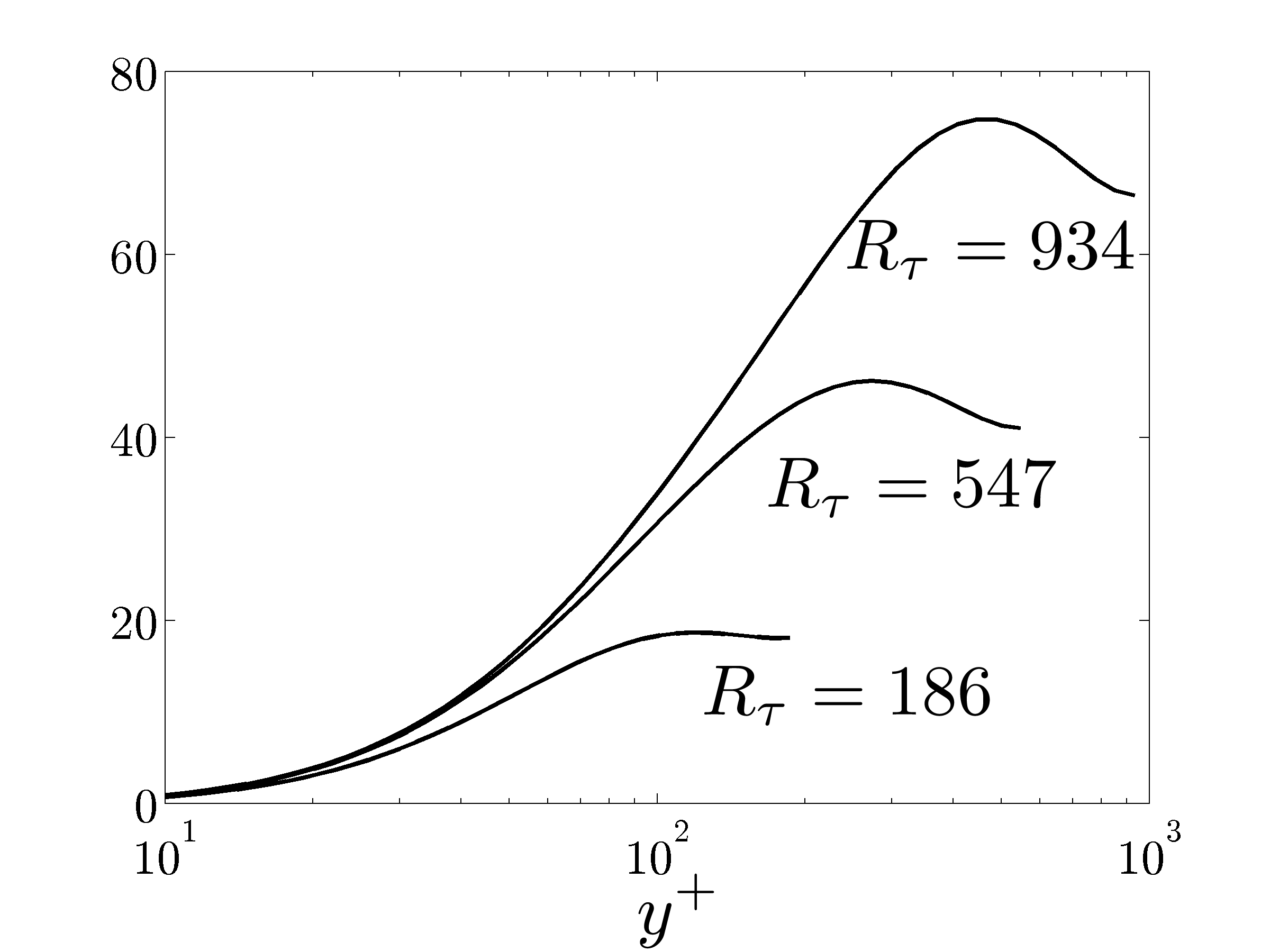}
    \label{fig.nuT0-R186-547-934}}
    &
    \subfigure[]{\includegraphics[width=0.49\columnwidth]
    {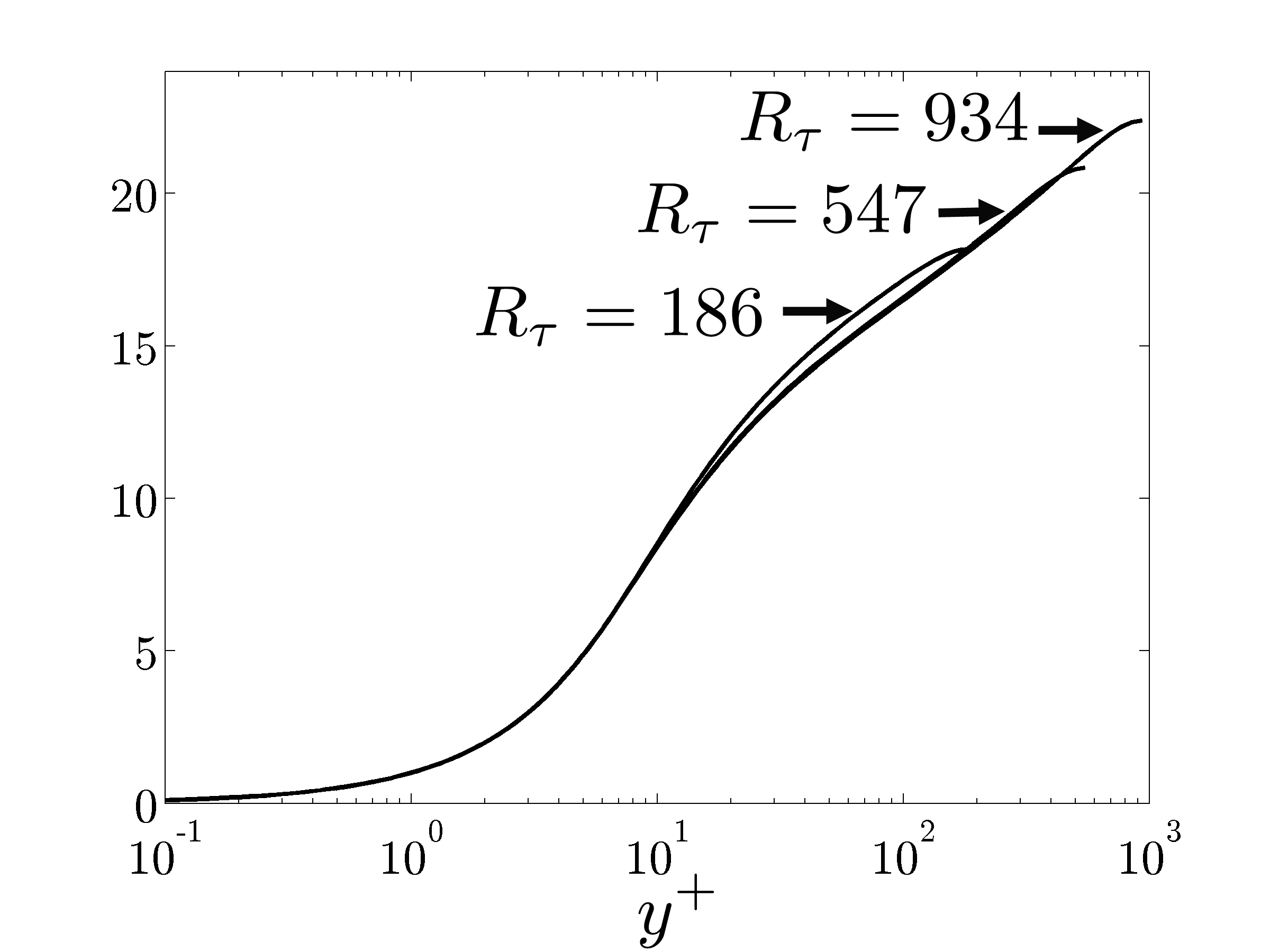}
    \label{fig.U0-R186-547-934}}
    \end{tabular}
    \end{center}
    \caption{
    (a) The turbulent viscosity, $\nu_{T0} (y^+)$; and
    (b) the turbulent mean streamwise velocity, $U_{0} (y^+)$, in the uncontrolled flows with $R_\tau = 186$, $547$, and $934$.
    }
    \label{fig.U0-nuT0-R186-547-934}
    \end{figure}

    \begin{figure}
    \begin{center}
    \begin{tabular}{cc}
    $\nu_{T2} (y^+; T^+)$
    &
    $U_{2} (y^+; T^+)$
    \\[-0.2cm]
    \subfigure[]{\includegraphics[width=0.49\columnwidth]
    {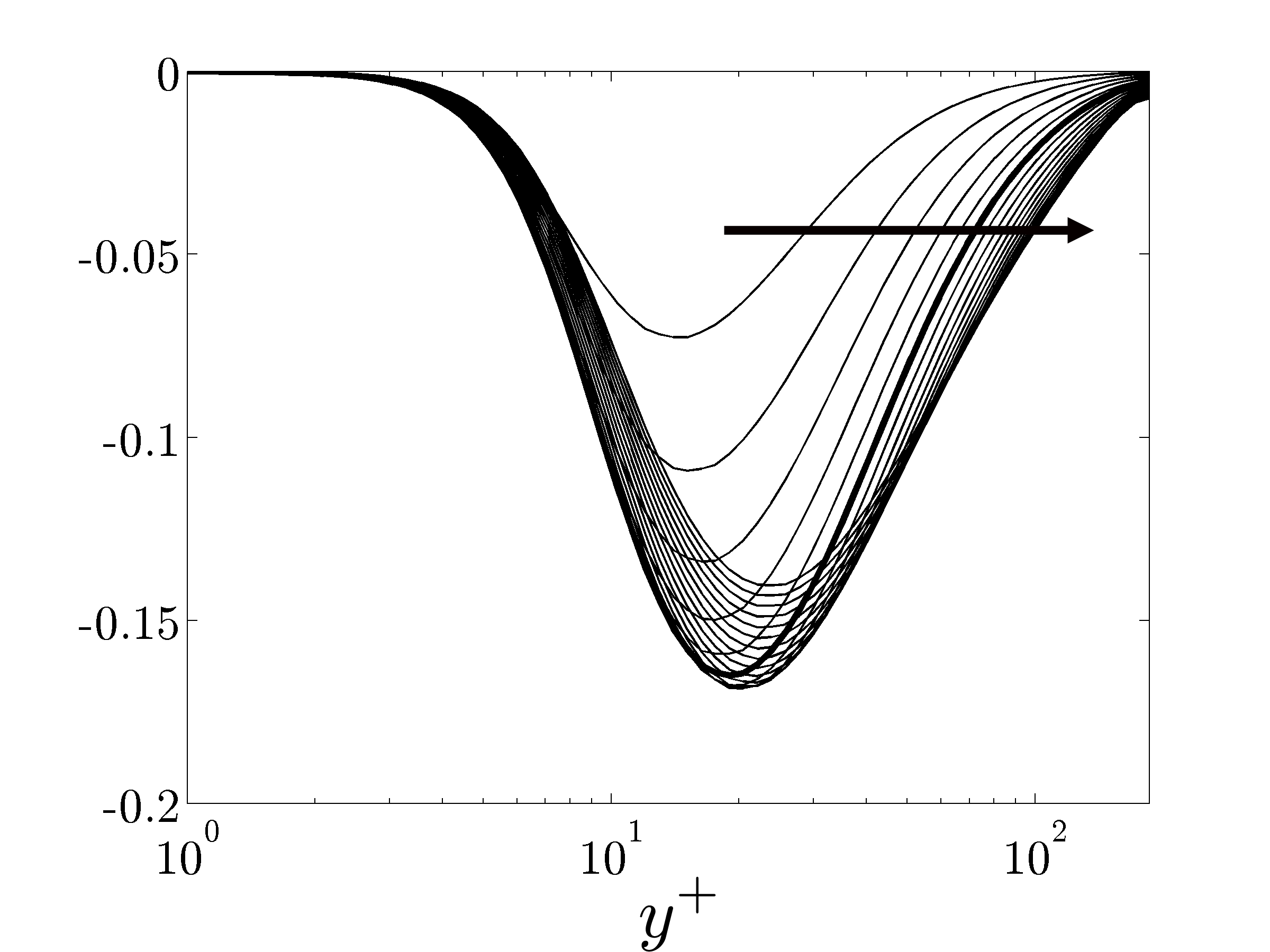}
    \label{fig.nuT2-Tplus-R186}}
    &
    \subfigure[]{\includegraphics[width=0.49\columnwidth]
    {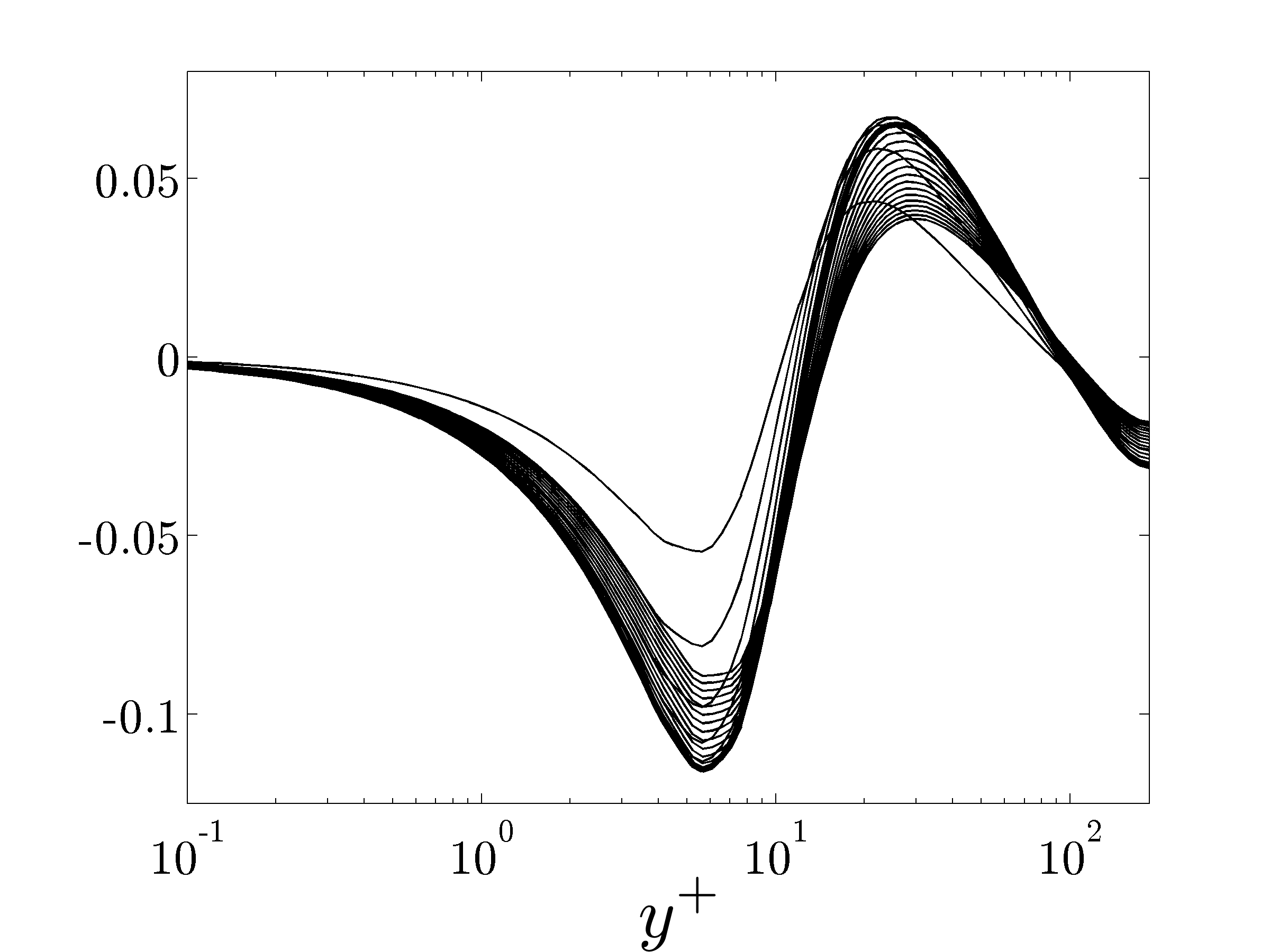}
    \label{fig.U2-Tplus-R186}}
    \end{tabular}
    \end{center}
    \caption{
    Second-order correction to (a) the turbulent viscosity, $\nu_{T2} (y^+; T^+)$; and
    (b) the mean streamwise velocity, $U_{2} (y^+; T^+)$, for $R_\tau = 186$ and different values of $30 \leq T^+ \leq 300$, where $T^+$ increases in the direction of the arrows. The thick curves correspond to the $T^+$ that yields the largest drag reduction (cf.\ figure~\ref{fig.U2-nuT2-maxT-R186-547-934}).
    }
    \label{fig.U2-nuT2-Tplus-R186}
    \end{figure}

\subsection{Effect of control on turbulent viscosity and turbulent mean velocity}
    \label{sec.effect-control-velocity}

We next examine the effect of wall oscillations on the turbulent viscosity and the turbulent mean velocity. Figure~\ref{fig.U0-nuT0-R186-547-934} shows $\nu_{T0} (y^+)$ and $U_{0} (y^+)$ for $R_\tau = 186$, $547$, and $934$. We note that the profiles of $U_0$ for different $R_\tau$ lie on the top of each other, and that $\nu_{T0}$ does not scale in wall units. In particular, the peak of $\nu_{T0}$ takes place at $y^+ \approx R_\tau/2$, approximately half way between the walls and the channel centerline. On the other hand, the effect of control on $\nu_{T}$ is strongest in the viscous wall region, $y^+ < 50$; see figure~\ref{fig.nuT2-Tplus-R186}.

Figure~\ref{fig.U2-nuT2-Tplus-R186} shows the second-order corrections to $\nu_{T2}$ and $U_{2}$ for $R_\tau = 186$ and $30 \leq T^+ \leq 300$. Since $\nu_{T2} < 0$ for all $T^+$, perturbation analysis up to second order in $\alpha$ predicts turbulence suppression for all periods of oscillations; see figure~\ref{fig.nuT2-Tplus-R186}. Furthermore, turbulence suppression region shifts away from the walls with increase in $T^+$. We note that suppression of the turbulent bursting process was observed in DNS for $25 \leq T^+ \leq 200$~\citep{junmanakh92}. Figure~\ref{fig.U2-Tplus-R186} shows that the oscillations reduce the mean velocity gradient in the immediate vicinity of the walls ($U_{2} < 0$ for $y^+ \lesssim 13$). On the other hand, the mean velocity is shifted upward in the log-law region.Both these trends were previously observed in the experiments~\citep{cho02} and DNS~\citep{chodebcla98,barqua96}.~\cite{cho02} argued that the negative spanwise vorticity introduces near wall modifications to the mean streamwise velocity, and that it suppresses the production of turbulence by weakening the vortex stretching mechanism. Our perturbation analysis underestimates the value of $y^+$ above which the shift-up in $U$ takes place; $y^+ > 13$ vs.\ $y^+ \gtrsim 30$ in DNS~\citep{barqua96,toules12} and experiments~\citep{chodebcla98,laaskamor94,ricwu04}.

    \begin{figure}
    \begin{center}
    \begin{tabular}{cc}
    $\nu_{T2} (y^+)$
    &
    $U_{2} (y^+)$
    \\[-0.2cm]
    \subfigure[]{\includegraphics[width=0.49\columnwidth]
    {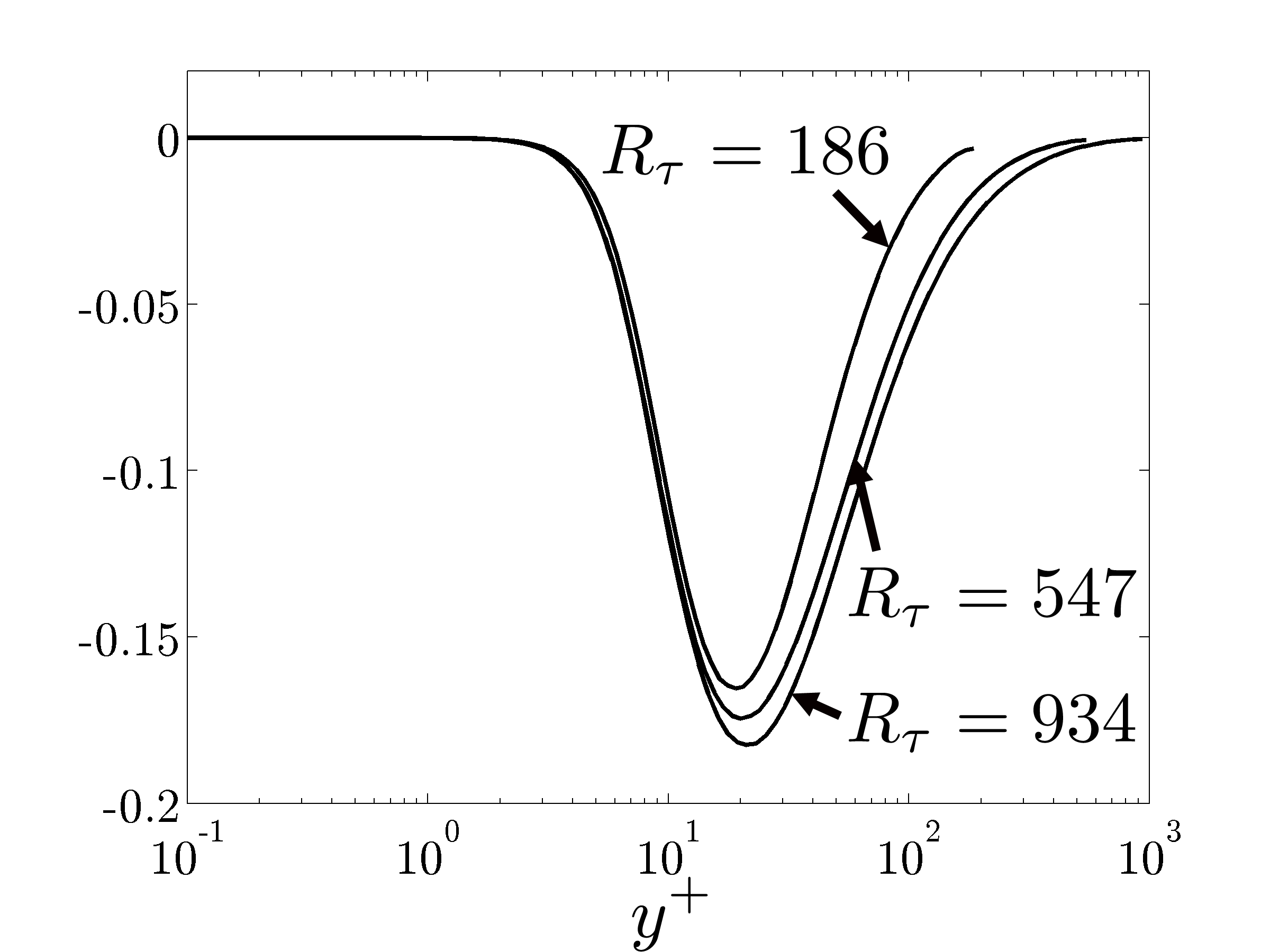}
    \label{fig.nuT2-maxT-R186-547-934}}
    &
    \subfigure[]{\includegraphics[width=0.49\columnwidth]
    {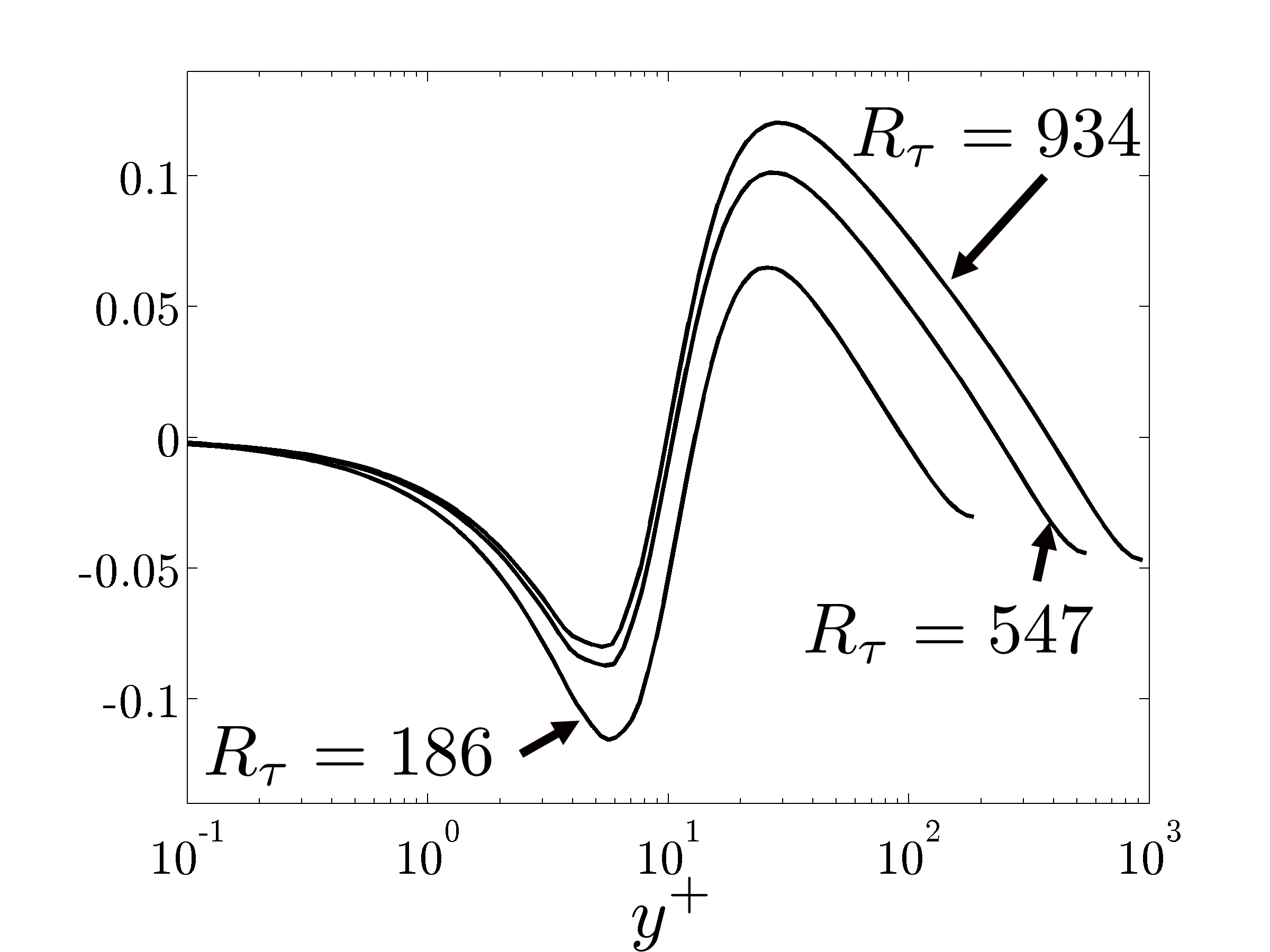}
    \label{fig.U2-maxT-R186-547-934}}
    \end{tabular}
    \end{center}
    \caption{
    The second-order correction to (a) the turbulent viscosity, $\nu_{T2} (y^+)$; and
    (b) the mean streamwise velocity, $U_{2} (y^+)$, for the values of $T^+$ that yield the largest drag reduction in the flows with $R_\tau = 186$, $T^+ = 102.5$; $R_\tau = 547$, $T^+ = 96$; and $R_\tau = 934$, $T^+ = 94$.
    }
    \label{fig.U2-nuT2-maxT-R186-547-934}
    \end{figure}

Figure~\ref{fig.U2-nuT2-maxT-R186-547-934} compares $U_{2}$ and $\nu_{T2}$ at three Reynolds numbers for the values of $T^+$ that induce the largest drag reduction. We see that $\nu_{T2}$ peaks at $y^+ \approx 20$ for all $R_\tau$. This suggests that the optimal drag-reducing frequency minimizes the turbulent viscosity near the interface of the buffer layer and the log-law region. Even though the negative peak of $\nu_{T2}$ increases with $R_\tau$ (cf.\ figure~\ref{fig.nuT2-maxT-R186-547-934}), the ratio of $\nu_{T2}$ and $\nu_{T0}$ decreases with $R_\tau$. Thus, wall oscillations are less effective in suppressing turbulence at larger Reynolds numbers. Finally, figure~\ref{fig.U2-maxT-R186-547-934} shows that the slope of $U_{2}$ decreases with $R_\tau$; this is in agreement with our earlier observation that smaller drag reduction is achieved at higher Reynolds numbers.

\subsection{Effect of control on turbulent kinetic energy}
	\label{sec.control-k-eps}

We next examine the effect of control on the kinetic energy of fluctuations. Figure~\ref{fig.E0-E2-R186} compares the premultiplied two-dimensional energy spectrum of the uncontrolled flow, $\kappa_x \kappa_z \bar{E} (\bkappa)$, with the premultiplied second-order correction to the energy spectrum, $\kappa_x \kappa_z$ $E_2 (\bkappa)$, in the flow subject to wall oscillations with the optimal drag-reducing period $T^+ = 102.5$. The energy spectra are premultiplied by the spatial wavenumbers such that the area under the log-log plot is equal to the total energy of fluctuations. We see that the most energetic modes of the uncontrolled flow take place at $\kappa_x \approx 2.5$ and $\kappa_z \approx 6.5$; cf.\ figure~\ref{fig.E0-R186}. In addition, wall oscillations further amplify the most energetic modes of the uncontrolled flow with small streamwise wavelengths (cf.\ red regions in figure~\ref{fig.E2-R186-T102p5}), while they suppress the most energetic modes of the uncontrolled flow with large streamwise wavelengths (cf.\ blue regions in figure~\ref{fig.E2-R186-T102p5}).
{This agrees well with the study of~\cite{ric11} that examined the effect of spanwise wall oscillations on free-stream vortical structures in a Blasius boundary layer. Over a range of spanwise wavelengths, it was found that wall oscillations amplify (attenuate) disturbances with small (large) streamwise wavelengths. Figure~\ref{fig.E2-R186-T102p5} shows that} the largest energy amplification takes place at $\kappa_x \approx 4.4$ and $\kappa_z \approx 10.2$, and the largest energy suppression occurs at $\kappa_x \approx 0.8$ and $\kappa_z \approx 8.8$. The total effect of control on the kinetic energy can be quantified by $\int_{\bkappa} E_2 (\bkappa) \, \mrd \bkappa / \int_{\bkappa} \bar{E} (\bkappa) \, \mrd \bkappa$, which for wall oscillations with $T^+ = 102.5$ is approximately $-1.5\%$. This yields $7.5\%$ reduction in the total energy of fluctuations for $\alpha = 2.25$.

    \begin{figure}
    \begin{center}
    \begin{tabular}{cc}
    $\kappa_x \kappa_z \bar{E} (\bkappa)$
    &
    $\kappa_x \kappa_z E_2 (\bkappa)$
    \\[-0.2cm]
    \subfigure[]{\includegraphics[width=0.49\columnwidth]
    {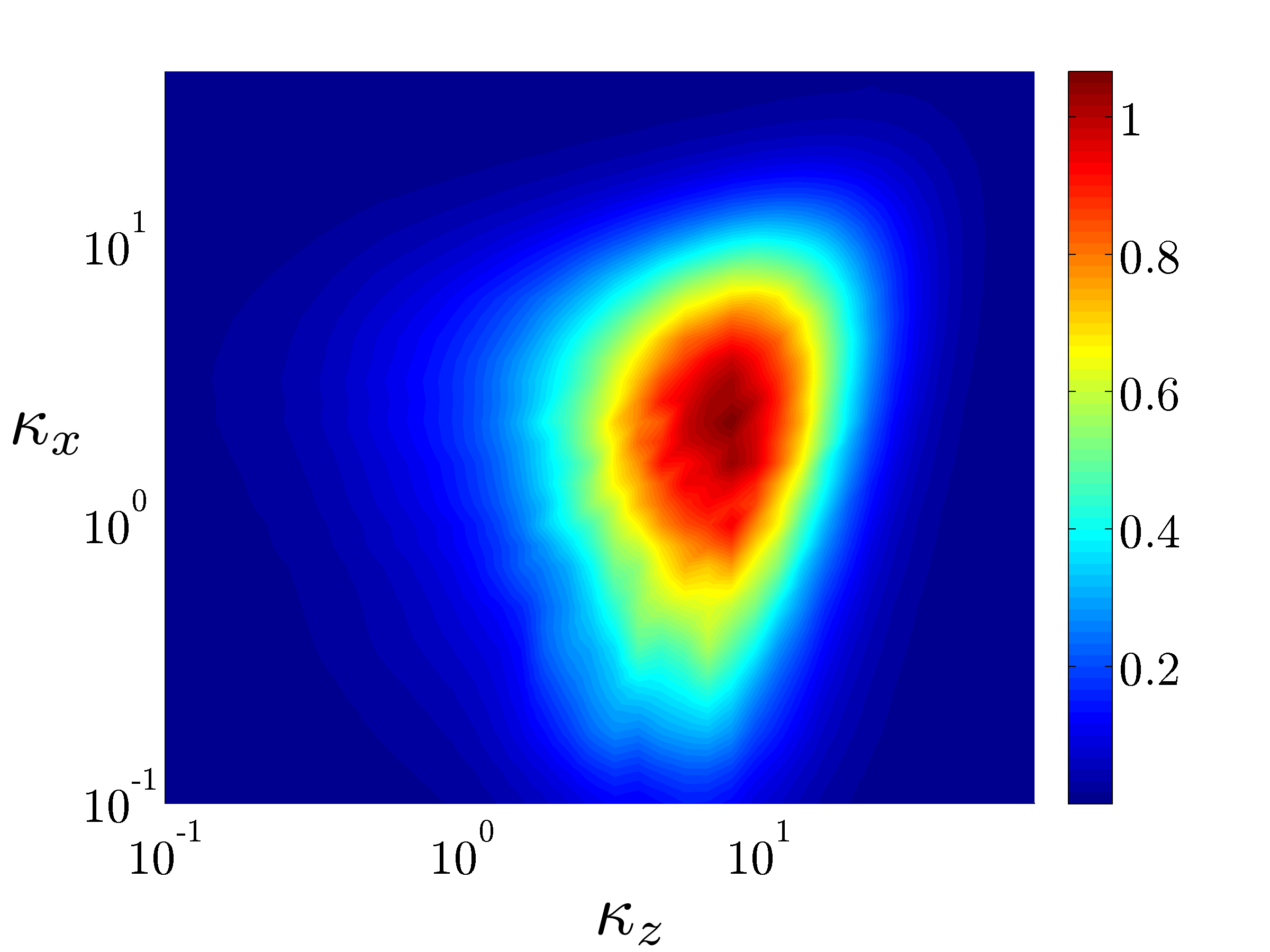}
    \label{fig.E0-R186}}
    &
    \subfigure[]{\includegraphics[width=0.49\columnwidth]
    {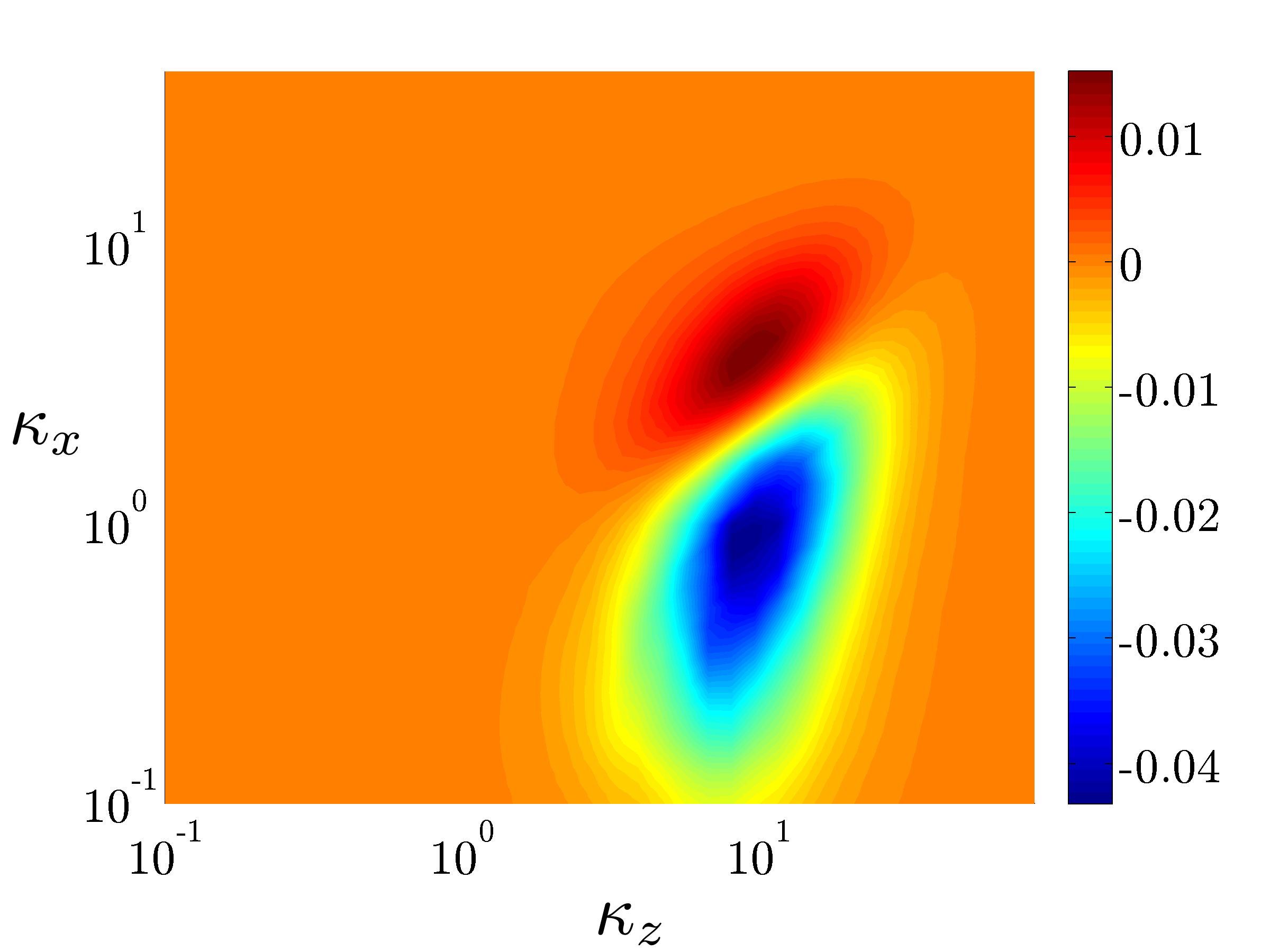}
    \label{fig.E2-R186-T102p5}}
    \end{tabular}
    \end{center}
    \caption{
    (Color online) (a) Premultiplied DNS-based energy spectrum of the uncontrolled flow, $\kappa_x \kappa_z \bar{E} (\bkappa)$, at $R_\tau = 186$~\citep{deljim03}; and (b) second-order correction to the energy spectrum, $\kappa_x \kappa_z E_2 (\bkappa)$, for the flow subject to wall oscillations with optimal drag-reducing period $T^+ = 102.5$.
    }
    \label{fig.E0-E2-R186}
    \end{figure}

The effect of wall oscillations on the turbulent kinetic energy and its rate of dissipation is shown in figure~\ref{fig.k2-epsilon2-R186}. We see that the second-order correction to the kinetic energy, $k_2$, is negative for $T^+ = 102.5$, suggesting that the turbulent kinetic energy is reduced with the largest suppression taking place at $y^+ \approx 8.6$; cf.\ figure~\ref{fig.k2-R186-T102p5}. In addition, the second-order correction to the rate of dissipation of turbulent kinetic energy, $\epsilon_2$, is negative almost everywhere (except for a small region $5.4 \lesssim y^+ \lesssim 8.9$) and the largest reduction occurs in the viscous sublayer $y^+ < 5$; cf. figure~\ref{fig.epsilon2-R186-T102p5}. Therefore, perturbation analysis up to second order in $\alpha$ captures previously made experimental and numerical observations that wall oscillations suppress both the production and dissipation of the turbulent kinetic energy~\citep{junmanakh92,chodebcla98,barqua96,cho02}. Figures~\ref{fig.k0-k0pk2-R186-T102p5-a2p25} and~\ref{fig.epsilon0-epsilon0pepsilon2-R186-T102p5-a2p25} show that, relative to the uncontrolled flow, the wall oscillations with $T^+ = 102.5$ and $\alpha = 2.25$ have more profound influence on the turbulent kinetic energy than on its rate of dissipation. This suggests that the turbulent production is suppressed more than the turbulent dissipation, which explains the reduced turbulent viscosity $\nu_T$ in the flow with control; cf.\ figure~\ref{fig.nuT2-maxT-R186-547-934}. We observe close agreement between the modification to the turbulent kinetic energy in figure~\ref{fig.k0-k0pk2-R186-T102p5-a2p25} and the modification to the dominant component of the Reynolds normal stress reported in the DNS of~\cite{toules12} at $R_\tau = 500$ (their figure 5(b)).

    \begin{figure}
    \begin{center}
    \begin{tabular}{cc}
    $k_2 (y^+)$
    &
    $k_0 (y^+)$; $k_0 (y^+) + \alpha^2 k_2 (y^+)$
    \\[-0.2cm]
    \subfigure[]{\includegraphics[width=0.49\columnwidth]
    {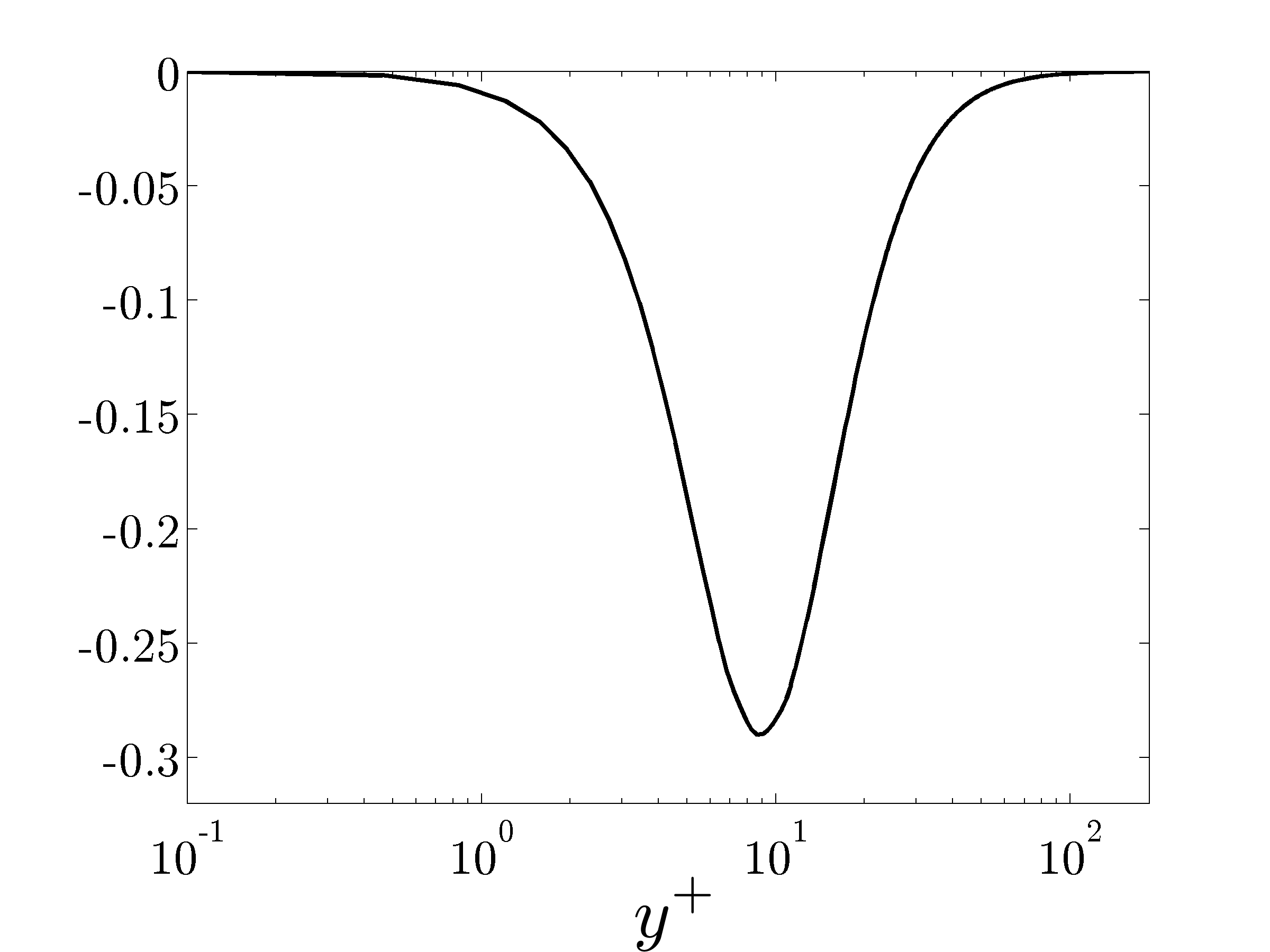}
    \label{fig.k2-R186-T102p5}}
    &
    \subfigure[]{\includegraphics[width=0.49\columnwidth]
    {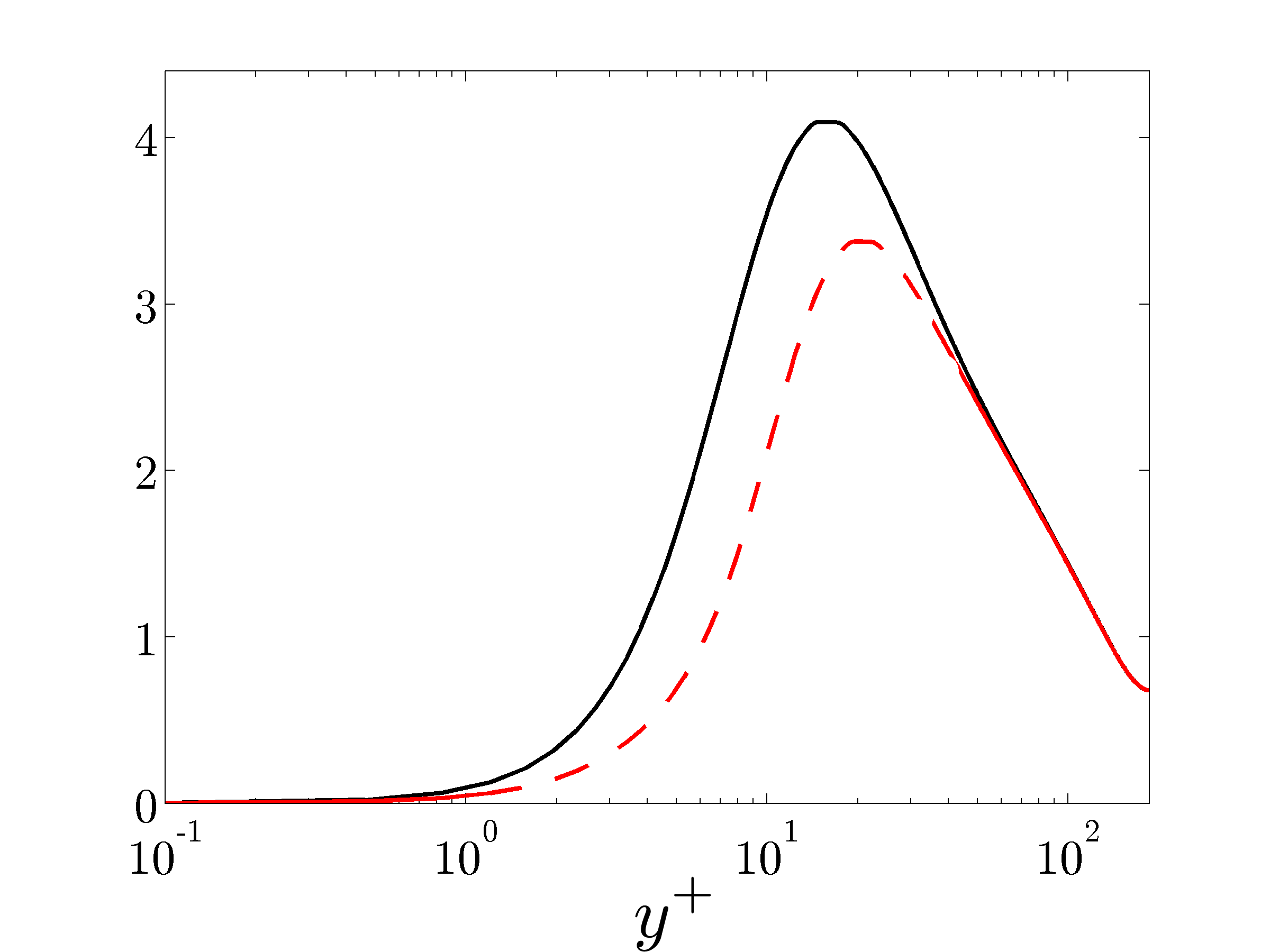}
    \label{fig.k0-k0pk2-R186-T102p5-a2p25}}
    \\[0.2cm]
    $\epsilon_2 (y^+)$
    &
    $\epsilon_0 (y^+)$; $\epsilon_0 (y^+) + \alpha^2 \epsilon_2 (y^+)$
    \\[-0.2cm]
    \subfigure[]{\includegraphics[width=0.49\columnwidth]
    {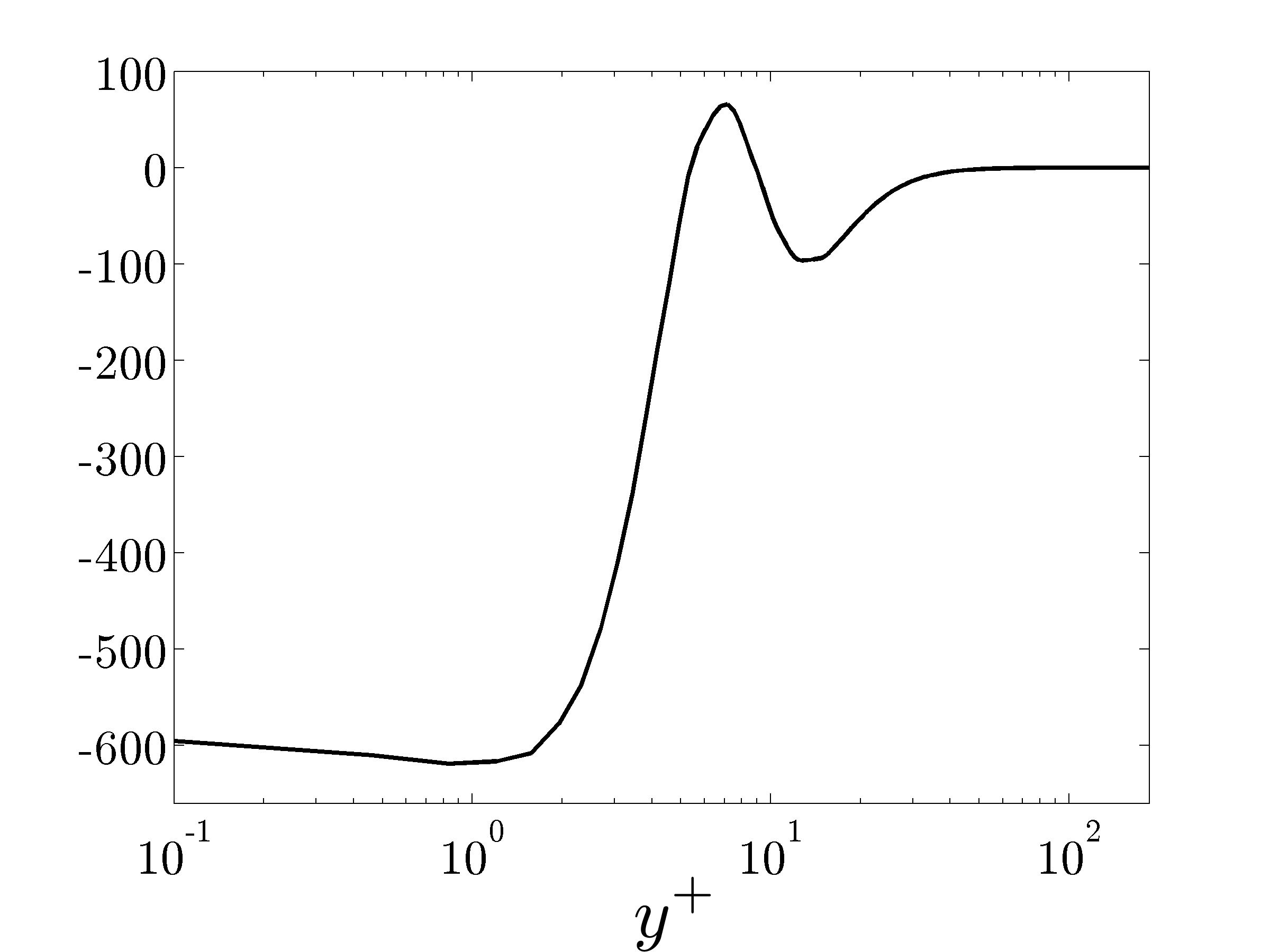}
    \label{fig.epsilon2-R186-T102p5}}
    &
    \subfigure[]{\includegraphics[width=0.49\columnwidth]
    {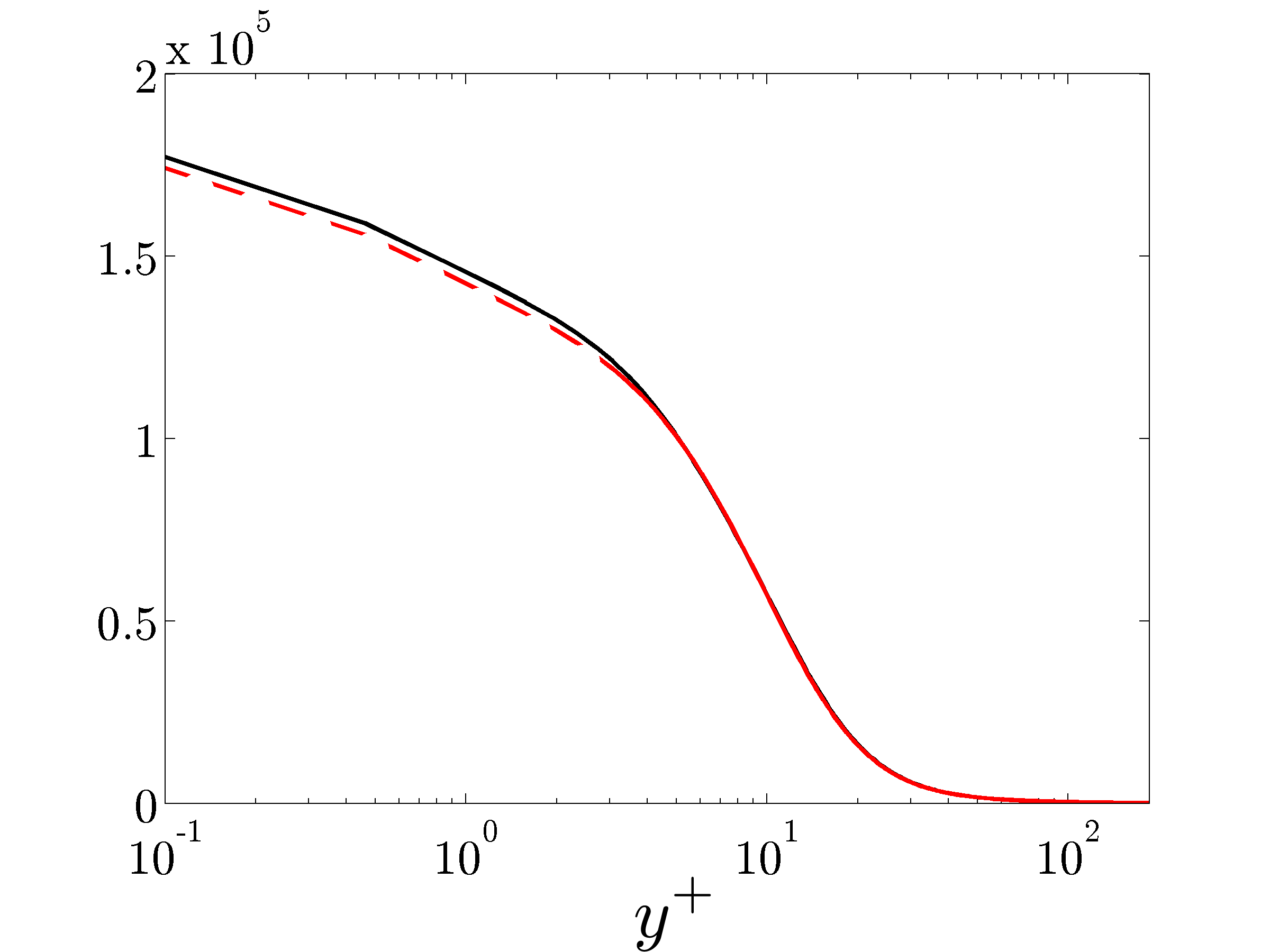}
    \label{fig.epsilon0-epsilon0pepsilon2-R186-T102p5-a2p25}}
    \end{tabular}
    \end{center}
    \caption{
    (Color online) (Left column) Second-order correction to {(a)} the turbulent kinetic energy $k_2 (y^+)$, and {(c)} its rate of dissipation $\epsilon_2 (y^+)$, for the flow subject to wall oscillations with optimal drag-reducing period $T^+ = 102.5$ at $R_\tau = 186$. (Right column) {(b)} Comparison between the turbulent kinetic energy in the uncontrolled flow $k_0$ (black) and in the flow with control $k_0 + \alpha^2 k_2$ (dashed red); {(d)} Comparison between the rate of dissipation of turbulent kinetic energy in the uncontrolled flow $\epsilon_0$ (black), and in the flow with control $\epsilon_0 + \alpha^2 \epsilon_2$ (dashed red), for $\alpha = 2.25$.
    }
    \label{fig.k2-epsilon2-R186}
    \end{figure}

\section{Turbulent flow structures}
	\label{sec.control-structures}

In this section, we use stochastically forced linearized model~(\ref{eq.NS-lin}) to examine the effect of wall oscillations on the turbulent flow structures. We only present results for $R_\tau = 186$ and note that similar flow structures are observed for all Reynolds numbers that we have considered.

We decompose the velocity field into \emph{characteristic eddies\/}~\citep{moimos89} by determining the spatial structure of the fluctuations that contribute most to the kinetic energy at a given $\bkappa = (\kappa_x,\kappa_z)${; see Appendix~\ref{sec.charac-eddy} for details}. It is worth noting that the dominant characteristic eddy resulting from the analysis of the stochastically forced linearized model in the uncontrolled flow qualitatively agrees with the results obtained using eigenvalue decomposition of the DNS-based autocorrelation matrices; compare figure~\ref{fig.u-omega_z-R186-T102p5-a2p25} of this paper with figure 15 in~\cite{moimos89}.

    \begin{figure}
    \begin{center}
    \begin{tabular}{cc}
    Uncontrolled
    &
    Controlled: $T^+ = 102.5$, $\alpha = 2.25$
    \\[-0.2cm]
    \subfigure[]{\includegraphics[height=4cm, width=0.49\columnwidth]
    {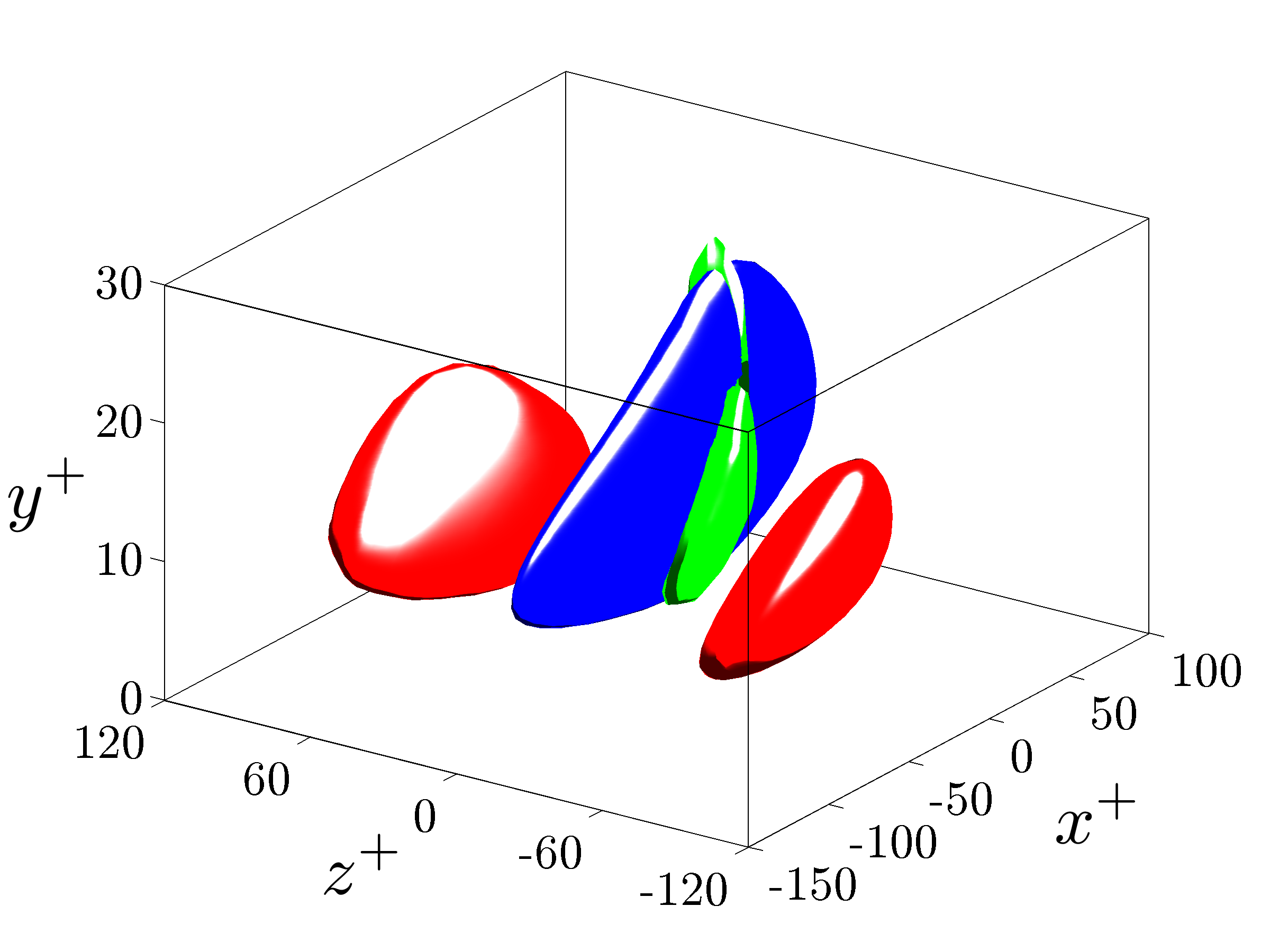}
    \label{fig.u-vortex-0-3D-R186}}
    &
    \subfigure[]{\includegraphics[height=4cm, width=0.49\columnwidth]
    {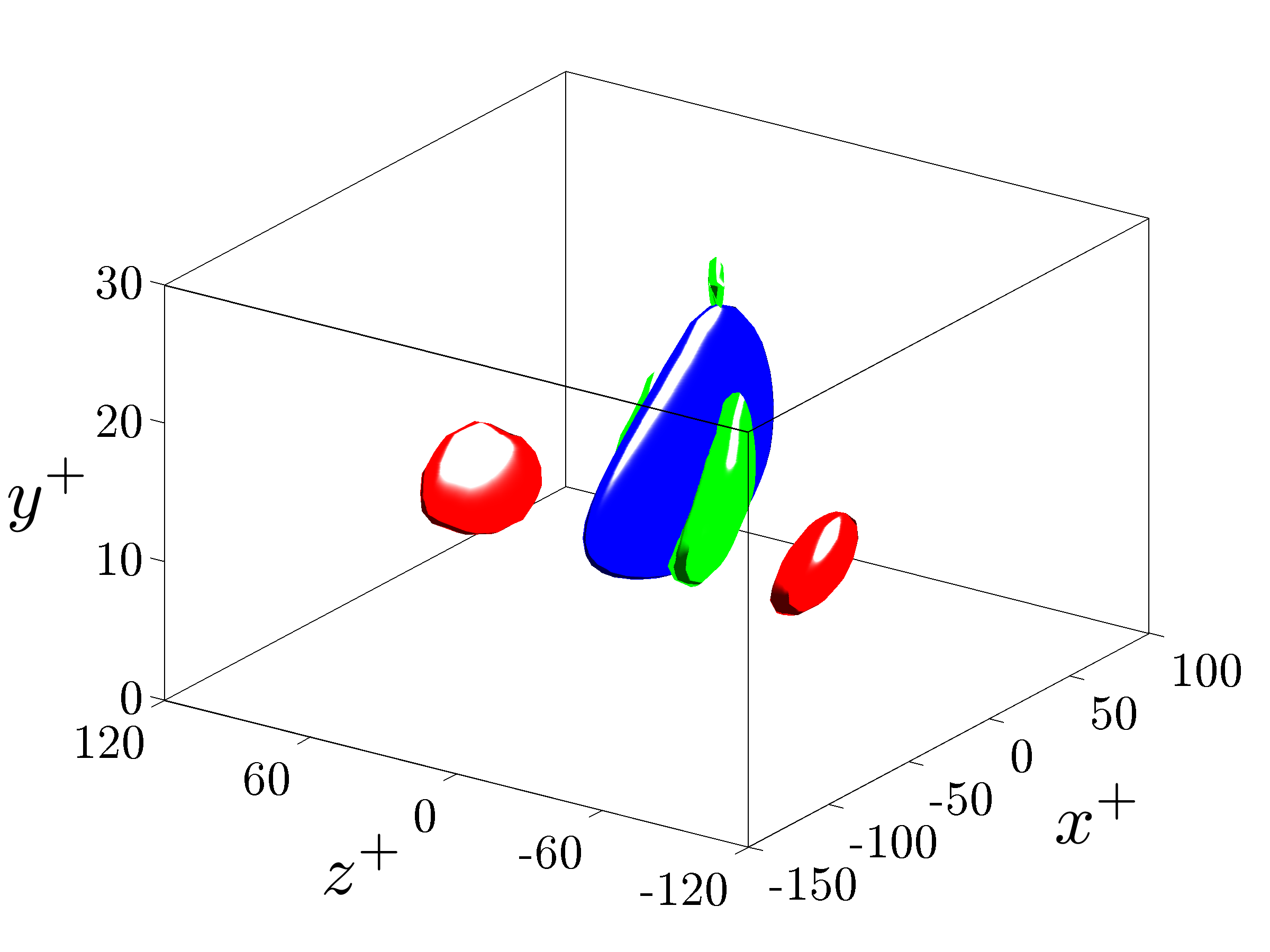}
    \label{fig.u-vortex-2-3D-R186-T102p5-a2p25-txt}}
    \\[0.2cm]
    \subfigure[]{\includegraphics[height=4cm, width=0.49\columnwidth]
    {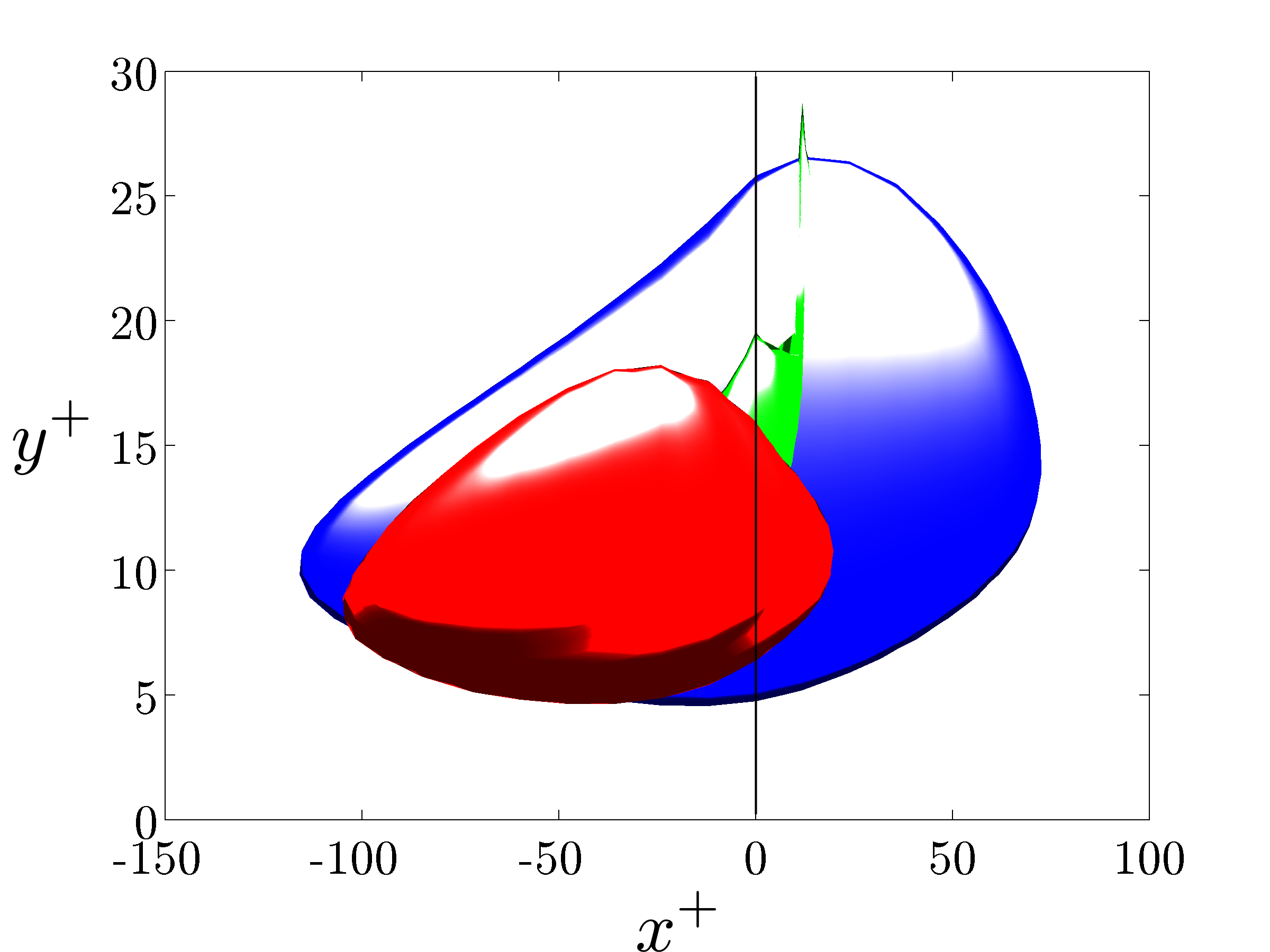}
    \label{fig.u-vortex-0-xy-R186}}
    &
    \subfigure[]{\includegraphics[height=4cm, width=0.49\columnwidth]
    {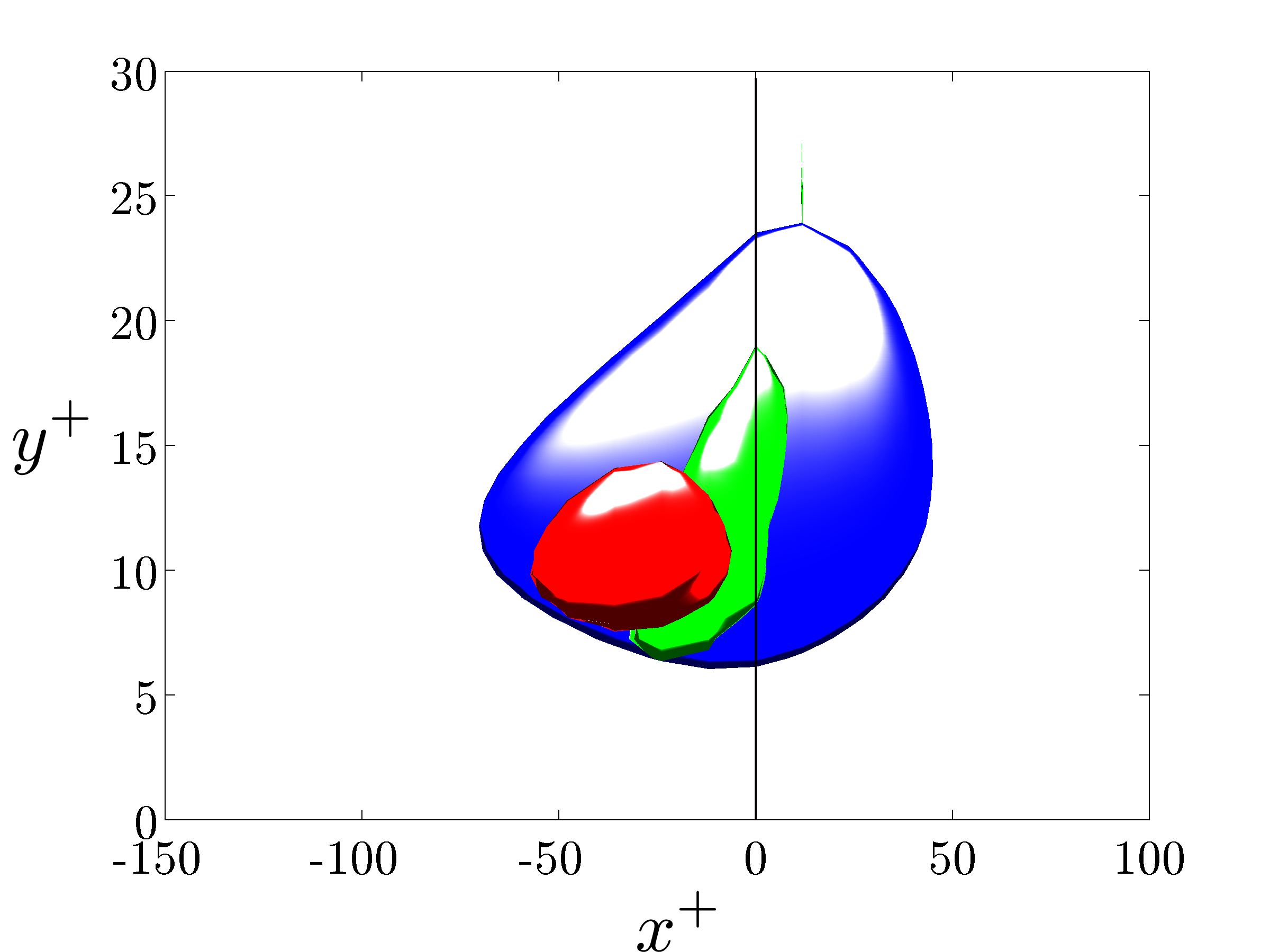}
    \label{fig.u-vortex-2-xy-R186-T102p5-a2p25}}
    \\[0.2cm]
    \subfigure[]{\includegraphics[height=4cm, width=0.49\columnwidth]
    {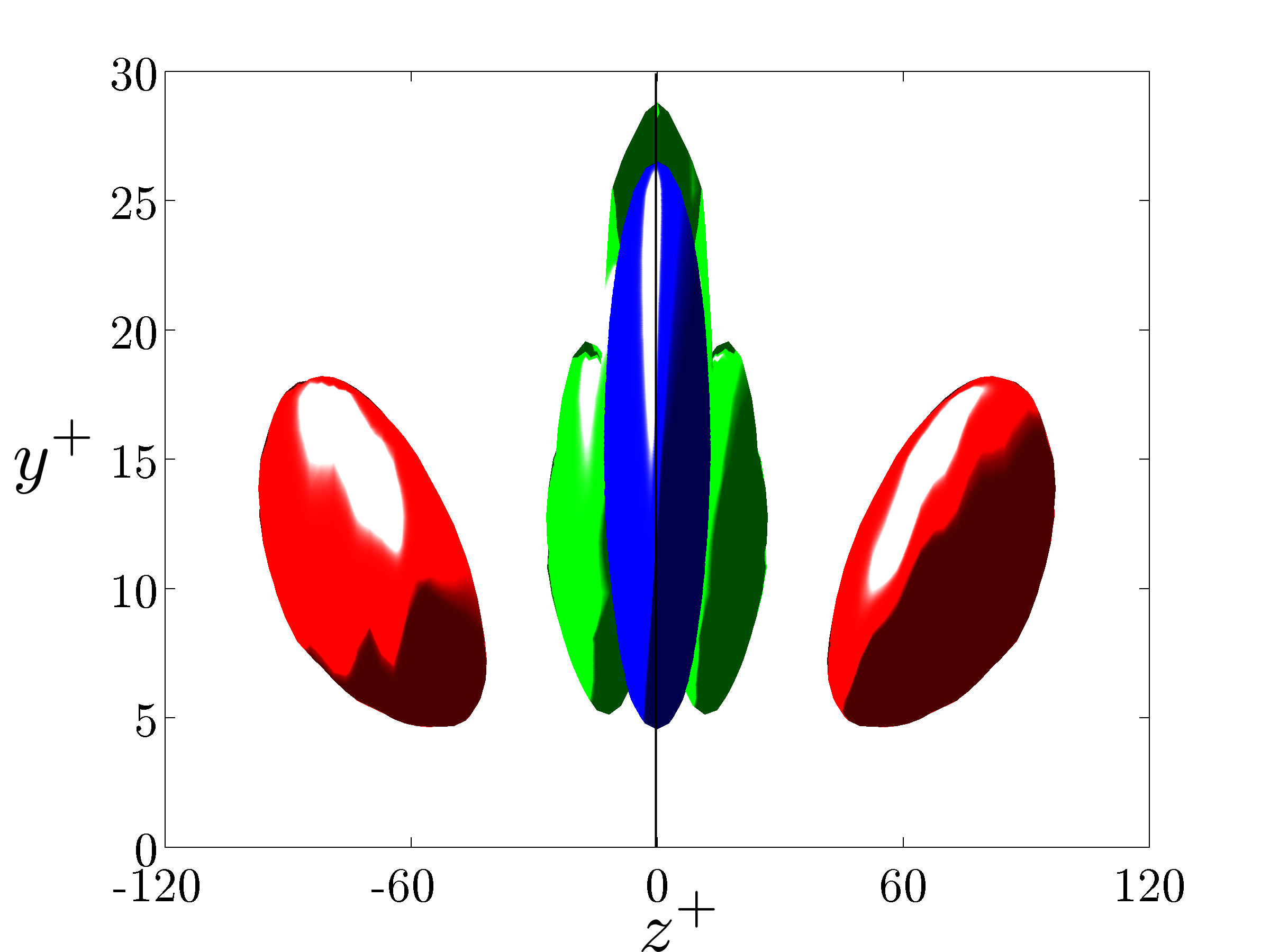}
    \label{fig.u-vortex-0-yz-R186}}
    &
    \subfigure[]{\includegraphics[height=4cm, width=0.49\columnwidth]
    {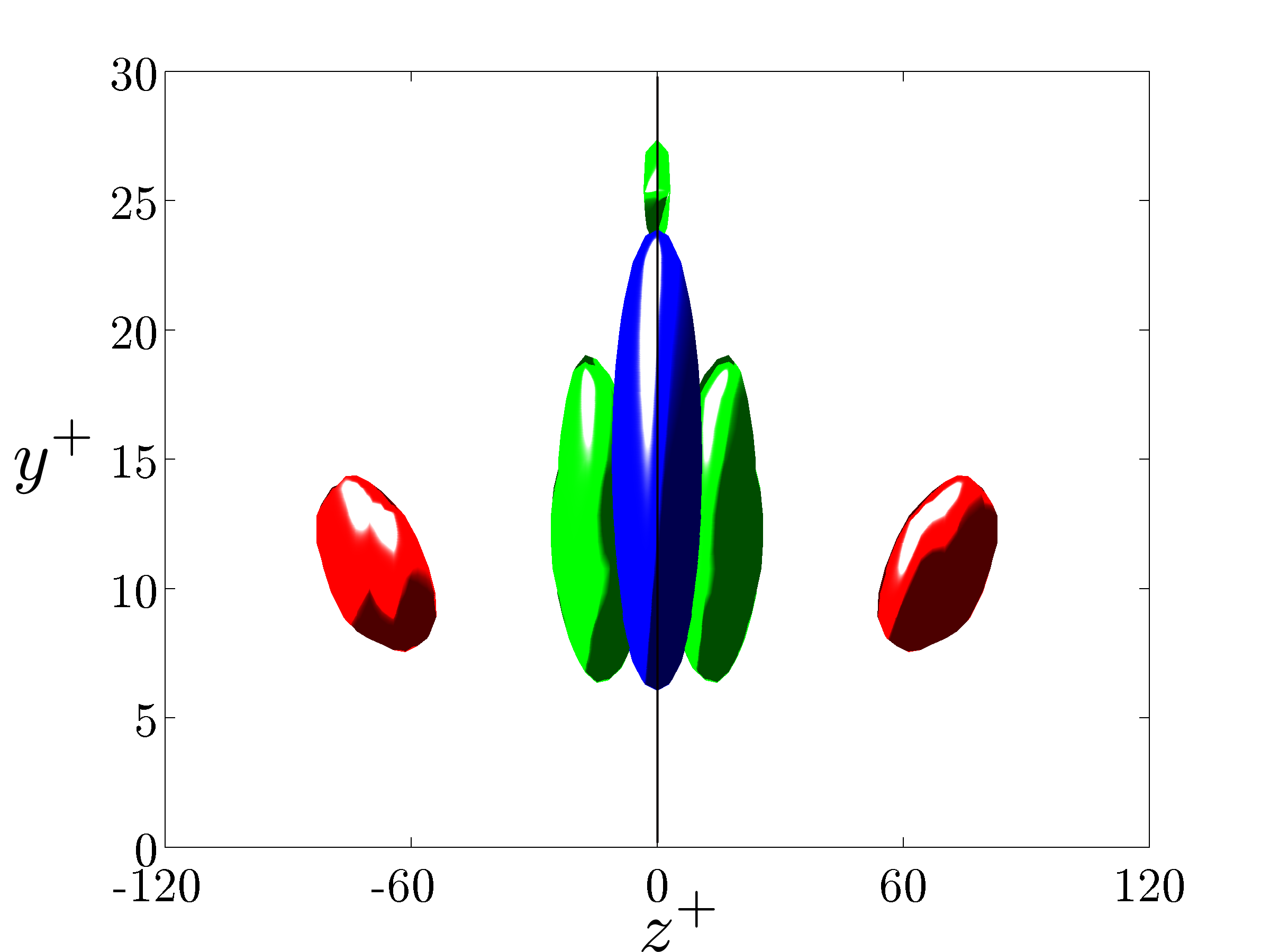}
    \label{fig.u-vortex-2-yz-R186-T102p5-a2p25}}
    \\[0.2cm]
    \subfigure[]{\includegraphics[height=4cm, width=0.49\columnwidth]
    {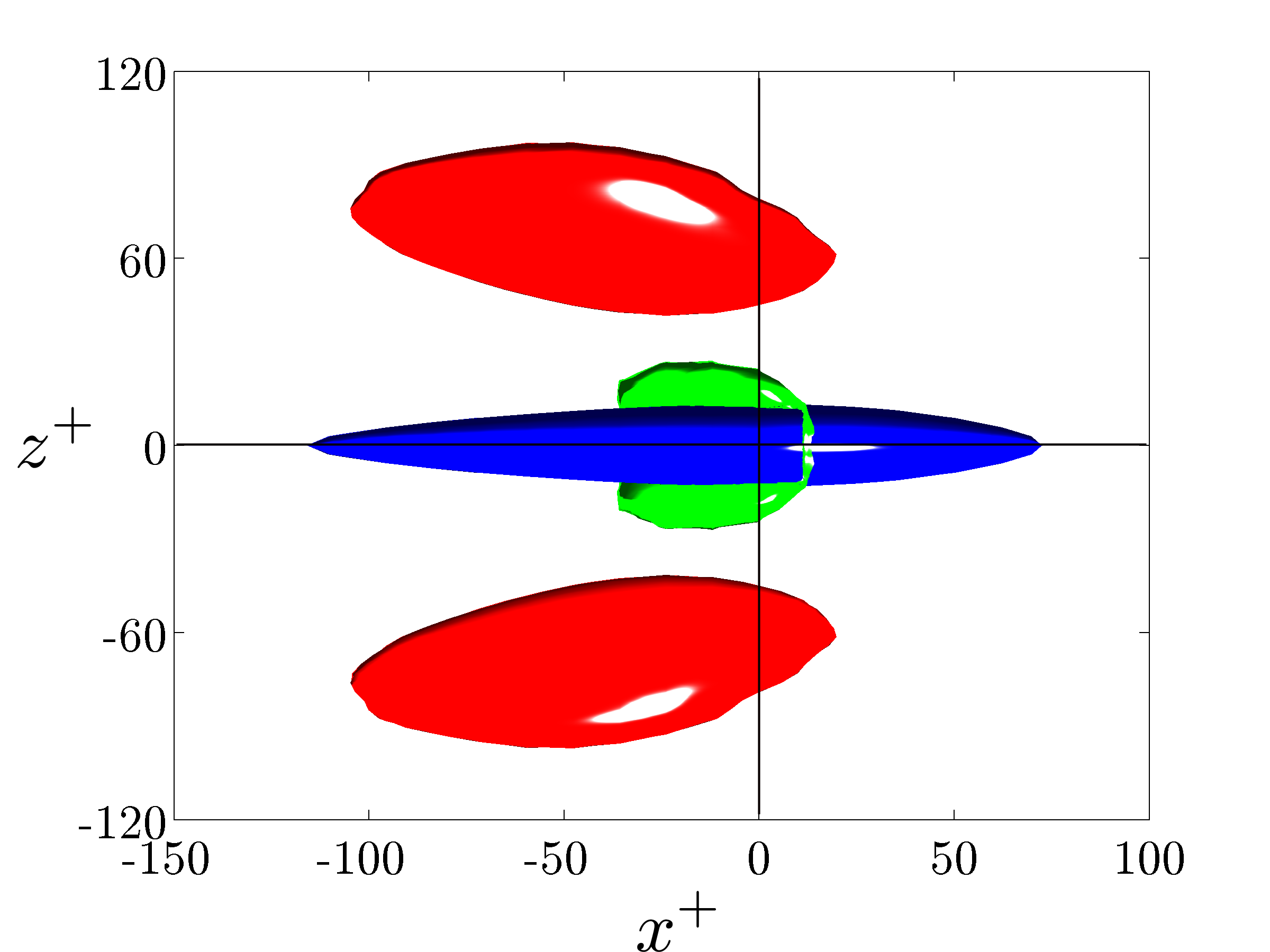}
    \label{fig.u-vortex-0-xz-R186}}
    &
    \subfigure[]{\includegraphics[height=4cm, width=0.49\columnwidth]
    {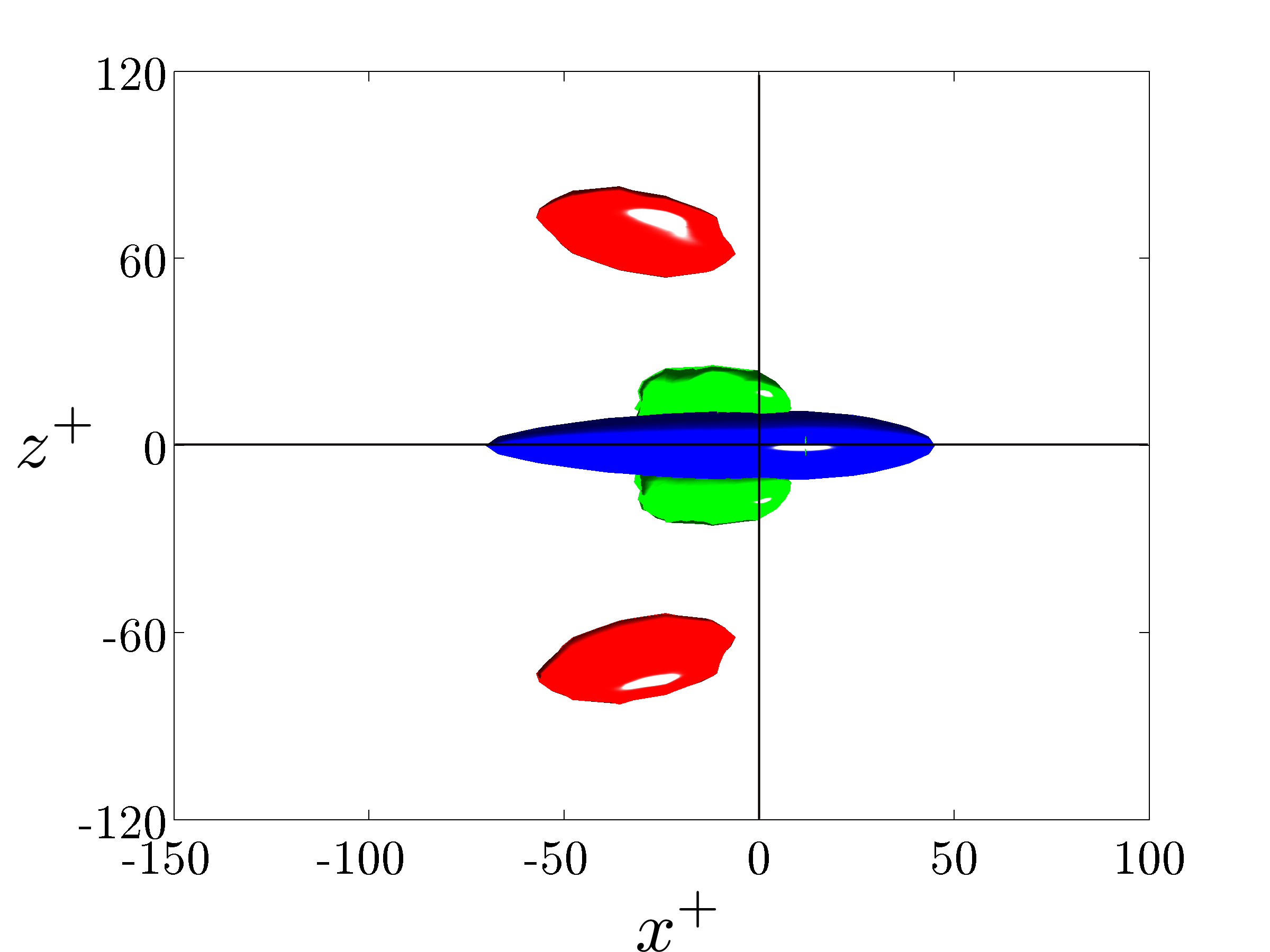}
    \label{fig.u-vortex-2-xz-R186-T102p5-a2p25}}
    \end{tabular}
    \end{center}
    \caption{
    (Color online) Three-dimensional iso-surfaces of the streamwise streaks (red and blue) and the vortex core (green) for the characteristic eddy in the uncontrolled flow (left column) and the flow subject to wall oscillations (right column) with optimal drag-reducing period $T^+ = 102.5$ at $R_\tau = 186$. (a)-(b): bird's-eye view; (c)-(d): side view; (e)-(f): front view; (g)-(h): top view. The fast- (red) and slow- (blue) moving streaks are respectively shown at $70\%$ and $60\%$ of their largest values in the uncontrolled flow, and the vortex core is obtained based on the `swirling strength' criterion~\citep{chabaladr05} with $\lambda_{ci} > 12$, $| \lambda_{cr}/\lambda_{ci} | < 0.4$, and $\lambda_{ci}/12 - |\lambda_{cr}/\lambda_{ci}|/0.4 = 1.2$.
    }
    \label{fig.u-vortex-R186-T102p5-a2p25}
    \end{figure}	

    \begin{figure}
    \begin{center}
    \begin{tabular}{cc}
    Uncontrolled
    &
    Controlled: $T^+ = 102.5$, $\alpha = 2.25$
    \\[-0.2cm]
    \subfigure[]{\includegraphics[height=4cm, width=0.49\columnwidth]
    {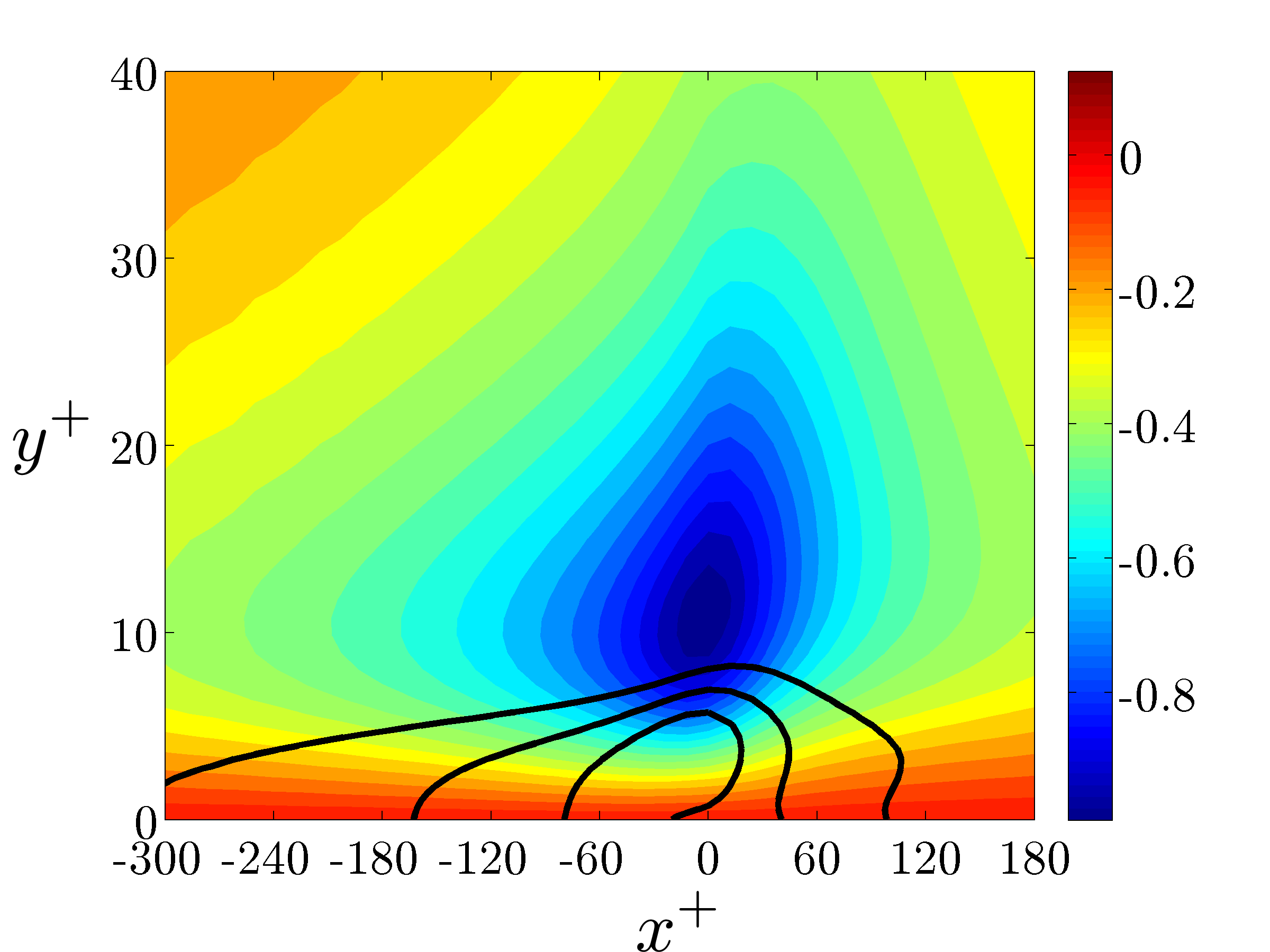}
    \label{fig.u-omega_z-0-R186-xy}}
    &
    \subfigure[]{\includegraphics[height=4cm, width=0.49\columnwidth]
    {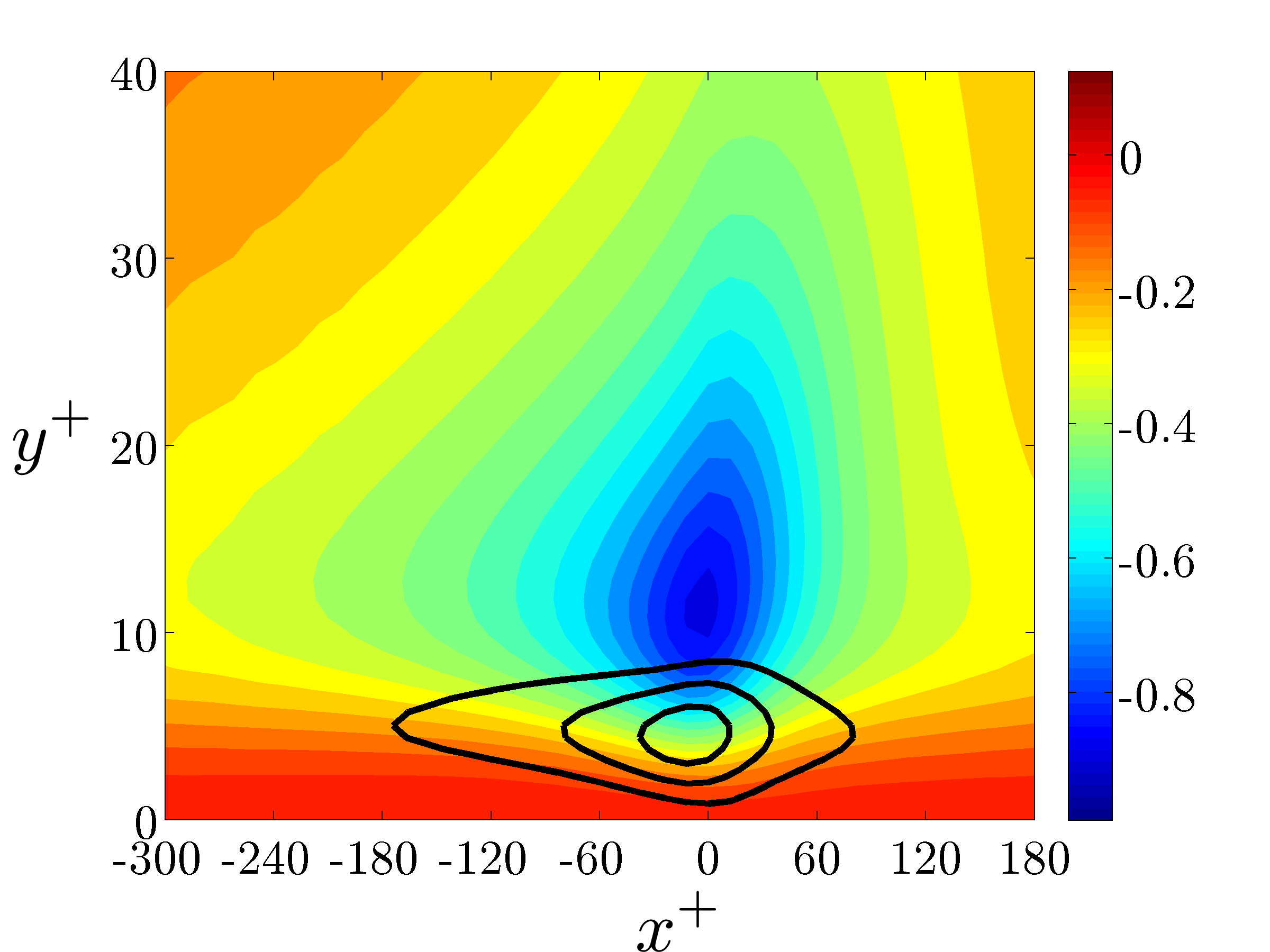}
    \label{fig.u-omega_z-2-R186-T102p5-a2p25-xy}}
    \\[0.2cm]
    \subfigure[]{\includegraphics[height=4cm, width=0.49\columnwidth]
    {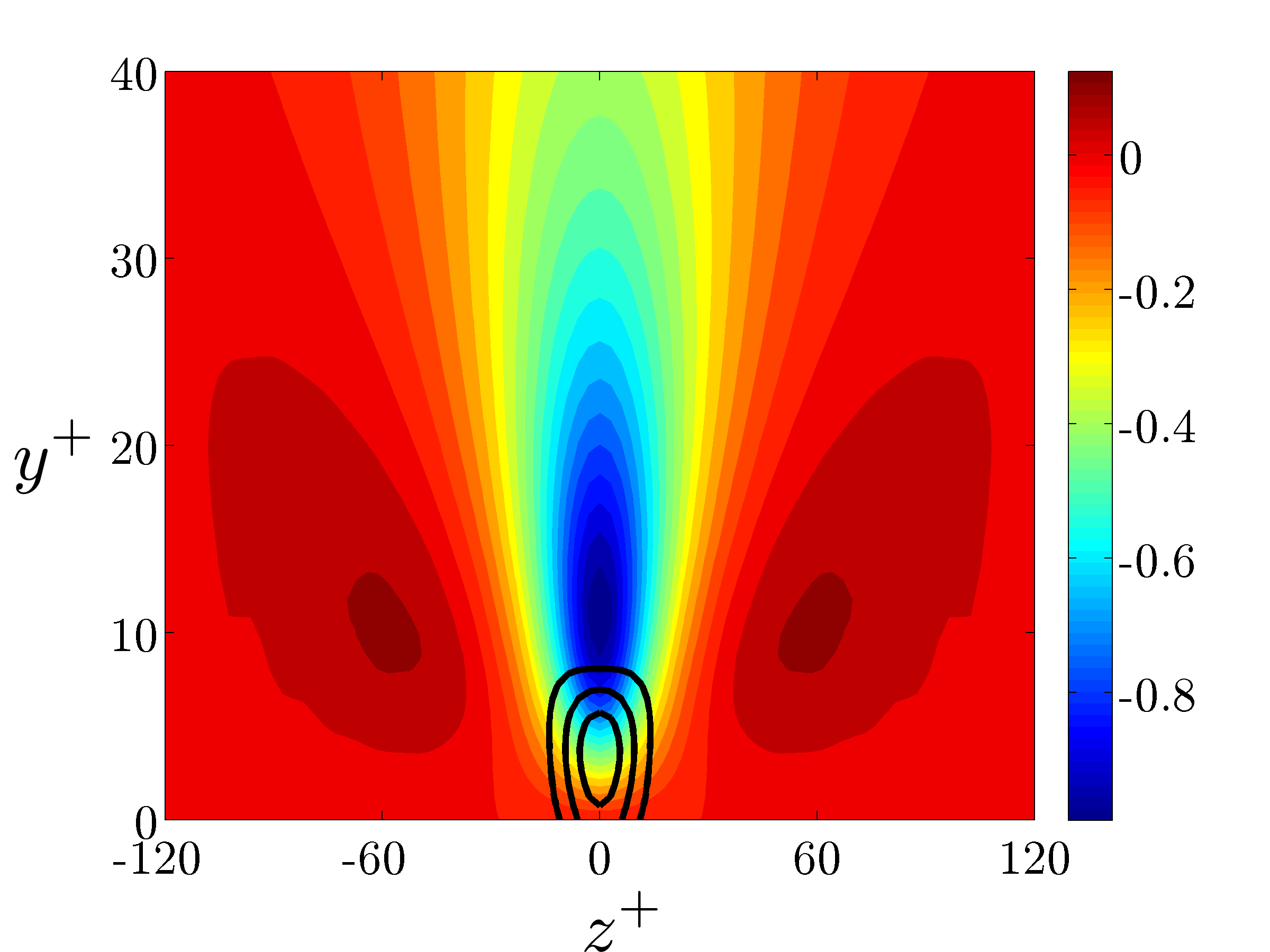}
    \label{fig.u-omega_z-0-R186-yz}}
    &
    \subfigure[]{\includegraphics[height=4cm, width=0.49\columnwidth]
    {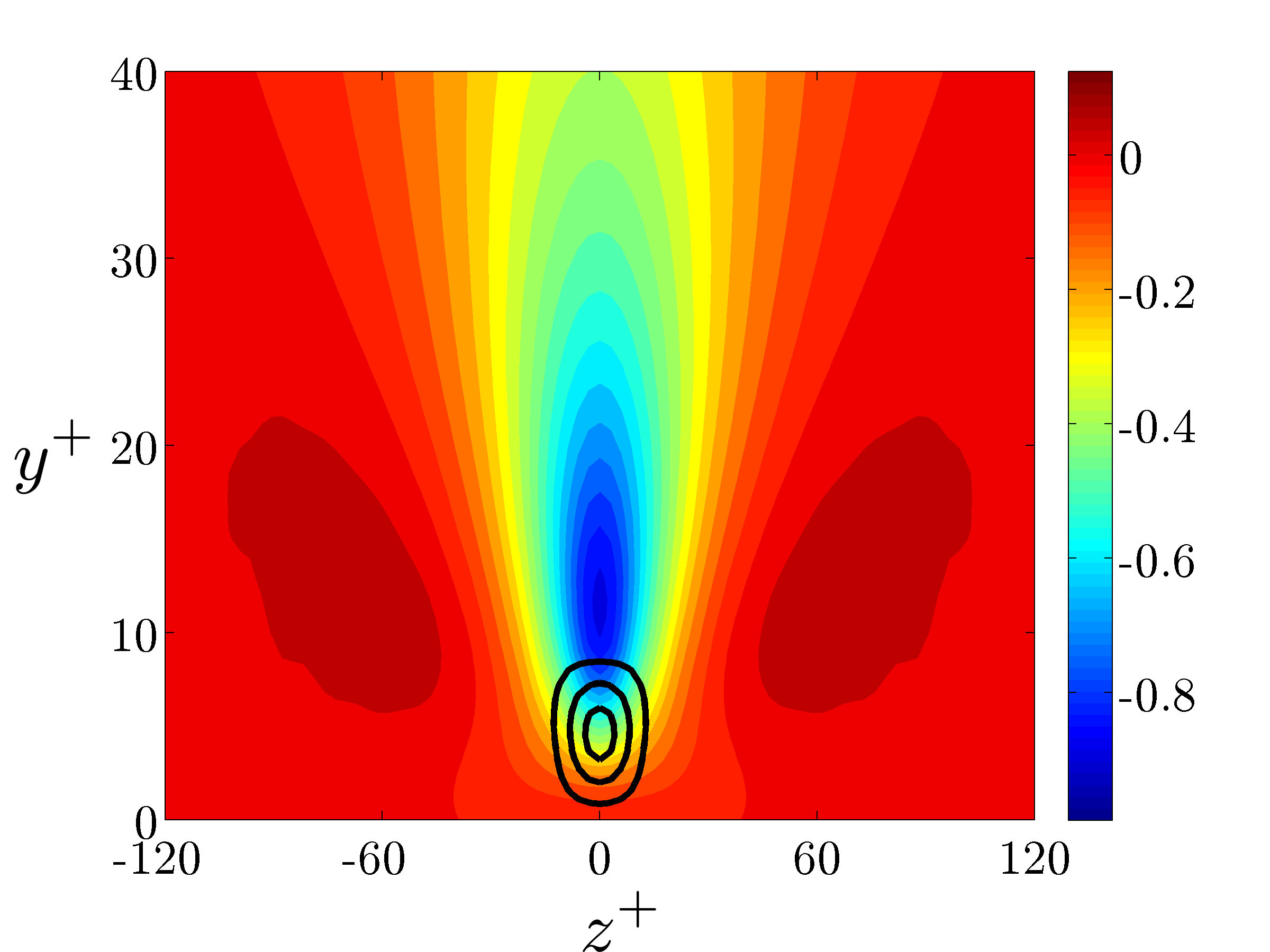}
    \label{fig.u-omega_z-2-R186-T102p5-a2p25-yz}}
    \\[0.2cm]
    \subfigure[]{\includegraphics[height=4cm, width=0.49\columnwidth]
    {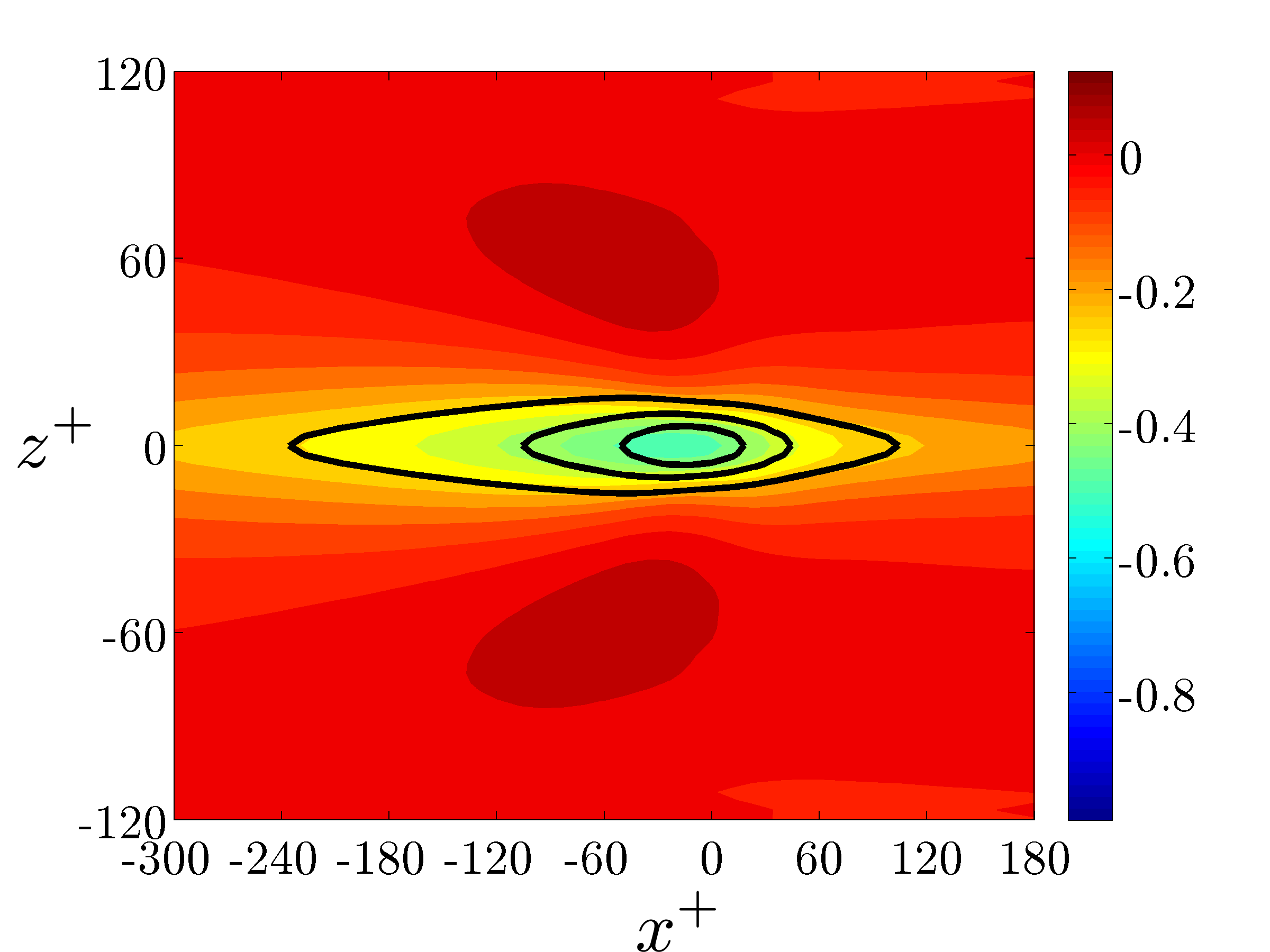}
    \label{fig.u-omega_z-0-R186-xz-yp3p8}}
    &
    \subfigure[]{\includegraphics[height=4cm, width=0.49\columnwidth]
    {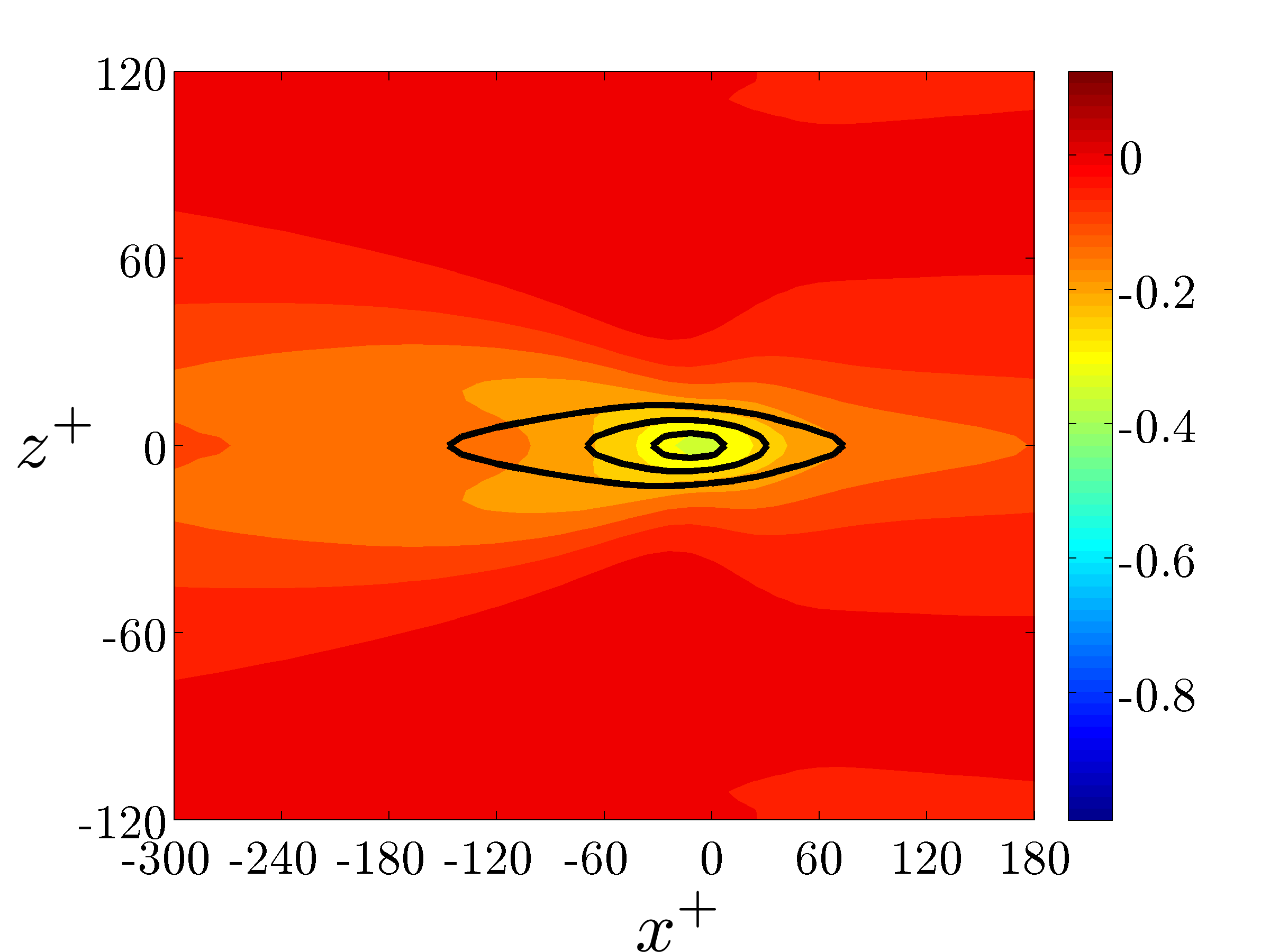}
    \label{fig.u-omega_z-2-R186-T102p5-a2p25-xz-yp3p8}}
    \\[0.2cm]
    \subfigure[]{\includegraphics[height=4cm, width=0.49\columnwidth]
    {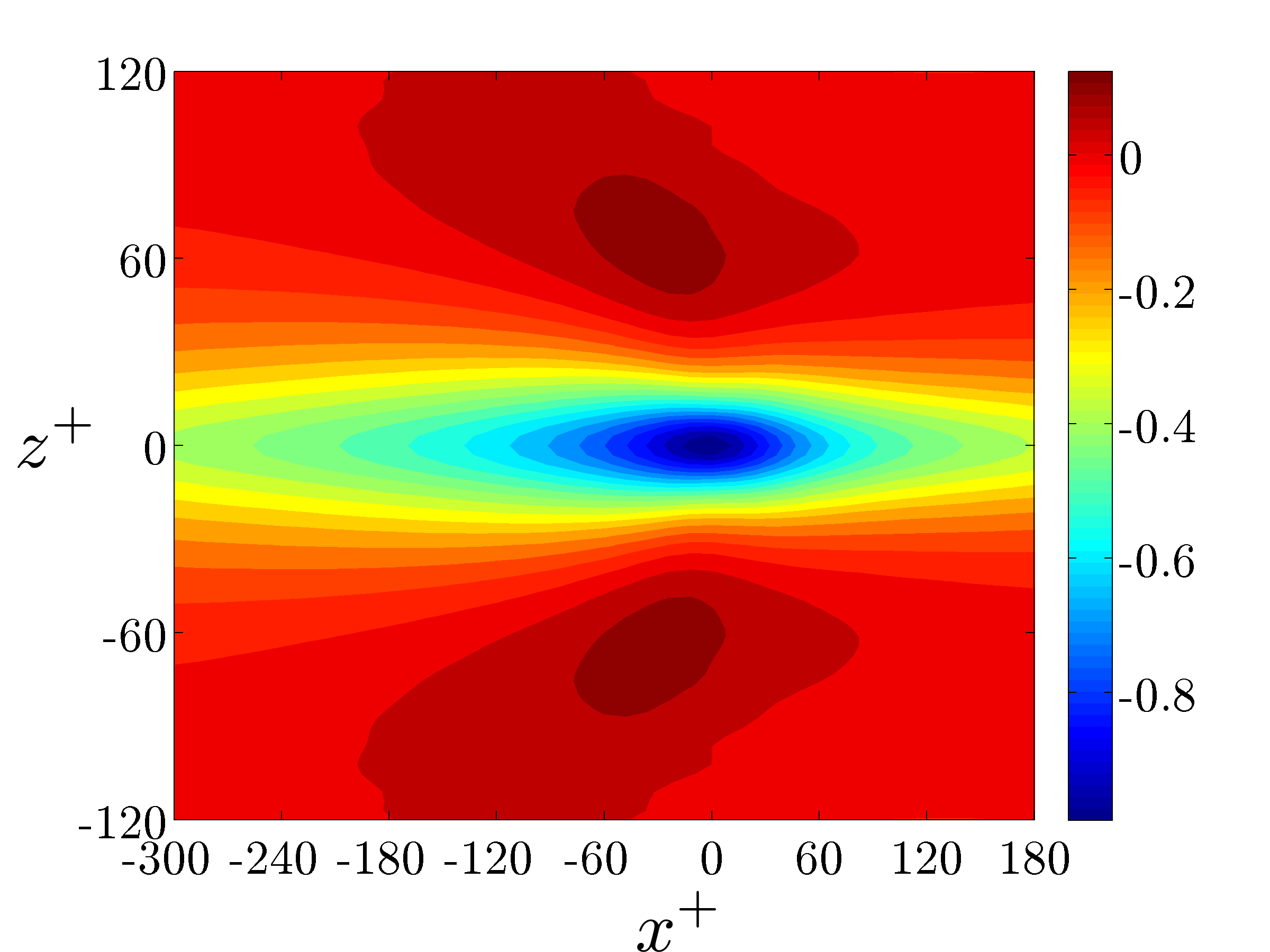}
    \label{fig.u-omega_z-0-R186-xz-yp10p8}}
    &
    \subfigure[]{\includegraphics[height=4cm, width=0.49\columnwidth]
    {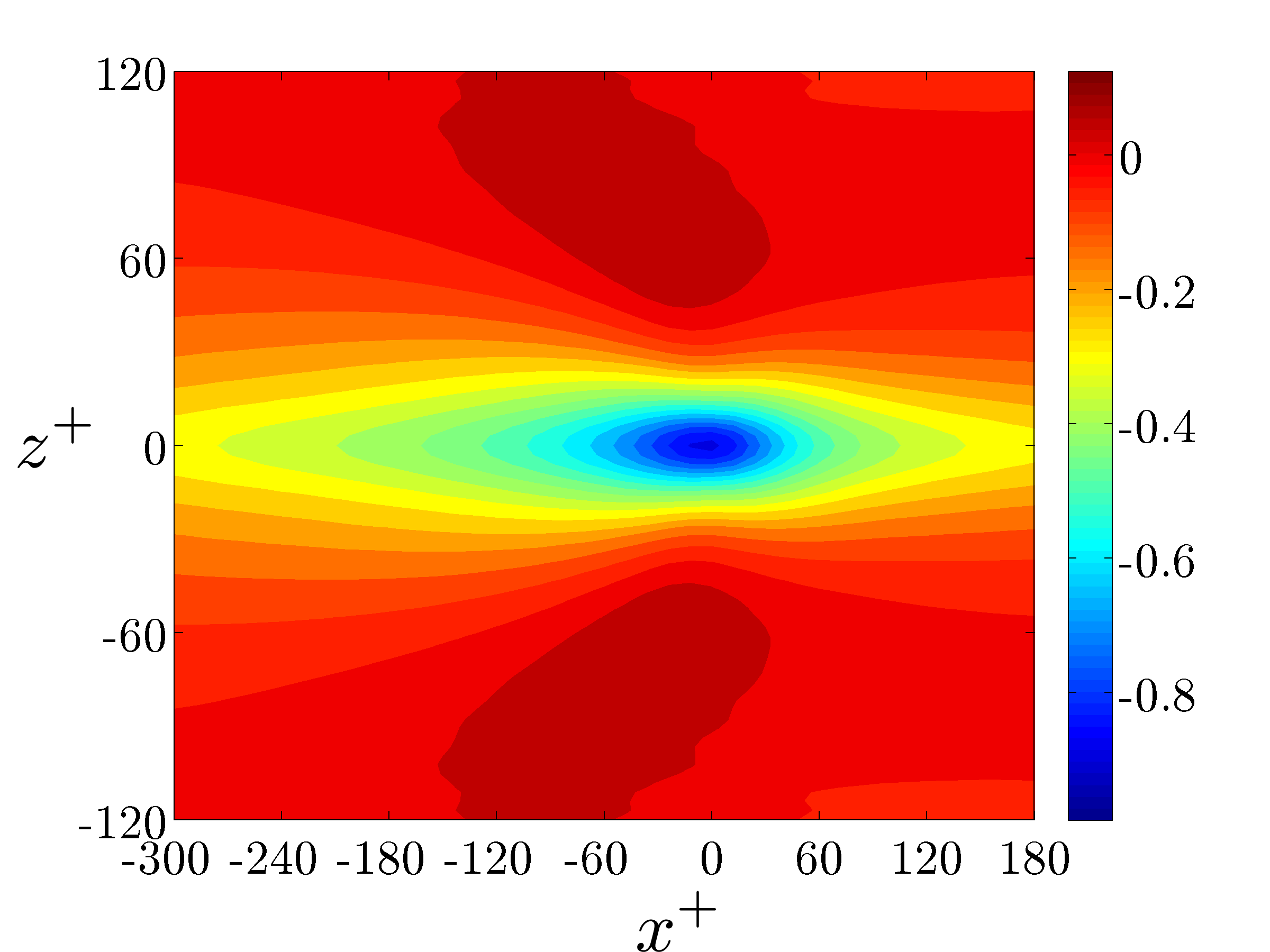}
    \label{fig.u-omega_z-2-R186-T102p5-a2p25-xz-yp10p8}}
    \end{tabular}
    \end{center}
    \caption{
    (Color online) Cross sections of the streamwise velocity (colored contours) and the spanwise vorticity (black contours) for the characteristic eddy in the uncontrolled flow (left column) and the flow subject to wall oscillations (right column) with optimal drag-reducing period $T^+ = 102.5$ at $R_\tau = 186$. (a)-(b): $z^+ = 0$; (c)-(d): $x^+ = 0$; (e)-(f): $y^+ = 3.8$; (g)-(h): $y^+ = 10.8$. The streamwise velocity is normalized by its largest absolute value in the uncontrolled flow, and the level sets for the spanwise vorticity correspond to $40\%$, $60\%$, and $80\%$ of its largest value in the uncontrolled flow.
    }
    \label{fig.u-omega_z-R186-T102p5-a2p25}
    \end{figure}

We next examine the effect of wall oscillations on the dominant characteristic eddy. Figure~\ref{fig.u-vortex-R186-T102p5-a2p25} compares the streamwise streaks and their surrounding vortex core for the dominant characteristic eddies in the uncontrolled flow (left column) and in the flow subject to wall oscillations with $T^+ = 102.5$ and $\alpha = 2.25$ (right column). In both cases, the iso-surfaces represent high- (red) and low- (blue) speed streaks at $70\%$ and $60\%$ of their largest absolute values in the uncontrolled flow, respectively. The vortex core (green surface) is obtained based on the `swirling strength' criterion which identifies motions with large rate of rotation and large orbital compactness~\citep*{chabaladr05}. This criterion can be expressed in terms of the real and imaginary parts of the complex eigenvalues, $\lambda_{cr} \pm \mri \lambda_{ci}$, of the rate of strain tensor at each point inside the channel. The large rate of rotation requires $\lambda_{ci} > b_1$, where $b_1$ determines the strength of the swirling motion. In addition, orbital compactness is guaranteed if $| \lambda_{cr}/\lambda_{ci} | < b_2$, where $b_2$ determines the compactness of the swirling motion in the plane spanned by the real and imaginary parts of the eigenvector corresponding to $\lambda_{cr} \pm \mri \lambda_{ci}$. For example, $b_2 = 0$ identifies a pure circular motion, and larger values of $b_2$ allow for inclusion of the in-plane converging or diverging spiral motions to the vortex core. In figure~\ref{fig.u-vortex-R186-T102p5-a2p25}, we use $b_1 = 12$ and $b_2 = 0.4$ to identify the strong vortex core that surrounds the slow-moving streaks. We see that wall oscillations reduce intensity and spatial spread of both the streamwise streaks and the vortex core, which is in agreement with experiments of~\citet{ric04}.

Figure~\ref{fig.u-omega_z-R186-T102p5-a2p25} shows the streamwise velocity, $u (x,y,z)$, and the spanwise vorticity, $\omega_z = v_x (x,y,z) - u_y (x,y,z)$, for the characteristic eddy in the uncontrolled flow (left column) and in the flow subject to wall oscillations with $T^+ = 102.5$ and $\alpha = 2.25$ (right column). The spanwise vortices are shown at $40\%$, $60\%$, and $80\%$ of their largest values (black contours). The streamwise streaks (colored contours) are normalized by their largest absolute value. We see that, in the viscous sublayer $y^+ < 5$, wall oscillations suppress the largest spanwise vorticity by approximately $6.5\%$. In addition, the magnitude of streamwise streaks is reduced by approximately $12\%$.~\cite{cho02} argued that transverse wall movements induce negative spanwise vorticity in the flow, thereby suppressing intensity of the streamwise streaks. In addition, wall oscillations slightly move the center of the spanwise vortices away from the wall ($y^+ = 3.8$ vs. $y^+ = 4.3$), and they significantly reduce the upstream extent of the spanwise vortices.

\section{Concluding remarks}
    \label{sec.remarks}

This paper has introduced a model-based approach to controlling turbulent flows. In contrast to standard practice that embeds turbulence models in numerical simulations, we have developed a simulation-free approach that enables computationally-efficient control design and optimization. This has been achieved by merging turbulence modeling with techniques from linear systems theory. In particular, we have used the turbulent viscosity hypothesis in conjunction with {a model equation for $\nu_T$ based on $k$ and $\epsilon$} to determine the influence of turbulent fluctuations on the mean velocity in the flow with control.

We have shown that the study of dynamics is of prime importance in designing drag-reducing wall oscillations. This has allowed us to determine the influence of control on the turbulent viscosity in a simulation-free manner. This contribution goes well beyond the problem that was used to demonstrate the predictive power of our model-based control design -- turbulent drag reduction by transverse wall oscillations. The computational complexity of determining the turbulent viscosity (in the flow with control) has been significantly reduced by obtaining $k$ and $\epsilon$ from the second-order statistics of {eddy-viscosity-enhanced} linearized model {with stochastic forcing}.

The first step in our control-oriented modeling involves augmentation of the molecular viscosity with the turbulent viscosity of the uncontrolled flow. The resulting model is then used to determine the turbulent mean velocity in the flow with control, and to study the dynamics of velocity fluctuations around it. By considering linearized equations in the presence of white-in-time stochastic forcing (whose spatial spectrum is selected to be proportional to the turbulent kinetic energy of the uncontrolled flow), we have quantified the influence of control on the second-order statistics of velocity fluctuations and thereby on the turbulent viscosity. Finally, the modifications to the turbulent viscosity determine the turbulent mean velocity and skin-friction drag in the flow with control.

Since the evolution model for flows subject to wall oscillations is time periodic, a wide-sense stationary stochastic forcing induces velocity fluctuations with cyclo-stationary statistics. Computing these statistics is challenging even in the linearized case. Motivated by the observation that large control amplitudes yield poor net efficiency, we have used perturbation analysis (in the amplitude of oscillations) to quantify the effect of control on the turbulent statistics in a computationally efficient manner.

We have shown that perturbation analysis up to second order reliably predicts the optimal period of drag-reducing oscillations. Furthermore, even though the required power obtained using the turbulent viscosity of the uncontrolled flow agrees well with the values obtained in DNS, this agreement has been further improved by accounting for the effect of control on fluctuations (and, consequently, on turbulent viscosity). In addition, the predicted net efficiency resulting from perturbation analysis qualitatively agrees with the DNS results. Perturbation analysis has also captured suppression of the turbulent kinetic energy and its rate of dissipation by wall oscillations, as well as modifications to the streamwise component of the turbulent mean velocity (reduction in the viscous sublayer and buffer layer and increase in the log-law region). Finally, the spatial spectral density tensors of velocity fluctuations obtained from the solution of the corresponding Lyapunov equations determine the effect of control on the dominant flow structures. As previously observed in experiments, and confirmed by our analysis, wall oscillations reduce the spatial spread and magnitude of the dominant characteristic eddies and suppress the spanwise vorticity in the viscous sublayer.

It is noteworthy that simple turbulence modeling (that relies on the turbulent viscosity hypothesis {with $\nu_T$ expressed in terms of $k$ and $\epsilon$}) in conjunction with {eddy-viscosity-enhanced} linearization of the flow with control has significant predictive power for capturing full-scale phenomena. Even though {this model} does not reveal all aspects of turbulent flow physics, we have shown that it is well-suited for control design and optimization. Development of more sophisticated control-oriented turbulence models may further reduce the gap between theoretical predictions and experiments/simulations. In addition, the predictive power of the proposed approach can be enhanced by optimization of the power spectrum of the forcing. We expect that our model-based approach will find use in designing feedback-based and sensor-less turbulence suppression strategies in other geometries, including pipes and boundary layers.

\section*{Acknowledgments}
    \label{sec.Ack}

Financial support from the National Science Foundation under CAREER Award CMMI-06-44793 and from the University of Minnesota Initiative for Renewable Energy and the Environment under Early Career Award RC-0014-11 is gratefully acknowledged. The University of Minnesota Supercomputing Institute is acknowledged for providing computing resources. The authors would also like to thank anonymous reviewers and the associate editor for their valuable comments.

\appendix

\section{The required power to maintain wall oscillations}
    \label{sec.Preq-A}

The wall oscillations require an input power to balance the spanwise shear stresses at the walls. The required power over one period $T$ per unit area of the channel walls is obtained from~\citep{cur03}
	\be
	\Pi_{\mathrm{req}}
	\; = \;
	\ds{
	\dfrac{1}{T}\,
	\left.
	\int_{0}^{T}\,
	(W (y,t) \, \tau_{23} (y,t))\,
	\mrd t\,
	\right|_{y \, = \, 1}\,
	}
	\,- \,
	\ds{
	\dfrac{1}{T}\,
	\left.
	\int_{0}^{T}\,
	(W (y,t) \, \tau_{23} (y,t))\,
	\mrd t\,
	\right|_{y \, = \, -1}\,
	},
	\non
	\ee
where $\tau_{23} (y,t) = \mu W' (y,t)$ denotes the spanwise shear stress and $\mu$ is viscosity. An equation for $\Pi_{\mathrm{req}}$ (normalized by $\rho u_\tau^2$) can be obtained by substituting $W$ from~(\ref{eq.U_3}) and using the boundary conditions on $W_p$ given in~(\ref{eq.U3-bar})
	\be
	\Pi_{\mathrm{req}}
	\; = \;
	\dfrac{2 \alpha^2}{R_\tau} \,
	\mbox{Im}
	\left(
	\left.
	W'_{p}
	\right|_{y \, = \, -1}
	\, - \,
	\left.
	W'_{p}
	\right|_{y \, = \, 1}
	\right).
	\non
	\ee
Relative to the power necessary for driving the uncontrolled flow, the required power is given by $\% \Pi_{\mathrm{req}} = 100 \,\Pi_{\mathrm{req}} / (2 U_B)$, which yields~(\ref{eq.delta-Preq}).

\section{Operators $A_0$, $A_{1}$, and $A_{-1}$ in~(\ref{eq.A-t})}
    \label{sec.A0-A1-Am1}

The operators $A_0$, $A_{1}$, and $A_{-1}$ in~(\ref{eq.A-t}) are obtained by substituting $W$ from~(\ref{eq.U_3}) into the expression~(\ref{eq.AC}) for $A$,
    \be
    \ba{rrl}
    A_0
    & \!\!=\!\! &
    \tbt{A_{0,11}}{0}{-\mri \kappa_{z} U'_{0}}{A_{0,22}},
    ~~
    A_{0,22}
    \; = \;
    (1/R_\tau)
    \left(
    (1+\nu_{T0})  {\Delta}
    \, + \,
    \nu'_{T0} \p_y
    \right)
    \, - \,
    \mri \kappa_{x} U_{0},
    \ea
    \non
    \ee
    \be
    \ba{rrl}
    A_{0,11}
    & \!\!=\!\! &
    {\Delta}^{-1}
    \left(
    (1/R_\tau)
    \left(
    (1+\nu_{T0})
    {\Delta}^2
    \,+\,
    2 \nu'_{T0} \Delta \p_y
    \,+\,
    \nu''_{T0} (\p_y^2+\kappa^2)
    \right)
    \,+
    \mri \kappa_{x} \big( U''_{0} - U_{0} {\Delta} \big)
    \right),
    \ea
    \non
    \ee
    \be
    \ba{rrl}
    A_{1}
    & \!\!=\!\! &
    \tbt
    {\mri \kappa_{z} \, {\Delta}^{-1} \, \left( W''_{p,0} - W_{p,0} {\Delta} \right)}
    {0}
    {\mri \kappa_{x} W'_{p,0}}
    {-\mri \kappa_{z} W_{p,0}},
    \ea
    \non
    \ee
    \be
    \ba{rrl}
    A_{-1}
    & \!\!=\!\! &
    \tbt
    {\mri \kappa_{z} \, {\Delta}^{-1} \, \left( W''_{m,0} - W_{m,0} {\Delta} \right)}
    {0}
    {\mri \kappa_{x} W'_{m,0}}
    {-\mri \kappa_{z} W_{m,0}}.
    \ea
    \non
    \ee
Here, $W_{m,0} (y) = W^*_{p,0} (y)$, and $U_0$, $W_{p,0}$, $\nu_{T0}$ and their $y$-derivatives denote multiplication operators in the wall-normal direction.

\section{Computing the velocity correlations}
\label{sec.variance}

For the time-periodic system~(\ref{eq.lnse-turb}), the normal modes are determined by Bloch waves~\citep*{odekel64,benliopap78}
	\be
	\ba{rcl}
	\fvec (y,\bkappa,t)
	& \!\! = \!\! &
	\ds{
	\sum_{n \, \in \, \bbZ}
	\fvec_{n} (y,\bkappa) \, \mre^{\mri \, (\theta + n \, \omega_t) \, t}
	},
	\\[0.4cm]
	\bpsi (y,\bkappa,t)
	& \!\! = \!\! &
	\ds{
	\sum_{n \, \in \, \bbZ}
	\bpsi_{n} (y,\bkappa) \, \mre^{\mri \, (\theta + n \, \omega_t) \, t}
	},
	\\[0.4cm]
	\bv (y,\bkappa,t)
	& \!\! = \!\! &
	\ds{
	\sum_{n \, \in \, \bbZ}
	\bv_{n} (y,\bkappa) \, \mre^{\mri \, (\theta + n \, \omega_t) \, t}
	},
	\ea
	\non
	\ee	
where $\theta \in [0,\omega_t)$ is the angular frequency. The frequency response of the time-periodic system~(\ref{eq.lnse-turb}) is an operator that maps the bi-infinite input column vector $\mbox{col} \, \{ \fvec_n \}_{n \in \bbZ}$, into the bi-infinite output column vector $\mbox{col} \, \{ \bv_n \}_{n \in \bbZ}$~\citep{jovfarAUT08,jovPOF08}. The system states can also be defined as a bi-infinite column vector $\mbox{col} \, \{ \bpsi_n \}_{n \in \bbZ}$.

As discussed in~\S~\ref{sec.compute-correlations}, $\bpsi (\,\cdot\,, \bkappa,t)$ is a cyclo-stationary process with second-order statistics given by~(\ref{eq.X-t}). The kernel representation $K_{X_r} (y,\xi,\bkappa)$ of the auto-correlation operator $X_r$ of $\bpsi (\,\cdot\,, \bkappa,t)$,
	\be
       	\left<
       	\bpsi (y,\bkappa,t) \, \bpsi^* (\xi,\bkappa,t)
       	\right>
	\; = \;
	\ds{\sum_{r \, \in \, \bbZ}}
	\;
	K_{X_r} (y,\xi,\bkappa) \, \mre^{\mri \, r \omega_t \, t},
	\non
	\ee
can be expressed in terms of $\{ \bpsi _n\}_{n \in \bbZ}$
	\be
	K_{X_r} (y,\xi,\bkappa)
	\; = \;
	\ds{\sum_{n \, \in \, \bbZ}}
	\;
	\bpsi_n (y,\bkappa) \, \bpsi_{n-r}^* (\xi,\bkappa).
	\non
	\ee
Furthermore, the frequency representation of the auto-correlation operator of $\bpsi (\,\cdot\,, \bkappa,t)$ is a self-adjoint bi-infinite block-Toeplitz operator that is parameterized by $\bkappa$
	\be
	\cX (\bkappa)
	\; = \;
	\mbox{Toep} \,
	\{
	\ldots, X^+_{2}, X^+_{1}, \fbox{$X_{0}$}, X_{1}, X_{2}, \ldots
	\},
	\non
	\ee
where the box denotes the element on the main block diagonal of $\cX$.

For the case where $\fvec$ is a zero-mean white process in $y$ and $t$ with second-order statistics given by~(\ref{eq.R}), we have $\fvec_0 = \fvec$, and $\fvec_n = 0$ for $n \neq 0$. Thus, the frequency representation of the spectrum of $\fvec$ is given by a bi-infinite block-diagonal operator $\cM (\bkappa)$ with block diagonals equal to $M (\bkappa)$. The auto-correlation operator of the state, $\cX (\bkappa)$, can be obtained from the following Lyapunov equation~\citep{jovfarAUT08,jovPOF08}
    \be
    \ba{c}
    \cF (\bkappa) \, \cX (\bkappa)
    \, + \,
    \cX (\bkappa) \, \cF^+ (\bkappa)
    \; = \;
    -\cM (\bkappa),
    \\[0.2cm]
    \cF (\bkappa)
    \; = \;
    \cA (\bkappa)
    \, - \,
    \cG (0).
    \ea
    \label{eq.X-lyap}
    \ee
Here, $\cG$ is a bi-infinite block-diagonal operator
	\be
	\cG (\theta)
	\; = \;
	\diag \, \{\mri \, (\theta \, + \, n\, \omega_t) \, I\}_{n \, \in \, \bbZ},
	\non
	\ee
and $\cA$ is a bi-infinite block-Toeplitz operator
	\be
	\cA (\bkappa)
	\; = \;
	\mbox{Toep} \,
	\{
	\ldots, 0, \alpha A_{1}, \fbox{$A_{0}$}, \alpha A_{-1}, 0, \ldots
	\}.
	\non
	\ee
The solution to~(\ref{eq.X-lyap}) can be efficiently computed using perturbation analysis in $\alpha$~\citep{jovfarAUT08,jovPOF08}. The operator $\cF$ is decomposed into a block-diagonal operator $\cF_{0}$ and an operator $\cF_{1}$ that contains the first upper and lower block sub-diagonals
	\be
	\ba{rcl}
	\cF
	& \!\! = \!\! &
	\cF_{0}
	\, + \,
	\alpha \, \cF_{1},
	\\[0.2cm]
	\cF_{0}
	& \!\! = \!\! &
	\diag \, \{ A_0 \, - \, \mri \, n \, \omega_t \, I\}_{n \, \in \, \bbZ},
	\\[0.2cm]
	\cF_{1}
	& \!\! = \!\! &
	\mbox{Toep} \,
	\{
	\ldots, 0, A_{1}, \fbox{$0$}, A_{-1}, 0, \ldots
	\}.
	\ea
	\label{eq.F-pert}
	\ee
For sufficiently small $\alpha$, the solution to~(\ref{eq.X-lyap}) can be written as~\citep{jovfarAUT08}
	\be
	\cX
	\; = \;
	\cX_{0}
	\, + \,
	\alpha \, \cX_{1}
	\, + \,
	\alpha^2 \, \cX_{2}
	\, + \,
	\alpha^3 \, \cX_{3}
	\, + \,
	\ldots.
	\label{eq.X-pert}
	\ee
Substituting~(\ref{eq.F-pert}) and~(\ref{eq.X-pert}) into~(\ref{eq.X-lyap}) and collecting equal powers of $\alpha$ yields the following set of Lyapunov equations
	\be
	\ba{lrcl}
	\alpha^0:
	&
	\cF_{0}  \, \cX_{0}
       \, + \,
       \cX_{0}  \, \cF_{0}^+
       & \!\! = \!\! &
       -\cM,
       \\[0.2cm]
       \alpha^n:
	&
	\cF_{0}  \, \cX_{n}
       \, + \,
       \cX_{n}  \, \cF_{0}^+
       & \!\! = \!\! &
       -\big(
       \cF_{1}  \, \cX_{n-1}
       \, + \,
       \cX_{n-1}  \, \cF_{1}^+
       \big).
	\ea
	\non
	\ee
Since $\cF_{0}$ is block-diagonal, $\cX_{n}$ inherits the structure of the right-hand-side of the equation at $\cO (\alpha^n)$. The structure of the above equations reveals that $\cX_{0}$ is a self-adjoint block-diagonal operator, $\cX_{1}$ is a self-adjoint block-Toeplitz operator where only the first upper and lower block sub-diagonals are non-zero, and $\cX_{2}$ is a self-adjoint block-Toeplitz operator where only the main block diagonal and the second upper and lower block sub-diagonals are non-zero
	\be
	\ba{rclcccccccc}
	\cX_{\theta0} (\bkappa)
	& \!\! = \!\! &
	\mbox{Toep} \,
	\{
	&\!\!&\!\!\ldots,&\!\! ~~0~~,&\!\! \fbox{$X_{0,0}$},&\!\! ~~0~~,&\!\! \ldots&\!\!&\!\!~~~
	\},
	\\[0.2cm]
	\cX_{\theta1} (\bkappa)
	& \!\! = \!\! &
	\mbox{Toep} \,
	\{
	&\!\!\ldots,&\!\! ~~0~~,&\!\! X^+_{1,1},&\!\! \fbox{$~~0~~$},&\!\! X_{1,1},&\!\! ~~0~~,&\!\! \ldots&\!\!~~~
	\},
	\\[0.2cm]
	\cX_{\theta2} (\bkappa)
	& \!\! = \!\! &
	\mbox{Toep} \,
	\{
	\ldots,&\!\! ~~0~~,&\!\! X^+_{2,2},&\!\! ~~0~~,&\!\! \fbox{$X_{0,2}$},&\!\! ~~0~~,&\!\! X_{2,2},&\!\! ~~0~~,&\!\! \ldots
	\}.
	\ea
	\non
	\ee
The above structure of the operator $\cX$ in conjunction with the fact that only the element on the main block diagonal of $\cX$ contributes to the averaged effect of forcing on the velocity correlations (cf.\ equation~(\ref{eq.X-mean})) reveal that, up to second order in $\alpha$, only $X_{0,0}$ and $X_{0,2}$ contribute to $X_0$
	\be
	X_0 (\bkappa)
	\; = \;
	X_{0,0} (\bkappa)
	\, + \,
	\alpha^2 \, X_{0,2} (\bkappa)
	\, + \,
	{\cal O}(\alpha^4).
	\non
	\ee
The operators $X_{0,0}$ and $X_{0,2}$ are obtained from a set of decoupled Lyapunov equations whose size is equal to the size of each block in the bi-infinite Lyapunov equation~(\ref{eq.X-lyap})~\citep{jovfarAUT08}	 \be
	\ba{rcl}
	A_{0}  \, X_{0,0}
       \, + \,
       X_{0,0}  \, A_{0}^+
       & \!\! = \!\! &
       -M,
       \\[0.2cm]
       (A_{0} + \mri \omega_t I)  \, X_{1,1}
       \, + \,
       X_{1,1}  \, A_{0}^+
       & \!\! = \!\! &
       -\big(
       A_{-1}  \, X_{0,0}
       \, + \,
       X_{0,0}  \, A_{1}^+
       \big),
       \\[0.2cm]
       A_{0}  \, X_{0,2}
       \, + \,
       X_{0,2}  \, A_{0}^+
       & \!\! = \!\! &
       -\big(
       A_{-1}  \, X_{1,1}^+
       \, + \,
       X_{1,1} \, A_{-1}^+
       \, + \,
       A_{1} \, X_{1,1}
       \, + \,
       X_{1,1}^+ \, A_{1}^+
       \big).
	\ea
	\non
	\ee
The decoupling between different harmonics of $X(\bkappa,t)$ for small $\alpha$ is used for efficient computation of the second-order statistics of the time-periodic system~(\ref{eq.lnse-turb}).
	
\section{Computing the modifications $k_2$ and $\epsilon_2$ to $k$ and $\epsilon$}
\label{sec.compute-k2-epsilon2}

We next show that the averaged effect (over one period $T$) of fluctuations around the mean velocity on $k_2$ and $\epsilon_2$ can be obtained from $X_{0,2} (\bkappa)$. Following~(\ref{eq.X0-pert}), the second-order correction (in $\alpha$) to the auto-correlation operator of velocity fluctuations $\bv$ averaged over one period $T$,
    $
    (1/T)
    \int_{0}^{T}
    \left<
    \bv (\, \cdot \,,\bkappa,t) \otimes \bv (\, \cdot \,,\bkappa,t)
    \right>
    \mrd t$,
is given by
    $
    C (\bkappa)
    X_{0,2} (\bkappa)
    C^+ (\bkappa)
    $.
The kinetic energy of fluctuations around the base flow and its rate of dissipation are given by (cf.\ equation~(\ref{eq.k-epsilon}))
	\be
	k_2 (y)
	\; = \;
	\ds{\int_{\bkappa}} \,
	K_{k} (y,y,\bkappa)
	\, \mrd \bkappa,
	~~~
	\epsilon_2 (y)
	\; = \;
	\ds{\int_{\bkappa}} \,
	K_{\epsilon} (y,y,\bkappa)
	\, \mrd \bkappa,
	\non
	\ee
where $K_{k} (y,\xi,\bkappa)$ and $K_{\epsilon} (y,\xi,\bkappa)$ are the kernel representation of the operators $N_{k}$ and $N_{\epsilon}$, respectively
	\be
	\ba{rcl}
	N_{k} (\bkappa)
	&\!\!\! = \!\!\!&
	(1/2)
	\left(
	C_u \, X_{0,2} \, C_u^+
	\, + \,
	C_v \, X_{0,2} \, C_v^+
	\, + \,
	C_w \, X_{0,2} \, C_w^+
	\right),
	\\[0.2cm]
	\!\!\!
	N_{\epsilon} (\bkappa)
	&\!\!\! = \!\!\!&
	2
	\left(
	\kappa_x^2 \, C_u \, X_{0,2} \, C_u^+
	\, + \,
	\p_y \, C_v \, X_{0,2} \, C_v^+ \, \p_y^+
	\, + \,
	\kappa_z^2 \, C_w \, X_{0,2} \, C_w^+
	\, - \,
	\right.
	\\[0.2cm]
	&&
	\left.
	\mri \kappa_x \, \p_y \, C_u \, X_{0,2} \, C_v^+
	\, + \,
	\kappa_x \kappa_z \, C_u \, X_{0,2} \, C_w^+
	\, + \,
	\mri \kappa_z \, C_v \, X_{0,2} \, C_w^+ \, \p_y^+
	\right)
	\, + \,
	\\[0.2cm]
	&\!\!\! \!\!\!&
	\p_y \, C_u \, X_{0,2} \, C_u^+ \, \p_y^+
	\, + \,
	\p_y \, C_w \, X_{0,2} \, C_w^+ \, \p_y^+
	\, + \,
	\kappa^2 \, C_v \, X_{0,2} \, C_v^+
	\, + \,
	\\[0.2cm]
	&\!\!\! \!\!\!&
	\kappa_x^2 \, C_w \, X_{0,2} \, C_w^+
	\, + \,
	\kappa_z^2 \, C_u \, X_{0,2} \, C_u^+.
	\ea
         \non
	\ee

\section{Computing the effect of fluctuations on the mean velocity and the required power in the flow with control}
\label{sec.compute-U2-tau2-W2-Psave2-Preq2}

Here, we show how the effect of fluctuations on the turbulent mean velocity is determined by the modification to the turbulent viscosity, $\nu_{T2}$, in the flow with control. The second-order correction to the mean streamwise velocity, $U_{2}$, is obtained by substituting $U$ and {$P_x$} from~(\ref{eq.U1-P1-pert}) into~(\ref{eq.turb-mean-x-z}), and collecting the terms quadratic in $\alpha$
	\be
	(1 + \nu_{T0} (y)) \, U''_{2} (y)
	\, + \,
	\nu_{T0}' (y) \, U'_{2} (y)
	\; = \;
	R_\tau {P_{x,2}}
	\, - \,
	\left(
	\nu_{T2} (y) \, U''_{0} (y)
	\, + \,
	\nu_{T2}' (y) \, U'_{0} (y)
	\right),
	\non
	\ee
which yields the following solution for $U_2$
	\be
	\ba{rcl}
	U_2 (y)
	&\!\! = \!\!&
	{-P_{x,2}} \, U_0 (y)
	\, - \,
	\ds{
	\int_{-1}^{y}
	}
	\dfrac
	{\nu_{T2} (\xi) \, U'_{0} (\xi)}
	{1\, + \, \nu_{T0} (\xi)}
	\, \mrd \xi.
	\ea
	\label{eq.U2}
	\ee
The second-order correction to {the driving pressure gradient, $P_{x,2}$}, is obtained from the requirement that the bulk flux of the flow with control remains constant, i.e.
	\be
	\ds{\int_{-1}^{1}}
	U_{2} (y) \, \mrd y
	\; = \;
	0.
	\non
	\ee	
Integrating $U_2 (y)$ in~(\ref{eq.U2}) from $-1$ to $1$, and enforcing the above requirement yields
	\be
	\ba{rcl}
	{P_{x,2}}
	&\!\! = \!\!&
	{-\dfrac{1}{2 \, U_B}
	\ds{
	\int_{-1}^{1}
	\,
	\int_{-1}^{y}
	}
	\dfrac
	{\nu_{T2} (\xi) \, U'_{0} (\xi)}
	{1\, + \, \nu_{T0} (\xi)}
	\, \mrd \xi
	\, \mrd y}.
	\ea
	\non
	\ee	
An equation for $W_{p,2}$ is determined by substituting $\nu_{T}$ from~(\ref{eq.nuT-f-pert}) into~(\ref{eq.U3-bar})	
	\be
	\ba{c}
	(1 + \nu_{T0} (y)) \, W''_{p,2} (y)
	\, + \,
	\nu_{T0}' (y) \, W'_{p,2} (y)
	\, - \,
	\mri R_\tau \, \omega_t \, W_{p,2} (y)
	\; = \;
	\\[0.15cm]
	-
	\left(
	\nu_{T2} (y) \, W''_{p,0} (y)
	\, + \,
	\nu_{T2}' (y) \, W'_{p,0} (y)
	\right).
	\ea
	\non
	\ee
Finally, the fourth-order correction to the required power, $\% \Pi_{\mathrm{req},2}$, is obtained by substituting $W_p$ from~(\ref{eq.U1-P1-pert}) into~(\ref{eq.delta-Preq})
	\be
	\ba{rcl}
	\% \Pi_{\mathrm{req},2}
	&\!\! = \!\!&
	100\, \dfrac{1}{R_\tau U_B}
	\mbox{Im}
	\left(
	\left.
	W'_{p,2}
	\right|_{y \, = \, -1}
	\, - \,
	\left.
	W'_{p,2}
	\right|_{y \, = \, 1}
	\right).
	\ea
	\non
	\ee
	
\section{{Computing the dominant characteristic eddies}}	
	\label{sec.charac-eddy}

The most energetic flow structures are obtained from the principal eigenfunctions of the averaged (over one period $T$) auto-correlation operator, $X_0 (\bkappa)$, defined in~\S~\ref{sec.compute-correlations}
	\be
	\left[ \, X_0 (\bkappa) \, \bphi (\,\cdot\,, \bkappa) \, \right] (y)
	~ = ~
	\lambda (\bkappa) \, \bphi (y, \bkappa).
	\non
	\ee
At each $\bkappa$, $\lambda (\bkappa)$ is the largest eigenvalue of $X_0 (\bkappa)$, and $\bphi (\,\cdot\,, \bkappa) = [\,v~~\eta \,]^T$ is the corresponding eigenfunction. The magnitude of $\bphi$ is determined by the requirement that the kinetic energy of fluctuations associated with $\bphi$ is equal to $\lambda (\bkappa)$. On the other hand, the phase of $\bphi$ is determined by requiring spatial compactness of $v (x,y,z)$ around $x = z = 0$ in the lower half of the channel~\citep{moimos89}. This is achieved by making sure that $\int_{-1}^{0} v (y,\bkappa) \, \mrd y$ is a positive real number for all $\bkappa$. We note that enforcing compactness on $u (x,y,z)$ yields similar results.

\begin{figure}
    \begin{center}
    \includegraphics[width=0.49\columnwidth]
    {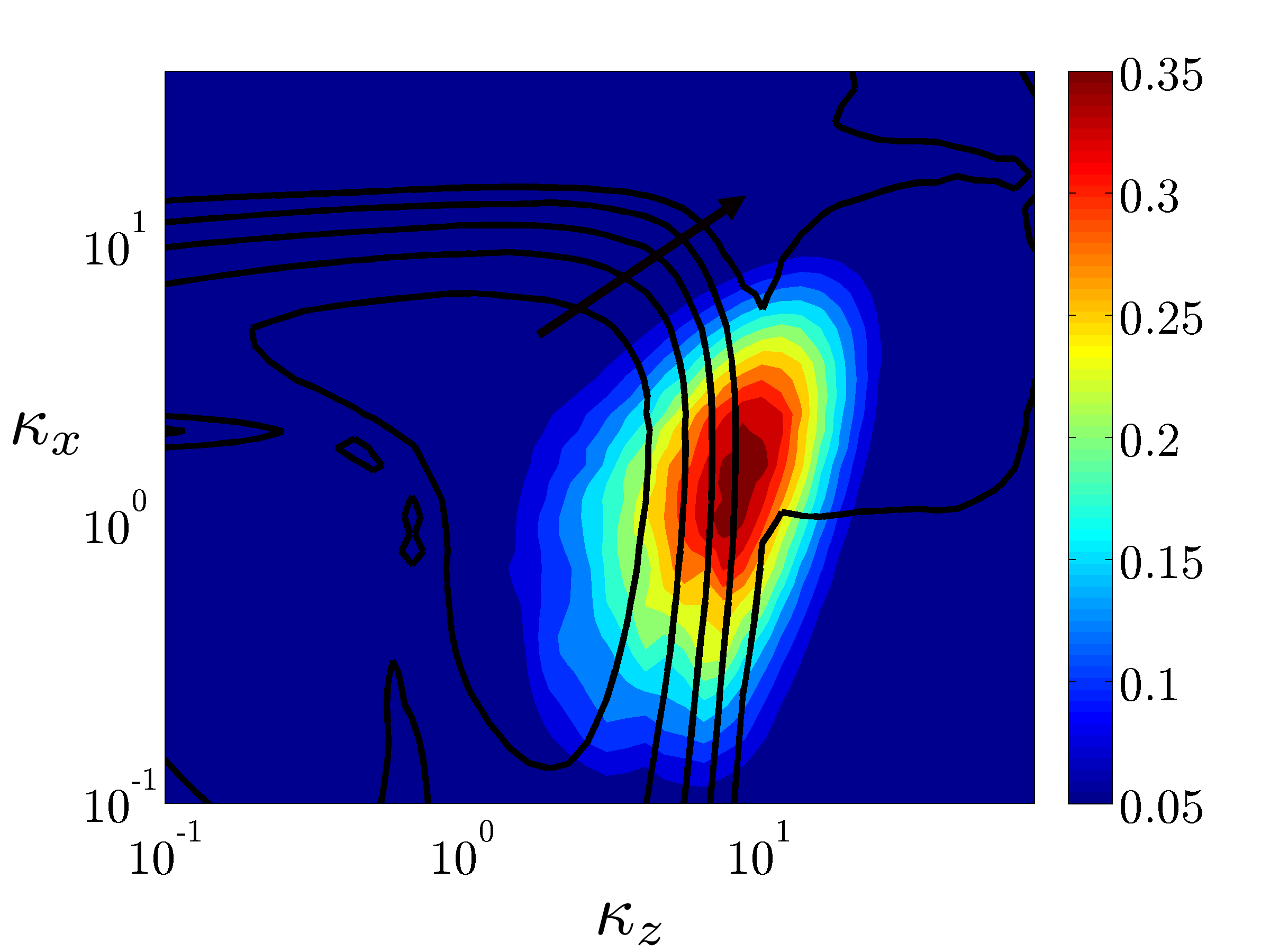}
    \end{center}
    \caption{
    {
    (Color online) Premultiplied largest eigenvalue, $\kappa_x \kappa_z \lambda (\bkappa)$, of the auto-correlation operator $X_0 (\bkappa)$ in the uncontrolled flow with $R_\tau = 186$. The contours show the relative difference between the premultiplied two largest eigenvalues of $X_0 (\bkappa)$. The contour levels $\{10^{-1}, 10^{-2}, \ldots, 10^{-5}\}$ decrease in the direction of the arrow.
    }
    }
    \label{fig.lambda1-R186}
    \end{figure}

In the uncontrolled flow, the principal eigenfunctions $\bphi (\,\cdot\,, \bkappa)$ account for approximately $29\%$ of the total kinetic energy; compare $\kappa_x \kappa_z \lambda (\bkappa)$ shown in figure~\ref{fig.lambda1-R186} with $\kappa_x \kappa_z \bar{E} (\bkappa)$ shown in figure~\ref{fig.E0-R186}. Furthermore, the two largest eigenvalues of $X_0 (\bkappa)$ are almost equal to each other. As shown in figure~\ref{fig.lambda1-R186}, the difference between them is negligible for the values of $\bkappa$ that correspond to the most energetic modes of the uncontrolled flow. In fact, the eigenfunctions corresponding to the second largest eigenvalue account for almost $28\%$ of the total energy. This indicates that examining the effects of the eigenfunction corresponding to the second largest eigenvalue is equally important. Figure~\ref{fig.efun-v-eta-1-2-kx2p5-kz6p5-R186} illustrates that the eigenfunctions corresponding to the two largest eigenvalues of $X_0$ at $\kappa_x = 2.5$ and $\kappa_z = 6.5$ are equal to each other in one half of the channel and are mirror image of each other in the other half. Imposing the $v$-compactness criterion on these two eigenfunctions aligns them in the lower half of the channel and it mirror-images them in the upper half. This implies that the flow structures that are obtained from the principal eigenfunction of $X_0 (\bkappa)$ in the lower half of the channel account for approximately $57\%$ of the total kinetic energy of fluctuations.

    \begin{figure}
    \begin{center}
    \begin{tabular}{cc}
    $v (y, \kappa_x = 2.5, \kappa_z = 6.5)$
    &
    $\eta (y, \kappa_x = 2.5, \kappa_z = 6.5)$
    \\[-0.2cm]
    \subfigure[]{\includegraphics[width=0.45\columnwidth]
    {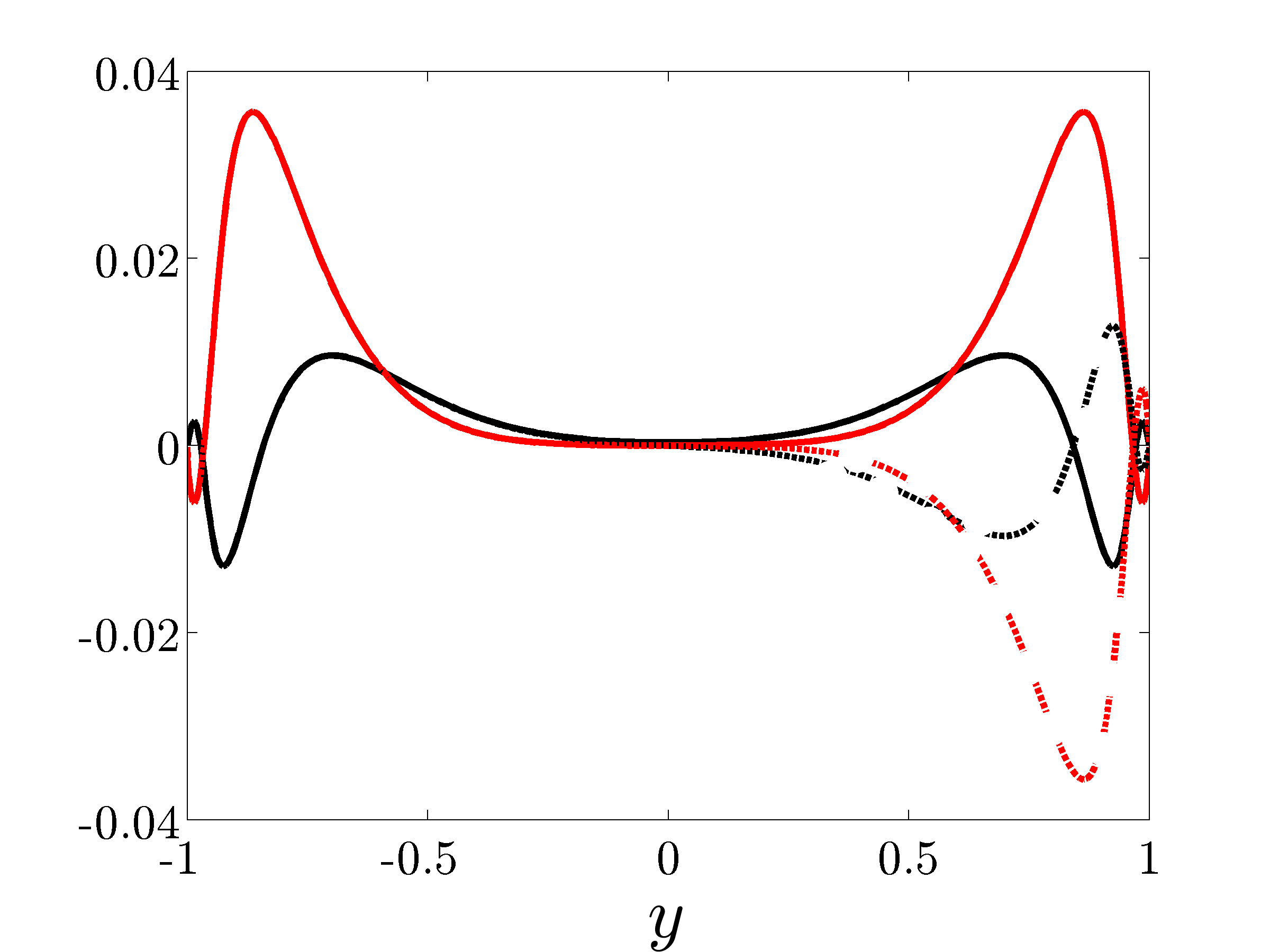}
    \label{fig.efun-v-1-2-kx2p5-kz6p5-R186}}
    &
    \subfigure[]{\includegraphics[width=0.45\columnwidth]
    {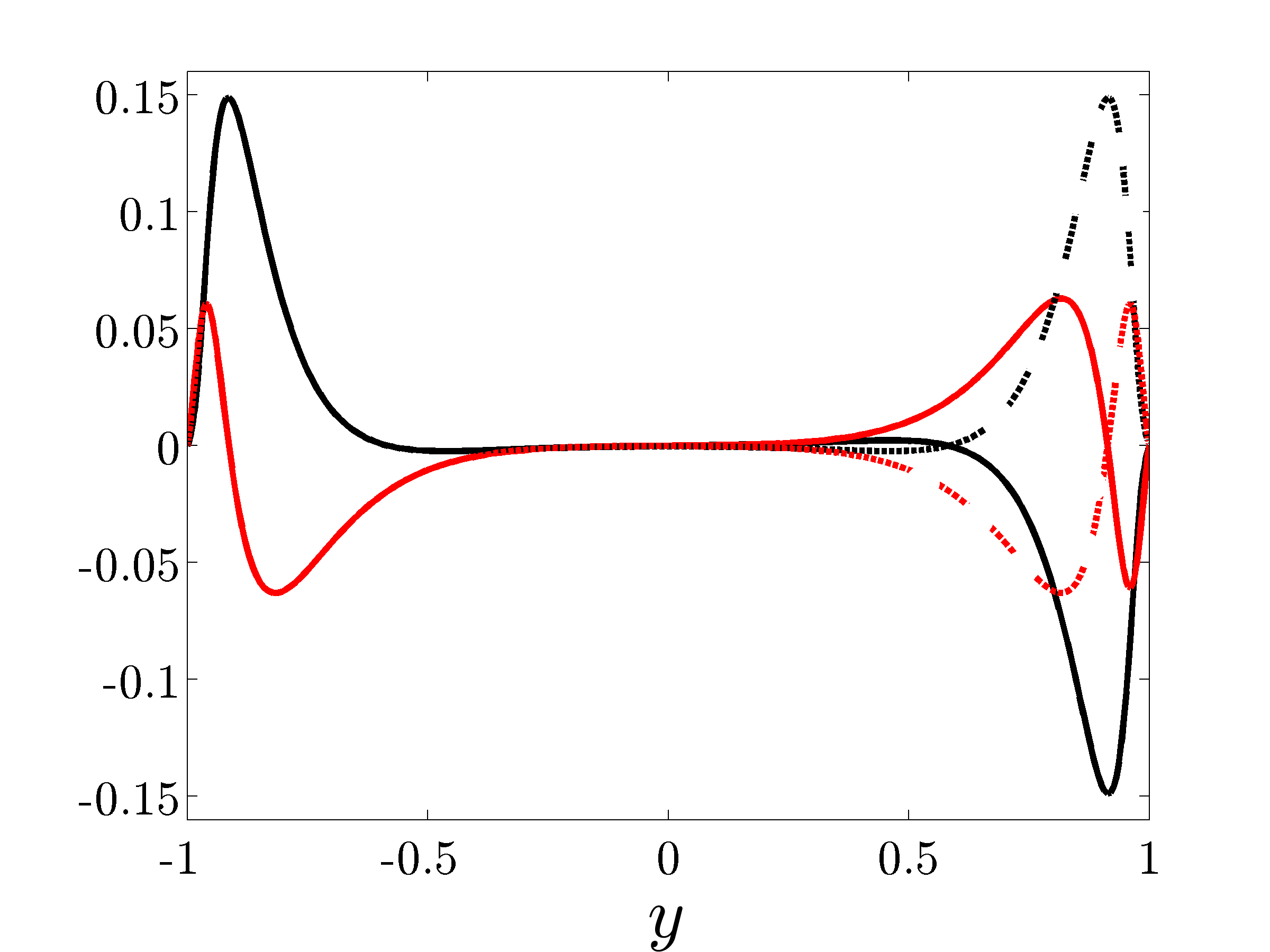}
    \label{fig.efun-eta-1-2-kx2p5-kz6p5-R186}}
    \end{tabular}
    \end{center}
    \caption{
    (Color online) The real (black) and imaginary (red) parts of the principal eigenfunctions ($v$, (a); $\eta$, (b)) corresponding to the two largest eigenvalues of the auto-correlation operator $X_0 (\bkappa)$ for the most energetic mode $\kappa_x = 2.5$ and $\kappa_z = 6.5$ in the uncontrolled flow with $R_\tau = 186$. The two eigenfunctions are differentiated by solid and dotted curves.
    }
    \label{fig.efun-v-eta-1-2-kx2p5-kz6p5-R186}
    \end{figure}

The velocity components $u$, $v$, and $w$ in the wavenumber space are obtained by acting with the operators $C_u$, $C_v$, and $C_w$ on the principal eigenfunction of the operator $X_0 (\bkappa)$; see~(\ref{eq.AC}) for the definition of these operators. We use the flow symmetries in the spanwise direction~\citep{moimos89} to determine the velocity profiles for the dominant characteristic eddy in the physical space
	\be
	\ba{rcl}
	u (x,y,z)
	& \!\! = \!\! &
	4 \ds{\int_{\kappa_x, \kappa_z \, > \, 0}}
	\mbox{Re}
	\left(
	u (y,\bkappa) \, \mre^{\mri \kappa_x x}
	\right)
	\cos (\kappa_z z)
	\, \mrd \bkappa,
	\\[0.5cm]
	v (x,y,z)
	& \!\! = \!\! &
	4 \ds{\int_{\kappa_x, \kappa_z \, > \, 0}}
	\mbox{Re}
	\left(
	v (y,\bkappa) \, \mre^{\mri \kappa_x x}
	\right)
	\cos (\kappa_z z)
	\, \mrd \bkappa,
	\\[0.5cm]
	w (x,y,z)
	& \!\! = \!\! &
	-4 \ds{\int_{\kappa_x, \kappa_z \, > \, 0}}
	\mbox{Im}
	\left(
	w (y,\bkappa) \, \mre^{\mri \kappa_x x}
	\right)
	\sin (\kappa_z z)
	\, \mrd \bkappa.
	\ea
	\non
	\ee

\end{document}